\documentclass[12pt]{article}
\usepackage{amsmath}
\usepackage{graphicx}
\usepackage{enumerate}
\usepackage{natbib}
\usepackage{url} % not crucial - just used below for the URL 

\newcommand{\blind}{0}

\usepackage{amsthm}
\usepackage{float}
\usepackage{bm}
\usepackage{enumitem}   
\usepackage[ruled,vlined,linesnumbered]{algorithm2e}
\usepackage{tikz}
\SetKwComment{Comment}{$\triangleright$\ }{}
\newtheorem{theorem}{Theorem}[section]
\theoremstyle{remark}

\usepackage{caption}

\makeatletter
\def\maketag@@@#1{\hbox{\m@th\normalfont\normalsize#1}}
\makeatother

\begin{document}

\def\spacingset#1{\renewcommand{\baselinestretch}%
{#1}\small\normalsize} \spacingset{1}

%%%%%%%%%%%%%%%%%%%%%%%%%%%%%%%%%%%%%%%%%%%%%%%%%%%%%%%%%%%%%%%%%%%%%%%%%%%%%%
% Inferring the Type of Phase Transitions Undergone in Epileptic Seizures via Network-Time-Series-Based Testing using Random Graph Hidden Markov Models for Percolation in Noisy Dynamic Networks 

\if1\blind
{
  \title{\bf Inferring the Type of Phase Transitions Undergone in Epileptic Seizures Using Random Graph Hidden Markov Models for Percolation in Noisy Dynamic Networks}
  \date{}
  \maketitle
} \fi

\if0\blind
{
  \title{\bf Inferring the Type of Phase Transitions Undergone in Epileptic Seizures Using Random Graph Hidden Markov Models for Percolation in Noisy Dynamic Networks}
  \author{Xiaojing Zhu\thanks{These authors contributed equally. Xiaojing Zhu (xiaojzhu@bu.edu), Eric D. Kolaczyk (kolaczyk@bu.edu), Mark A. Kramer (mak@bu.edu), Department of Mathematics and Statistics, Boston University. Heather Shappell (hshappel@wakehealth.edu), Department of Biostatistics and Data Science, Wake Forest School of Medicine. Catherine J. Chu (cjchu@mgh.harvard.edu), Massachusetts General Hospital, Harvard Medical School. This research was supported by ARO award W911NF1810237 and NIH award 1R01NS095369-01.
    }, 
  Heather Shappell\footnotemark[1], 
  Mark A. Kramer, \\
  Catherine J. Chu, Eric D. Kolaczyk
\hspace{.2cm}
  }
  \date{}
  \maketitle
} \fi

\begin{abstract}
In clinical neuroscience, epileptic seizures have been associated with the sudden emergence of coupled activity across the brain. The resulting functional networks – in which edges indicate strong enough coupling between brain regions – are consistent with the notion of percolation, which is a phenomenon in complex networks corresponding to the sudden emergence of a giant connected component. Traditionally, work has concentrated on noise-free percolation with a monotonic process of network growth, but real-world networks are more complex. We develop a class of random graph hidden Markov models (RG-HMMs) for characterizing percolation regimes in noisy, dynamically evolving networks in the presence of edge birth and edge death, as well as noise. This class is used to understand the type of phase transitions undergone in a seizure, and in particular, distinguishing between different percolation regimes in epileptic seizures. We develop a hypothesis testing framework for inferring putative percolation mechanisms. As a necessary precursor, we present an EM algorithm for estimating parameters from a sequence of noisy networks only observed at a longitudinal subsampling of time points. Our results suggest that different types of percolation can occur in human seizures. The type inferred may suggest tailored treatment strategies and provide new insights into the fundamental science of epilepsy. 
\end{abstract}

\noindent%
{\it Keywords: Erdos-Renyi model, Achlioptas’ process, particle filtering, data augmentation}  

%3 to 6 keywords, that do not appear in the title
\vfill

\newpage
\spacingset{1.5} % DON'T change the spacing!
\section{Introduction}
Epilepsy is a common neurological syndrome, affecting over 70 million people worldwide, and a major burden with respect to quality of life, morbidity, and risk of premature mortality (\cite{thijs2019epilepsy}). It is well known now that synchronization of neurons (and on the macro-scale - groups of neurons that make up larger brain regions) play a pivotal role in maintaining normal brain function. The abnormal synchronization of neurons, however, is a defining characteristic of some neurological disorders, such as epilepsy. 

In recent years, it has become increasingly clear that epilepsy and seizures result not only from isolated brain areas, but from networks of interacting brain regions 
%(Kramer & Cash (2012)). 
(\cite{kramer2012epilepsy}).
More specifically, epileptic seizures have been associated with the sudden emergence of coupled/synchronized activity across the brain (\cite{guye2006role, ponten2007small, schindler2007assessing, schindler2007increasing, schindler2010peri, kramer2010coalescence, martinet2020robust}). This synchronized activity may be defined via functional brain networks, where the nodes of the network represent distinct brain regions, and the edges between the nodes indicate strong coupling of brain activity (\cite{rubinov2010complex}). Moreover, the emergence of coupled activity (or increased network edges), aligns with the notion of \textit{percolation} – the sudden emergence of a giant connected component (GCC) in a network (\cite[Chapter~3]{grimmett2018probability}). 

In percolation theory, the behavior of the GCC in a network is studied as a function of its evolution over time. Two popular models of percolation are the Erdos-Renyi (ER) model (\cite{erdHos1960evolution}) and the Achlioptas’ process (\cite{achlioptas2009explosive}). They are both random graph models assuming single-edge change over time, but they differ in the choice of which edge changes. The choice of edge addition in the ER model is uniform over all non-edges, while that in the Achlioptas’ process model is based on a ‘product rule’ (PR), which slows down the growth of the GCC by favoring the creation of edges between smaller connected components. Percolation in the ER model is considered a classical archetype, while that in the PR process is of a more rapid form. As such, these two percolation models represent prototypical extremes.

Figure 1 depicts an example of functional brain network behavior during seizure onset. As can be seen in the figure, the size of the largest connected component of the brain network increases dramatically following the clinically determined onset of the epileptic seizure.  The behavior of the curve in this figure is qualitatively similar to that of a percolation curve, in which a network is undergoing a transition from a large collection of small networks to a single large connected network.

 %Percolation in the ER model constitutes one of the first examples of a fully characterized mathematical phase transition (\cite{erdHos1960evolution, alon2004probabilistic}). While the ER model of percolation is an example of a (second order) continuous phase transition, recent efforts have focused on identifying the conditions under which a random network process can yield a (first order) discontinuous percolation or approximations thereof (\cite{riordan2011explosive}).  One of the most popular attempts to model discontinuous percolation has been the Achlioptas’ process and its variants (\cite{achlioptas2009explosive}).  Although this particular  percolation  model  has  been  shown  to  be,  in  fact,  continuous  and  therefore  of second order (\cite{da2010explosive, riordan2011explosive, riordan2012achlioptas}); it nonetheless provides an interesting – and now relatively standard – alternative to the ER model.% Percolation in the ER model is considered a classical archetype, while that in the PR process is of a more rapid form. As such, these two percolation models represent prototypical extremes.

Increased interest in network percolation has recently been fueled by its relevance to epileptic seizures. While prior work has shown an explosive density increase (i.e. more edges) in functional connectivity networks in epilepsy patients during seizure onset, aligning with the notion of percolation, our work delves deeper to provide methods to uncover the underlying network evolution behavior behind the density increase. We aim to answer the question:  How can we distinguish between different percolation regimes in practice? Understanding such phenomena, corresponding with the transition between normal brain function, seizure propagation, and seizure termination, may be critical for both the fundamental science of epilepsy and developing improved strategies for treatment.
%, improving treatment, and providing early warning signs. 

In this paper, we propose a framework to distinguish between different types of phase transitions undergone in a seizure, and in particular, to distinguish between different percolation regimes. Traditionally, work in this area has concentrated on noise-free percolation with a monotonic process of network growth, but real-world networks are more complex.  It is more realistic to consider both edge creation and dissolution in network evolution.  Furthermore, we should expect observed networks to be contaminated by noise, such as measurement error. The presence of edge death and noise significantly confounds the distinction between percolation regimes and makes the two percolation models indistinguishable using heuristic statistics, e.g. the size of the largest component, as shown in Figure 2.  Therefore, a framework for the statistical testing of competing hypotheses of percolation regimes, under these conditions, is needed.

We develop a class of random graph hidden Markov models (RG-HMMs) and the necessary inferential methodologies, for characterizing percolation regimes in noisy, dynamically evolving networks in the presence of edge birth and edge death, as well as noise. Our model class builds on the framework proposed (with only preliminary inferential machinary) by \cite{viles2016percolation}, where the nonstationary process characterized by birth and death of edges was modeled in a hidden/latent layer in discrete time, assuming the true underlying networks evolve by a single edge change per time step.  We extend the model to the continuous-time setting where the process may stay in different states for differing (continuous) amounts of time. This is critical for making the framework applicable to epilepsy, as well as other real-world contexts, in which, even if observed at regular intervals, a dynamically evolving network can almost never practically be observed at the resolution of changes in individual edge status.

The remainder of this paper is organized as follows. In Section 2, we provide model definitions for continuous-time percolation models and our HMM set-up. An expectation-maximization (EM) algorithm for obtaining maximum likelihood (ML) estimates of the model parameters, along with its asymptotic properties are described in Section 3. Section 4 outlines a statistical testing framework for competing percolation regimes, and Section 5 reports simulation results for our estimation and testing algorithms. An application for the proposed framework to real epileptic seizure data is given in Section 6, followed by a discussion in Section 7.

\begin{figure}[!htb]
    \centering
    \begin{minipage}{0.45\textwidth}
        \centering
        \includegraphics[width=0.9\textwidth]{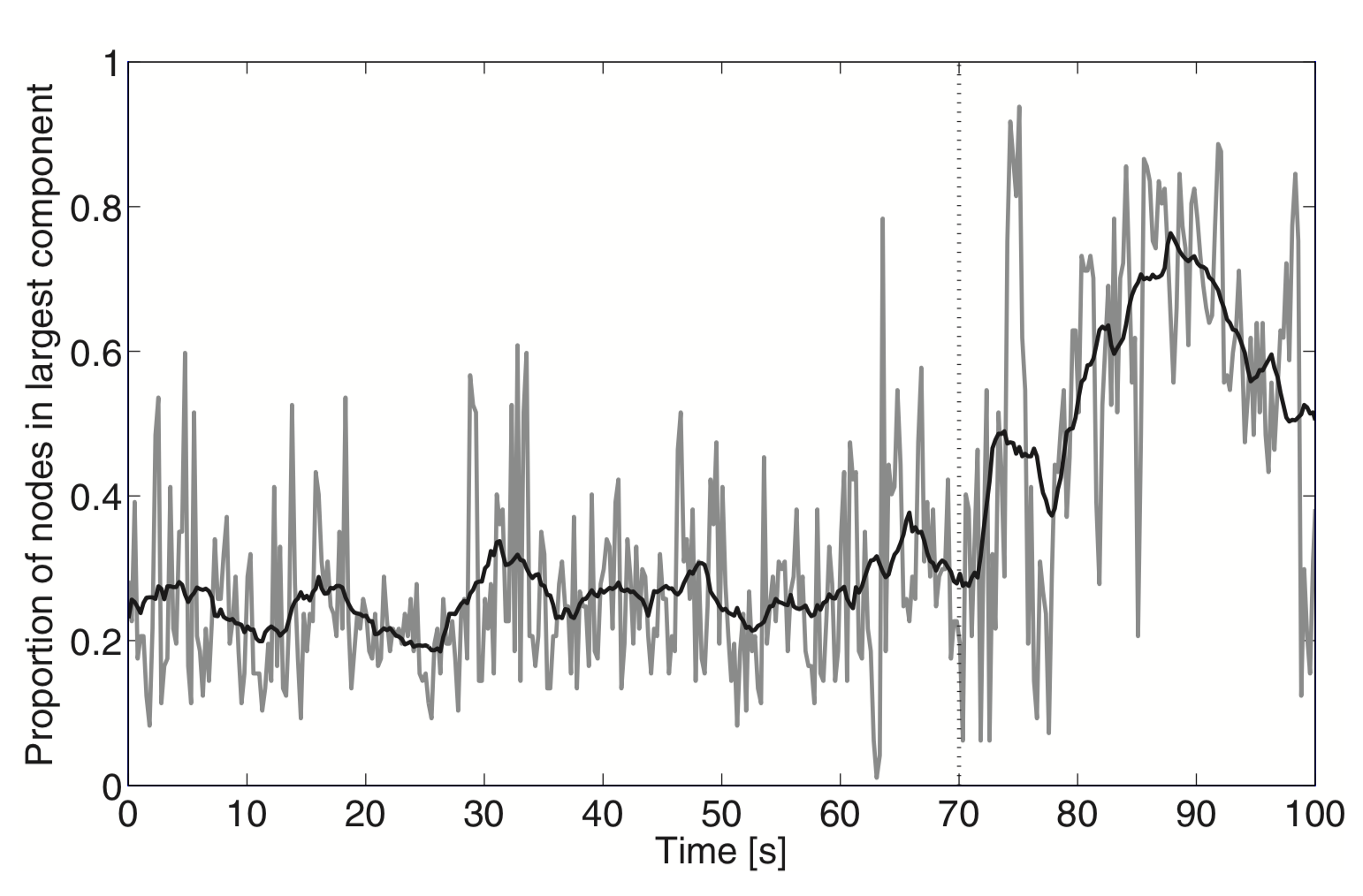} % first figure itself
        \caption{Proportion of nodes in the largest component as a function of time for a functional connectivity network deduced from the electrocorticogram of a single patient with epilepsy during a seizure. The black trace is a smoothed version of this process. Dotted vertical line indicates seizure onset. Source: \cite{viles2016percolation}}
        \label{fig-1-viles}
    \end{minipage}\hfill
    \begin{minipage}{0.45\textwidth}
        \centering
        \includegraphics[width=0.9\textwidth]{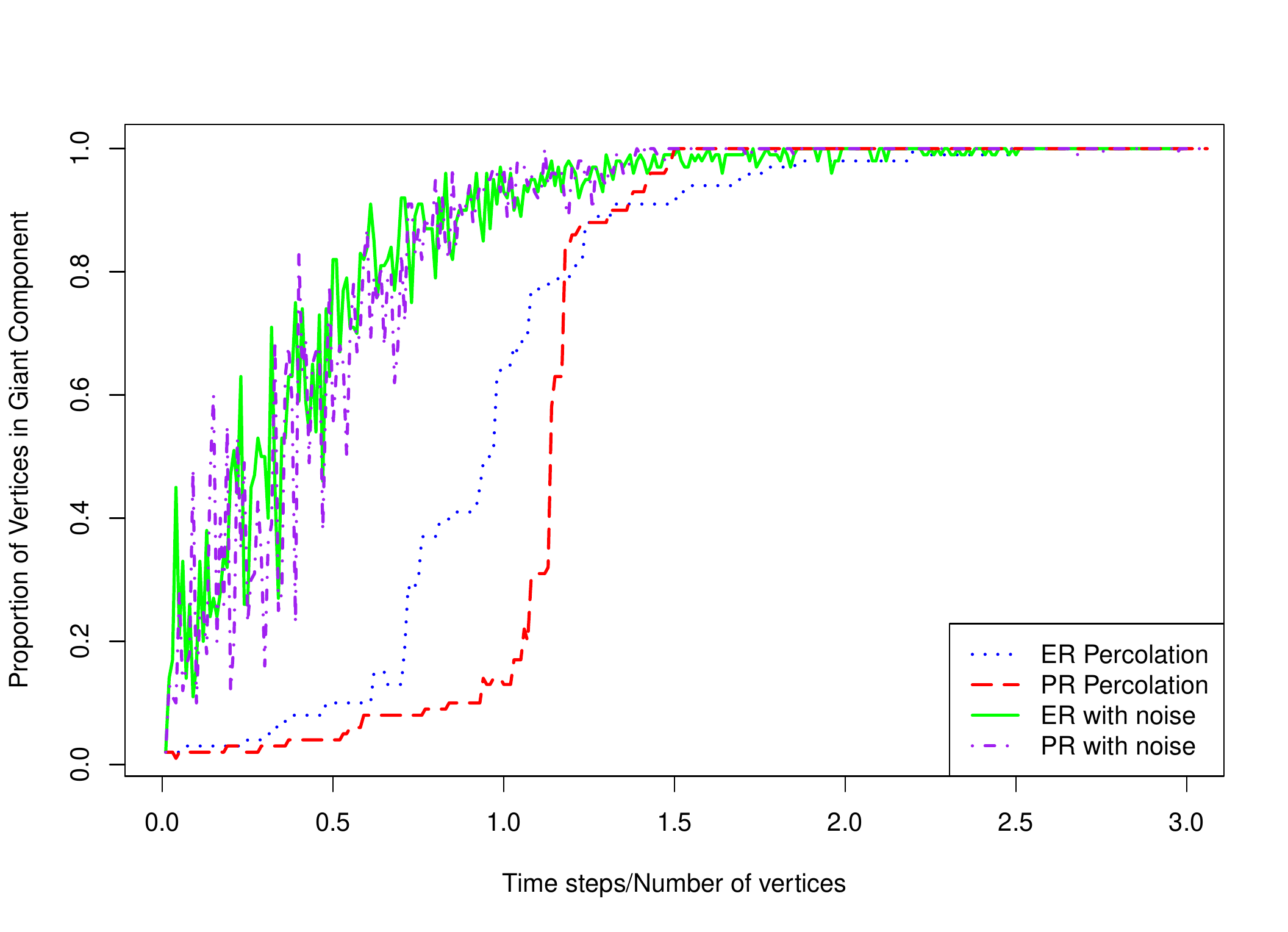} % second figure itself
        \caption{Proportion of nodes in the largest component as a function of scaled time for ER and PR birth and death process with and without noise on network with 100 vertices. Observe that the processes are virtually indistinguishable with noise.}
        \label{fig-2-ERPR-noise}
    \end{minipage}
\end{figure}

\section{Model Definition}
We first introduce two continuous-time percolation models allowing for both birth and death of edges: an Erdos-Renyi (ER) process and a product rule (PR) process, which build off of the work of 
%Viles et al.
\cite{viles2016percolation} on discrete-time percolation processes. Both models represent  continuous-time network evolution as the result of many one-edge changes over time. We then overlay the hidden Markov model framework on top of the percolation models to capture the presence of noise in the observations, which gives us the random graph hidden Markov model (RG-HMM). Throughout the paper, we use capitals to denote random variables. Additionally, we only consider networks with a fixed vertex set.

\subsection{Continuous-time Percolation Models}

%\subsubsection*{Continuous-time birth and death Erdos-Renyi (ER) process}
The \textbf{\textit{continuous-time birth and death Erdos-Renyi (ER) process}} is a network-valued bivariate continuous-time Markov chain $\{ W(t), G(t), t \geq 0 \}$ taking on values in a finite set $\mathcal{X}$. $G(t)$ is an undirected-graph variable with $N$ nodes. 
%and can be represented as a $N \times N$ adjacency matrix, of which the elements denote whether there is an edge between node $i$ and node $j$ ($G_{ij}(t) = 1$) or not ($G_{ij}(t) = 0$). 
The state space of this network-valued variable is the space of all simple graphs on $N$ nodes, which is of cardinality $2^{N \choose 2}$. $W(t)$ is a binary variable with state space $\{0, 1\}$, indicating a single edge is either added ($W(t)=1$) or deleted ($W(t) = 0$) to produce $G(t)$ at the most recent transition time. This process has the following properties:
\begin{enumerate}[label=(\roman*)]
\item
at each transition time $\tau_1$ with state $\{W(\tau_1) = w, G(\tau_1) = g\}$, the amount of time it spends in that state before making a transition into a different state is exponentially distributed with rate parameter $\lambda > 0$,

\item 
for two consecutive transition times $\tau_1< \tau_2$, the network $G(\tau_2)$ differs from $G(\tau_1)$ by a single edge, and the edge is chosen to be added or deleted uniformly at random on non-edge set or edge set of $G(\tau_1)$, the choice of which depends on the binary variable $W(\tau_2)$. The transition matrix for the binary variable, when $G(\tau_1)$ is neither empty nor complete, is shown in Table \ref{tab1}. $p$ and $q$ are respectively birth and death rates, and are not required to sum to $1$.  When $G(\tau_1)$ is a complete graph, $W(\tau_2) = 0$ with probability 1 (i.e. the growth of edges are not possible since all possible edges exist). When $G(\tau_1)$ is an empty graph, $W(\tau_2) = 1$ with probability 1 (i.e. the death of edges are not possible since none exist).
\end{enumerate}

\begin{table}[!htb]
\centering
\begin{tabular}{lll}
\hline
 &  $W(\tau_2)=0$&  $W(\tau_2)=1$\\\hline
$W(\tau_1)=0$ &  $1-p$ &  $p$\\
$W(\tau_1)=1$ &  $q$ & $1-q$\\\hline
\end{tabular}
\caption{Transition probability matrix for $W(t)$ at two consecutive transition time $(\tau_1, \tau_2)$, $\tau_1 < \tau_2$, when $G(\tau_1)$ is neither empty nor complete. }
\label{tab1}
\end{table}

%%%%%%PR%%%%%%%%%%%%%%
The \textbf{\textit{Continuous-time birth and death product rule (PR) process}} is analogous to the aforementioned birth and death ER process, except for the choice of which edge is added or deleted at each transition time. In the case of the ER model, such choice is uniform over current non-edge set or edge set. However, in the case of the PR model, the choice depends on the modular structure of the current network. 

Let $E_t$ denote the edge set of current network $G(t)$, and $E_t^C$ the non-edge set. Suppose that there exist $m_t = \big \rvert E_t\big \rvert$ edges for network $G(t)$ at time $t$. For the upcoming transition time $\tau$, if $W(\tau) = 1$, the choice of which edge to add is done in the following manner.
\begin{enumerate}[label=(\roman*)]
    \item Uniformly choose two candidate vertex pairs among all edges in $E_t^C$ (i.e. among all non-edges). Denote the two vertex pairs by $e_1 = (v_{11}, v_{12})$ and $e_2 = (v_{21}, v_{22})$.
    \item 
    Evaluate the size of the connected components to which $v_{11}, v_{12}, v_{21}, v_{22}$ belong, and denote them by $C_{11}, C_{12}, C_{21}, C_{22}$, respectively.
    \item 
    Apply the following product rule in %Achlioptas et al.
    \cite{achlioptas2009explosive}: If $\big \rvert C_{11}\big \rvert\big \rvert C_{12}\big \rvert < \big \rvert C_{21}\big \rvert\big \rvert C_{22}\big \rvert$, then add edge $e_1$. Otherwise, add edge $e_2$.
\end{enumerate}
The death of the an edge is handled in an analogous manner. If $W(\tau) = 0$, we uniformly choose two candidate vertex pairs among all edges in $E_t$, and similarly evaluate the size of the connected components to which $v_{11}, v_{12}, v_{21}, v_{22}$ would belong if their edges were absent. If $\big \rvert C_{11}\big \rvert\big \rvert C_{12}\big \rvert < \big \rvert C_{21}\big \rvert\big \rvert C_{22}\big \rvert$, then delete edge $e_2$. Otherwise, delete edge $e_1$. 
%Additional details can be found in 
%Viles et al. 
%\cite{viles2016percolation}.
Note that the Achlioptas product rule slows down the growth of the GCC by favoring the creation of edges between small connected components. 

%Both aforementioned processes represent the continuous-time network evolution as the result of many small one-edge changes occurring between consecutive transition times. 

\subsection{Random Graph Hidden Markov Model (RG-HMM)}
\begin{figure}[!htb]
    \centering
    \includegraphics[width=0.75\textwidth]{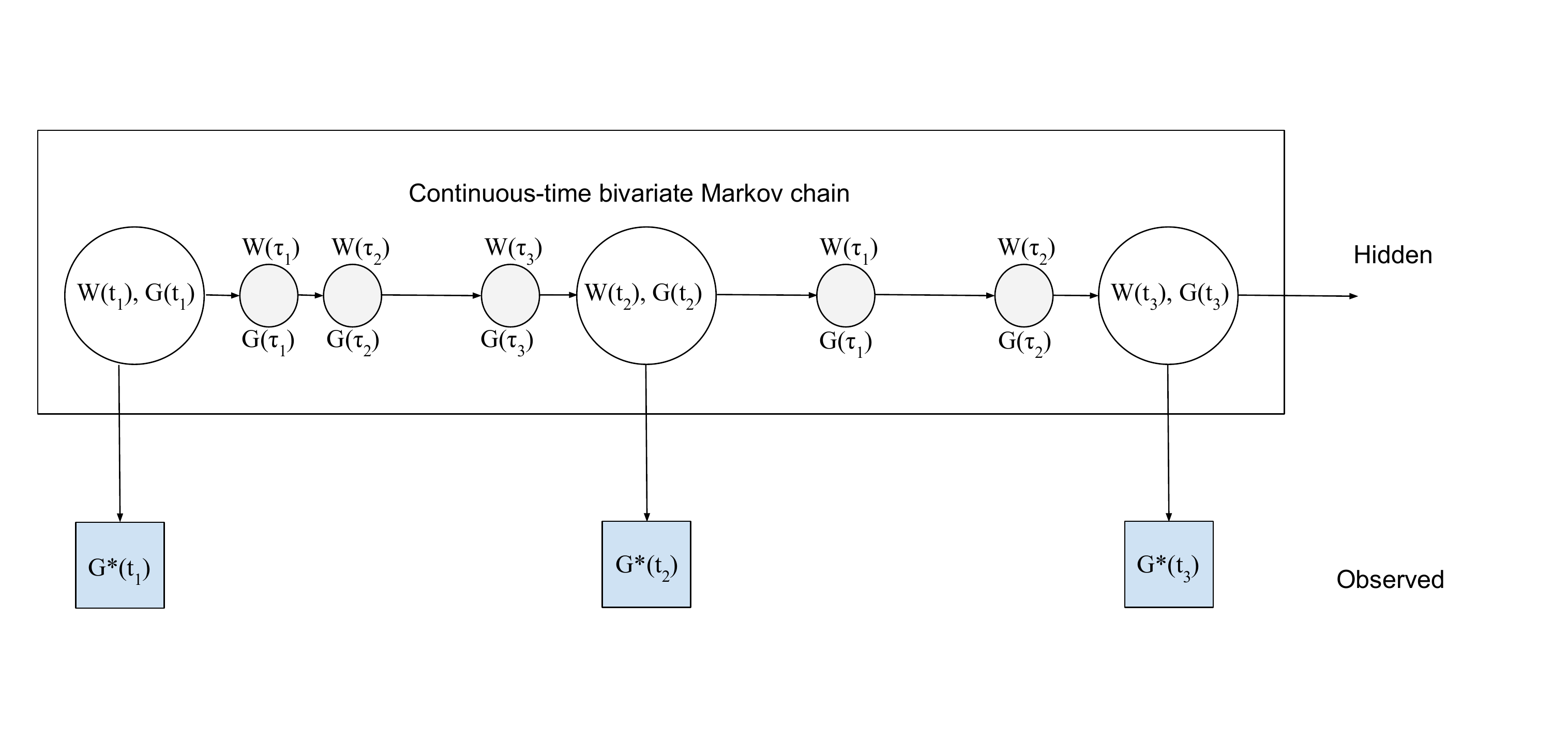}
    \caption{RG-HMM set-up. The unobserved hidden variables evolve according to a continuous-time Markov process, and we observe networks at discrete observation times with error. The unobserved small changes occurring between the consecutive observation times are represented in gray.}
\label{fig1}
\end{figure}
We now consider observing the continuous-time network evolution at discrete time points with errors. Assume that we have $M$ repeated network-valued observations $\bm g^\star_{1:M} = [g^\star(t_1)$, $g^\star(t_2), \cdots, g^\star(t_M)]$, where $t_1 < t_2 < ... < t_M$ are $M$ observation time points. 
% Denote the vector of observed network variables [$G^\star(t_1), \cdots, G^\star(t_M)$] by $\bm G^\star_{1:M}$. 
Let $\bm G^\star_{1:M} = [G^\star(t_1), \cdots, G^\star(t_M)]$ denote the random vector of observed networks, of which  $\bm g^\star_{1:M}$ is a realization.
We represent a true/hidden network variable underlying the observed network at a particular observation time by $G(t_m)$, and the hidden binary variable at the observation time by $W(t_m)$. We assume that 
\begin{enumerate}[label=(\roman*)]
    \item the latent networks evolve according to a percolation model. Then the true/hidden variables at discrete observation times $\{W(t_m), G(t_m), m = 1,2,...,M\}$ are embedded in the continuous-time Markov chain (ER/PR process), thus constituting a discrete-time Markov chain, 
    %Such discrete-time Markov chain has stationary transition probabilities when the observation times are evenly spaced. (Details are shown in Appendix \ref{discrete-time_mc})
    
    \item there is a time-independent error process, which corrupts the observing process by type-I error rate and type-II error rate, denoted respectively by $\alpha$ and $\beta$. Specifically, for any $\alpha,\beta \in [0, 1]$, we have 
\iffalse
\begin{equation}
    \begin{split}
        P(G^\star(e) &= 0 \big \rvert G(e) = 0) = 1 - \alpha \\
        P(G^\star(e) &= 1 \big \rvert G(e) = 0) = \alpha \\
        P(G^\star(e) &= 1 \big \rvert G(e) = 1) = 1 - \beta \\
        P(G^\star(e) &= 1 \big \rvert G(e) = 0) = \beta, \\
    \end{split}
\end{equation}
\fi 
\begin{equation}
    \begin{split}
        P(G^\star(e) &= 0 \big \rvert G(e) = 0) = 1 - \alpha , 
        P(G^\star(e) = 1 \big \rvert G(e) = 0) = \alpha \\
        P(G^\star(e) &= 1 \big \rvert G(e) = 1) = 1 - \beta, 
        P(G^\star(e) = 1 \big \rvert G(e) = 0) = \beta, \\
    \end{split}
\end{equation}
where $G^\star(e) = 1 \text{ or } 0$ represent a specific edge is present or not in observed variable, and $G(e) = 1 \text{ or } 0$ represent a specific edge is present or not in hidden variable,
\item 
the first observation $G^\star(t_1)$ is error-free.
\end{enumerate}

The last assumption is for convenience and standard.  Combining the network-valued latent Markov chain with the error process, we obtain a random graph hidden Markov model (RG-HMM). The schematic representation is shown in Figure \ref{fig1}.

\section{MLE for RG-HMM}
Although our ultimate goal is testing for two competing hypotheses of percolation regimes, learning the RG-HMM from a sequence of noisy networks observed only at a longitudinal subsampling of times is a necessary precursor and of independent interest.

Throughout this section, we assume $\alpha < 0.5$ and $\beta < 0.5$. Under such an assumption, the parameters in the model are identifiable. (A proof is provided in section 1 of the Supplement.) 
%More details are provided in Appendix \ref{identifiability}. 
We present an Expectation-Maximization (EM) algorithm for estimating the parameters in an RG-HMM (i.e. $p, q, \gamma, \alpha, \beta$) with a given sequence of noisy networks observed only from a longitudinal subsampling of time points, i.e. $\bm g^\star_{1:M}$. We assume that the network in the hidden layer changes faster than the observing rate so that there are likely many non-observed small changes occurring between the consecutive observation times, as is typically the case for functional connectivity networks used in the study of epilepsy. 

%Functional connectivity networks studied in computational neuroscience, for example, including those described in Figure 6, typically are of this type. Those networks we have used in the study of epilepsy rely on brain voltage data with micro-second resolution, but the network construction itself requires the use of sliding windows that typically are chosen to have at least milli-second resolution. I

Estimation is done conditional on the latent state $W(t_1) = 1, G(t_1) = g^\star(t_1)$ at the first observation time $t_1$. As in %Snijders et al.
\cite{snijders2010maximum}, this has the advantage that no initial distribution assumption is needed for the latent Markov chain, and the estimated parameters refer exclusively to the dynamics of the network. 

The algorithm consists of an E-step, wherein we calculate the expected value of the complete data log-likelihood $l(\alpha, \beta, \gamma, p, q) = \log f(\bm W_{2:M}, \bm G_{2:M}, \bm g^\star_{2:M} )$, with respect to the distribution of unknown latent variables $\bm W_{2:M}, \bm G_{2:M}$, given the observed networks $\bm g^\star_{1:M}$ and the current parameter estimates. We then maximize the expected log-likelihood in the M-step. 

The E-step is non-trivial in that there are two levels of unknown latent variables in our model: we do not know the true networks and binary variables $W(t_m), G(t_m)$ at the observation times, nor do we know the intermediate path that connects one true state $W(t_{m-1}), G(t_{m-1})$, to the next $W(t_m), G(t_m)$. EM is standard for parameter estimation in general HMMs. However, we face some unique challenges: (1) direct calculation of the expectation is computationally infeasible due to the enormously large state space of size $2^{n \choose 2}$; (2) the observed-data likelihood is difficult to calculate due to the unobserved changes occurring between consecutive observation times. Particle filtering and augmented data sampling using MCMC are used to tackle these challenges.

\subsection{Complete-data Log-likelihood}
For the complete data scenario, we assume we have samples $\bm w_{2:M}, \bm{g}_{2:M}$ for the hidden layer. Then the complete data log-likelihood for the RG-HMM, conditional on $W(t_1) = 1, G(t_1) = g^\star(t_1)$, can be written as 
\begin{equation}
\begin{split}
l^\star(\alpha,\beta,p,q,\gamma) &= \log f(\bm w_{2:M}, \bm{g}_{2:M}, \bm{g^\star}_{2:M}) \\
%& = \log f(\bm{g^\star}_{2:M}\big \rvert\bm w_{2:M}, \bm{g}_{2:M}) + \log f(\bm w_{2:M}, \bm{g}_{2:M})\\
%& = \log f(\bm{g}^\star_{2:M}\big \rvert\bm{g}_{2:M}) + \log f(\bm w_{2:M}, \bm{g}_{2:M})\\
& =  \sum_{m=2}^{M}\log f_{\alpha,\beta}(g^\star_m\big \rvert g_{m}) +  \sum_{m=2}^{M} \log f_{p, q, \gamma}(w_m, g_m\big \rvert w_{m-1}, g_{m-1}), \\
\end{split}
\label{3.1}
\end{equation}
where $g_m$, $g^\star_m$ is the abbreviation for $g(t_m)$ and $g^\star(t_m)$, respectively, $\alpha$ is the type I error rate, $\beta$ is the type II error rate, $p$ and $q$ are birth and death rate, and $\gamma$ is the rate parameter for the continuous-time percolation model in which the hidden chain in RG-HMM was embedded. 

%\begin{enumerate}
%    \item 
The first term in \eqref{3.1} is based on the conditional distribution of the observed network given the truth, which takes the form 
\begin{equation}
\label{3.2}
    f_{\alpha,\beta}(g^\star_m\big \rvert g_{m}) = \alpha^{c_m} (1-\alpha)^{d_m} \beta^{b_m} (1-\beta)^{a_m},
\end{equation}
where $a_m, b_m, c_m, d_m$ are counts corresponding to type I and type II errors in the error process from $g_m$ to $g^\star_m$ at observation time $t_m$. Details are shown in Table \ref{tab2}.
\begin{table}[H]
\centering
\begin{tabular}{lll}
\hline
True/Observed        & $E$ & $E^c$ \\\hline
$E$                    & $a$ & $b$                    \\
$E^c$ & $c$ & $d$   \\ 
\hline
\end{tabular}
\caption{Counts corresponding to type I and type II errors}
\label{tab2}
\end{table}

% \item 
The second term in \eqref{3.1} is the log-likelihood of the embedded discrete-time Markov chain in the hidden layer, which cannot be calculated in closed from due to the non-observed small changes occurring between consecutive observation times in the hidden layer. Therefore, we consider the log-likelihood for augmented data in order to obtain a more easily computed likelihood. The details are as follows.

%\end{enumerate}

\subsubsection*{Augmented data}
The data augmentation in the hidden layer can be done for each period $(t_{m-1}, t_m)$, $m=2,...,M$. Consider period $t_1$ to $t_2$. We assume there are $R_1$ transition time points between $t_1$ and $t_2$, denoted as $\tau_1,...,\tau_{R_1}$, which are ordered increasingly. Setting $\tau_0 = t_1$, we have $t_1=\tau_0<\tau_1<\tau_2,...,<\tau_{R_1} \leq t_2$.
The sample path from $G(t_1)$ to $G(t_2)$ is characterized by
\begin{equation}
\begin{split}
\bm S_1 &= (G(\tau_1), G(\tau_2),...,G(\tau_{R_1}))\\
\bm V_1 &= (W(\tau_1), W(\tau_2),...,W(\tau_{R_1})),
\end{split}
\end{equation}
which specify the sequence of unobserved change that brings the network from $G(t_1), W(t_1)$ to $G(t_2), W(t_2)$. Thus, the feasible sample path $\bm S_1$, $\bm V_1$ from $G(t_1)$ to $G(t_2)$ should satisfy two conditions: 1) $G(\tau_j), G(\tau_{j+1})$ differ by only one edge for $j=1,...,R_1-1$; and 2) $G(\tau_{R_1}) = G(t_2)$, $W(\tau_{R_1}) = W(t_2)$. The details of the probability mass function of a sample path $\bm S_1, \bm V_1$ conditional on $G(t_1), W(t_1)$ can be found in section 2 of the Supplement.
%\ref{pmf_from_main}.

\subsection{EM for Incomplete-data }
The goal is to find $\arg\max_{\Gamma} f(\bm g^\star_{2:M}\big \rvert G_1 = g^\star_1, W_1 = w_1)$, where $\Gamma = (p, q, \gamma, \alpha, \beta)$.
\subsubsection{E-step}
The expected value of the complete data log-likelihood with respect to the true networks and true binary variables $\bm G_{2:M}, \bm W_{2:M}$, given the observed networks $\bm g_{1:M}^\star$ and current parameter estimates $\Gamma^{(k-1)}$ is 
\begin{equation}
\begin{split}
    Q(\Gamma; \Gamma^{(k-1)}) 
 %   &= E \left[ \log f(\bm W_{2:M}, \bm{G}_{2:M}, \bm{g^\star}_{2:M}) \big \rvert  \bm{g^\star}_{1:M}, \Gamma^{(k-1)}\right] \\
    & = E \left[\log f(\bm g^\star_{2:M}\big \rvert\bm G_{2:M})\big \rvert \bm g^\star_{1:M}, \Gamma^{(k-1)} \right] + E\left[\log f(\bm W_{2:M}, \bm G_{2:M})\big \rvert \bm g^\star_{1:M}, \Gamma^{(k-1)} \right]. \\
\end{split}
\label{exp_log}
\end{equation}

\subsubsection{M-step}
The first term in \eqref{exp_log} can be easily maximized since it has a closed-form log-likelihood in \eqref{3.2}, which yields the following:
\begin{equation}
\label{alpha}
\begin{split}
\hat \alpha &= \frac{E[C\big \rvert \bm g^\star_{1:M},\Gamma^{(k-1)}]}
{E[C+D\big \rvert \bm g^\star_{1:M},\Gamma^{(k-1)} ]} \\
\hat \beta & = \frac{E[B\big \rvert \bm g^\star_{1:M},\Gamma^{(k-1)}]}
{E[A+B\big \rvert \bm g^\star_{1:M},\Gamma^{(k-1)} ]}.
\end{split}
\end{equation}
The formula for $\hat \alpha $ is the expected number of false edges in the observed network divided by the expected number of non-edges in the true network, given the observed networks $\bm g^\star_{1:M}$ and current parameter estimates $\Gamma^{(k-1)}$. The formula for $\hat \beta $ is the expected number of false non-edges in the observed network divided by the expected number of edges in the true network, given the observed networks $\bm g^\star_{1:M}$ and current parameter estimates $\Gamma^{(k-1)}$.

The second term in \eqref{exp_log} can be maximized by setting the score function to be zero, 
\begin{equation}
     E \left[ S(\bm \theta; \bm W_{2:M},  \bm G_{2:M})\big \rvert \bm g^\star_{1:M}, \Gamma^{(k-1)} \right] = 0,
\label{score}
\end{equation}
where $\bm \theta = [p, q, \gamma]$, $S(\bm \theta; \bm w_{2:M},  \bm g_{2:M}) = \frac{\partial}{\partial \bm \theta}
 \log f(\bm w_{2:M},  \bm g_{2:M})$ is the partial data score function, which cannot be calculated in closed form. Note that the score function for the augmented data is in closed form and by the Missing Information Principal used in
 %of 
%Orchard and Woodbury 
% \cite{woodbury1970missing} and 
%Louis 
% \cite{louis1982finding}, which also was used in 
%Snijders, et al. 
\cite{snijders2010maximum}, we have that 
\begin{equation}
     S(\bm \theta; \bm w_{2:M},\bm g_{2:M})  =  E \big[ S(\bm \theta; \bm w_{2:M},  \bm g_{2:M}, \bm S, \bm V)\big \rvert\bm W_{2:M} = \bm w_{2:M}, \bm G_{2:M} =\bm g_{2:M} \big]. 
\end{equation}
In other words, the partial data score function can be calculated by taking the expectation of the total data score function. 
%(See appendix \ref{MissingInformationPrincipal} for proof.) 
The proof is provided in section 4 of the Supplement. Therefore, solving \eqref{score} yields the following closed-form solution for $\gamma$, $p$ and $q$,
\begin{equation}
\label{gamma}
   \begin{split}
       \hat \gamma = \frac{1}{t_M - t_1} E \left[
           E \left[
       \sum_{m =2}^M R_{m-1}
         \big \rvert   W_{2:M} = \bm w_{2:M}, \bm G_{2:M} =\bm g_{2:M} \right]
       \Big \rvert \bm g^\star_{1:M}, \Gamma^{(k-1)} \right]
   \end{split} 
\end{equation}

\begin{equation}
\scriptsize
\label{p}
       \hat p =  
\frac{
       E \left[
           E \left[
       \sum_{m =2}^M \sum_{r = 1}^{R_{m-1}}
       I \big\{
       W(\tau_{r-1}^{m-1}) =0, W(\tau_{r}^{m-1}) =1
       \big\}
         \big \rvert   W_{2:M} = \bm w_{2:M}, \bm G_{2:M} =\bm g_{2:M} \right]
       \Big \rvert \bm g^\star_{1:M}, \Gamma^{(k-1)} \right]
}
{
              E \left[
           E \left[
       \sum_{m =2}^M \sum_{r = 1}^{R_{m-1}}
       I \big\{
       W(\tau_{r-1}^{m-1}) =0 
       \big\}
         \big \rvert   W_{2:M} = \bm w_{2:M}, \bm G_{2:M} =\bm g_{2:M} \right]
       \Big \rvert \bm g^\star_{1:M}, \Gamma^{(k-1)} \right]
}
\end{equation}

\begin{equation}
\scriptsize
\label{q}
       \hat q =  
\frac{
       E \left[
           E \left[
       \sum_{m =2}^M \sum_{r = 1}^{R_{m-1}}
       I \big\{
       W(\tau_{r-1}^{m-1}) =1, W(\tau_{r}^{m-1}) =0
       \big\}
         \big \rvert   W_{2:M} = \bm w_{2:M}, \bm G_{2:M} =\bm g_{2:M} \right]
       \Big \rvert \bm g^\star_{1:M}, \Gamma^{(k-1)} \right]
}
{
              E \left[
           E \left[
       \sum_{m =2}^M \sum_{r = 1}^{R_{m-1}}
       I \big\{
       W(\tau_{r-1}^{m-1}) =1
       \big\}
         \big \rvert   W_{2:M} = \bm w_{2:M}, \bm G_{2:M} =\bm g_{2:M} \right]
       \Big \rvert \bm g^\star_{1:M}, \Gamma^{(k-1)} \right]
}.
\end{equation}
where $R_{m-1}$ is the total number of transition times between $t_{m-1}$ and $t_{m}$ in the augmented sample path and $\tau_{r}^{m-1}$ is the $r$-th transition time between $t_{m-1}$ and $t_{m}$. 

\subsection{Particle Filtering\label{subsection3.3}}
The expectation step of our E-M algorithm is difficult to calculate. Since the state space for latent variables is prohibitively large, it is not feasible to directly calculate the conditional expectations in the aforementioned MLEs by summing over all possible network sequences, binary variable sequences and sample paths in the hidden layer. Instead, we propose to estimate those conditional expectations in \eqref{alpha}, \eqref{gamma}, \eqref{p}, \eqref{q} by sampling from the state space of latent variables. 

Note that there are two levels of the expectation in \eqref{gamma}, \eqref{p}, and \eqref{q}. The first is the outer expectation, which is taken with respect to the conditional distribution of $\bm W_{2:M}, \bm G_{2:M}$ in the hidden layer given the observed networks $\bm g_{1:M}^\star$, which we propose to sample from by a particle filtering based sampling scheme described in \cite{doucet2001introduction} (i.e. a Sequential Monte Carlo method). This sampling scheme involves first approximating $f(w(t_m), g(t_m) \,|\, \bm g^\star_{1:m-1}, \bm \Gamma^{(k-1)})$ with $B$ particles (i.e. samples for the true network and binary variable pair) at each observation time $t_m$, then tracing the ancestral lines, or particle paths, for the particles as samples from $f(\bm w_{2:M}, \bm g_{2:M} \,|\, \bm g^\star_{1:M}, \bm \Gamma^{(k-1)})$. The $B$ particles at each observation time reduces the space from which we will sample from for the particle paths, thus allowing us to effectively sample some number $\Psi$ network and binary variable sequences that are more likely to generate the observed networks in the error process. A visual representation for this procedure is provided in Figure \ref{pf}. More details are provided in section 5 of the Supplement.

\begin{figure}[!htb]
\centering
    \includegraphics[width=0.45\textwidth]{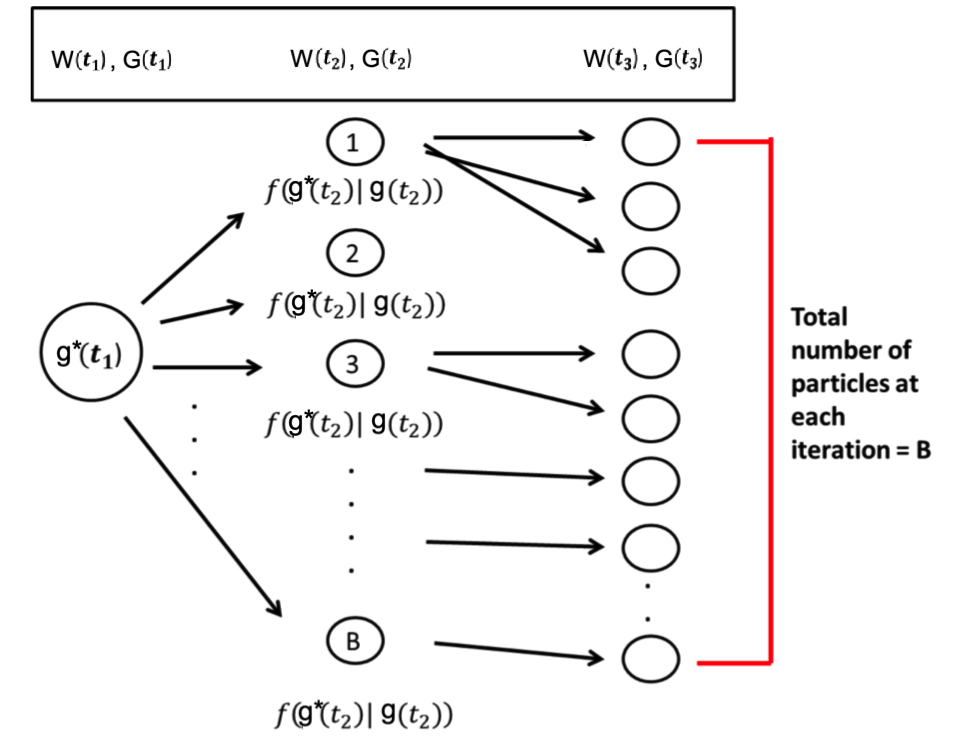}
     \caption{Particle filtering sampling scheme. $B$ particles are sampled at each observation moment following a two-stage process: 1) particles are sampled at the previous moment with probability proportional to the conditional distribution of the true network given the observed network; 2) then a true network and binary variable at the current observation moment, i.e. a new particle, is simulated starting from one of the sampled 
     %network and binary variable pair 
     particle in step 1, according to ER/PR process. $\Psi$ ancestral lines are obtained by tracing back the ancestral lines for $\Psi$ particles at time $t_M$, sampled with probability proportional to the conditional distribution of the observed network given the true network at time $t_M$.}
     \label{pf}
\end{figure} 

\subsection{Simulating the Sample Path}
The second level of the expectation in \eqref{gamma}, \eqref{p}, and \eqref{q} is the inner expectation, which is taken with respect to the conditional distribution of sample path $\bm S, \bm V$ that connects the given true sequences $\bm w_{2:M}, \bm g_{2:M}$ at the observation times in the hidden layer. We sample from this conditional distribution by simulating paths using an MCMC method. We draw upon the work of 
%Snijders et al. 
\cite{snijders2010maximum} 
where draws are generated by the Metropolis-Hastings algorithm, using a proposal distribution consisting of small changes in a sample path that brings one network to the next. More details are provided in section 6 of the Supplement.

\subsection{Putting it together - the EM Algorithm}
The EM algorithm resulting from the combination of all elements described in the previous subsections, for maximum likelihood estimation of the parameters $p$, $q$, $\gamma$, $\alpha$, $\beta$ in our RG-HMM, is shown in Algorithm \ref{EM}. 

\begin{algorithm}[!htbp]
\SetAlgoLined
\DontPrintSemicolon
 \scriptsize
 \KwIn{Network snapshots: $\bm g^\star = \big (g^\star(t_1),g^\star(t_2),..., g^\star(t_M) \big)$, $B$, $H$, $\Psi$}
 \KwOut{$\hat \Gamma = [\hat\alpha, \hat \beta,\hat p, \hat q, \hat \gamma]$}
 Initialization: $\Gamma^0$\;
 \While{! stopping condition for EM}{
\Comment*[l]{At $(l+1)^{th}$ EM iteration with estimates $\bm \Gamma^{l}= [p^{l}, q^{l}, \gamma^{l}, \alpha^{l}, \beta^{l}]$)}
\Comment*[l]{Sample $B$ particles $\{(g^{(b)}(t_m), w^{(b)}(t_m))\}_{b=1}^B$ at each $t_m$ (particle filtering)}

        Set $g^{(b)}(t_1) = g^\star(t_1)$, $g^{(b)}(t_1) = 1$ for each $b$. \\
        \For{$m=1,2,...,M-1$}{
            \For{$b = 1, ..., B $}{
                 Calculate the conditional probability $p^m_b := P(G^\star (t_{m}) = g^\star (t_m)| G(t_m) = g^{(b)} (t_m))$

            }
            Sample $B$ index $\eta_1, \eta_2,...\eta_B$ from  set $(1,2,...,B)$ with replacement with probability proportional to $p^m_1,p^m_2,...,p^m_B$.\\
            \For{$b = 1, ..., B$}{
                 Start from $g^{(\eta_b)}(t_{m}), w^{(\eta_b)}(t_{m})$, simulate according to ER/PR up until time $t_m - t_{m-1}$, then set $g^{(b)}(t_{m+1}), w^{(b)}(t_{m+1})$ to be the current simulated network and binary variable.
            }
          }
        %Create $\Psi$ particle paths $\{ \bm w^\psi_{2:M}, \bm g^\psi_{2:M}\}_{\psi = 1}^\Psi$ by tracing ancestral lines for the above particles.
\Comment*[l]{Sample ancestral lines and simulate sample paths}
     \For{$h = 1, 2,.., H$}{    
          Sample an index $\eta$ from $1,2,...,B$ with probability proportional to $p^M_1,p^M_2,...,p^M_B$, trace back the ancestral line for the $\eta$-th particle at time $t_M$, then obtain one sequence of networks and binary variables, $\bm g^{(h)}_{2:M}, \bm w^{(h)}_{2:M}$ \\
          \For{$d=1,2,...,D$}{
               Sample one path $\bm s^{(d)}_h, \bm v^{(d)}_h$ from $\bm S,\bm V|\bm W = \bm w^{(h)}_{2:M}, \bm G = \bm g^{(h)}_{2:M}$\ by Metropolis–Hastings. \\
               Calculate and store the sufficient statistics. 
          }
      }
\Comment*[l]{Update parameters}
    Update $\hat p^{l+1}, \hat q^{l+1}, \hat \gamma^{l+1}$ using formulas in \eqref{gamma}, \eqref{p}, \eqref{q} with sampled paths $\{\bm s^{(d)}_h, \bm v^{(d)}_h\}_{d=1}^{D}$ that interpolate $\bm w^{(h)}_{2:M}, \bm g^{(h)}_{2:M}$, $h=1, \cdots, H$. \\
    Draw $\Psi$ ancestral lines, then update $\hat \alpha^{l+1}, \hat \beta^{l+1}$ using formulas in \eqref{alpha} with sampled network sequences $\{\bm g^{(\psi)}_{2:M}\}_{\psi= 1}^{\Psi}$.
 }
\caption{EM Algorithm for MLEs}
\label{EM}
\end{algorithm}
%In our EM algorithm, we iterate the E-step and M-step until convergence, and the final estimators approximately maximize the observed likelihood function.
The convergence criteria we have used in practice is $\frac{||\bm\hat{\Gamma}^{l+1}-\bm\hat{\Gamma}^{l}||_2}{||\bm\hat{\Gamma}^{l}||_2} < 0.10$. Based on our numerical experience, the proposed EM algorithm usually converges within 5 iterations when all parameters are initialized as $0.5$. Less iterations will be needed if parameters are initialized somewhere close to the true values. 
% The final estimates from the two ways of initialization are similar, so it is not very sensitive to the initial values of parameters. 
The choice of $B$ (i.e. the number of particles sampled at each observation moment), the choice of $\Psi$ (i.e. the number of true network series to sample for the maximization of error rates $\alpha$ and $\beta$) and the choice of $H$ (i.e. the number of true network series to sample for the maximization of $p, q, \gamma$) may vary depending on the size of network and the number of observation time points. Our recommended starting point is to set $B = 50,000$, $\Psi = 40,000$ and $H=10$. It is important to keep in mind that the maximization of $p, q, \gamma$ involves a computationally expensive MCMC procedure. Setting $H$ to a small number helps efficiently reduce the computational burden without compromising much on the accuracy of parameter estimates, in our experience.

In our EM algorithm, time complexity is $\mathcal{O}(BM)$ for the particle filtering step, $\mathcal{O}(HM)$ for the MCMC step, and $\mathcal{O}(\Psi M)$ for sampling $\Psi$ number of ancestral lines. Although the estimation procedure scales linearly in $B$, $M$ and $H$, it is still a computationally expensive procedure for large networks because simulating the true networks in the hidden layer involves sampling from edges or non-edges of a $N$-node network, which scales quadratically in $N$ using standard techniques.

\subsection{Asymptotic Properties of MLE for RG-HMM}

We provide a convergence result for the MLE in
%that justifies the use of our EM 
our proposed RG-HMM in its stationary period as time progresses. Let $\bm \Gamma_0$ and $\bm{\hat \Gamma}$ denote the true values of parameters and the MLE of RG-HMM, respectively. To simplify the notation, let $\bm X(t_m)$ denote the pair $(W(t_m), G(t_m))$.

The asymptotic behavior of $\bm{\hat \Gamma}$ in our RG-HMM can be established using results from
%Bickel et al.
\cite{bickel1998asymptotic} on asymptotic normality of the MLE for a general stationary HMM. 
\begin{theorem}
Let $\pi_{\bm \Gamma}(\bm x)$ be the stationary distribution of a Markov chain in RG-HMM $\{\bm X(t_m), G^\star(t_m)\}$ and $\mathcal{L}_0$ be the limiting covariance matrix of 
%$m^{-1/2} \frac{\partial}{\partial \bm{\Gamma}} \log p_{\bm \Gamma}( G^\star(t_1), \cdots,  G^\star(t_m))$. 
$m^{-1/2} \frac{\partial}{\partial \bm{\Gamma}} \log p_{\bm \Gamma}( \bm G^\star_{1:m})$. 
Assume (C1) that $t_m - t_{m-1} = t_{m+1} - t_{m} = \Delta t, \forall \, m \geq 2$, and stationarity is reached at $t_1$, (C2) that $\forall \, \bm x \in \mathcal{X}$, $\bm \Gamma \rightarrow \pi_{\bm \Gamma}(\bm x)$ has two continuous derivatives in some small neighborhood of $\bm \Gamma_0$, and (C3) that $\mathcal{L}_0$ is nonsingular. Then $m^{1/2}(\bm{\hat \Gamma} - \bm{ \Gamma_0}) \xrightarrow{p} N(0, \mathcal{L}_0^{-1})$ as $m \rightarrow \infty$.
\label{theorem}
\end{theorem}
The proof of Theorem \ref{theorem} is given in 
%Appendix \ref{appendex_proof}. 
section 7 of the Supplement. 
In the E-step of our EM algorithm, we apply particle filtering to approximate the expected log-likelihood of the complete-data. Although a central limit theorem (\cite{chopin2004central}) exists for the particle estimate of the likelihood, which assures the asymptotic unbiasedness of the approximation in E-step as $B\rightarrow \infty$, it is still difficult to theoretically investigate the convergence property of our EM algorithm with the rather complicated likelihood function in the RG-HMM. Therefore, we resort to simulation to evaluate the empirical performance of our EM in the later section.

\section{Hypothesis Testing of Putative Percolation Regimes}

In this section, we present a hypothesis testing framework using Bayes factors for distinguishing between two percolation models, ER and PR.
The testing problem can be formulated as a test of separate families of hypotheses, i.e. 
\begin{equation}
H_{ER}:  \bm g^\star \sim f(\bm g^\star, \bm \Gamma_{ER}) \text{   vs   }  
H_{PR}:  \bm g^\star \sim g(\bm g^\star, \bm \Gamma_{PR}),
\label{test_h}
\end{equation}
where $\bm g^\star = [g^\star(t_1), g^\star(t_2),..., g^\star(t_M)]$ is the observed sequence of networks, and $f(\cdot, \bm \Gamma_{ER})$ and $g(\cdot, \bm \Gamma_{PR})$ are the probability functions of $\bm G^\star$ under the ER process with parameters $\bm \Gamma_{ER}$ and PR process with parameters $\bm \Gamma_{PR}$. $H_{ER}$ and $H_{PR}$ are the hypotheses that the observed sequences $\bm g^\star$ is from the ER process and PR process, respectively.

\subsection{Bayes Factor}
Following 
%Cox 
\cite{cox1961tests} on tests of separate families of hypotheses, we adopt a Bayesian approach to this testing problem. The posterior odds for $H_{ER}$ vs $H_{PR}$ are, by Bayes's theorem,
\begin{equation}
    \frac{pr(H_{ER}|\bm g^\star)}{pr(H_{PR}|\bm g^\star)} = \frac{\tilde \omega_f \int f(\bm g^\star, \bm \Gamma_{ER}) p_f(\bm \Gamma_{ER}) d \bm \Gamma_{ER}}
    {\tilde \omega_g \int g(\bm g^\star, \bm \Gamma_{PR}) p_g(\bm \Gamma_{PR}) d \bm \Gamma_{PR}},
\end{equation}
where $\tilde \omega_f$ and $\tilde \omega_g$ are the prior probabilities of $H_{ER}$ and $H_{PR}$ being true, respectively. $p_f(\bm \Gamma_{ER})$ is the prior p.d.f of $\bm \Gamma_{ER}$ under $H_{ER}$, and $p_g(\bm \Gamma_{PR})$ is the prior p.d.f of $\bm \Gamma_{PR}$ under $H_{PR}$. 

The Bayes factor ($BF$) is the posterior odds for $H_{ER}$ vs $H_{PR}$ when the prior probabilities of the two hypotheses are equal, i.e. when $\tilde \omega_f = \tilde \omega_g = 0.5$. Exact analytic calculation of the Bayes factor for the testing problem in \eqref{test_h} is not tractable, so we resort to numerical methods. 
%Kass and Raftery 
\cite{kass1995bayes} proposed two large-sample approximations for Bayes factors - Laplace's Method and the Schwarz Criterion. 
%The details of each method are presented in section 8 of the Supplement. 
The approximation from Laplace's method requires specifying the prior distributions $p_f(\bm \Gamma_{ER})$ and $p_g(\bm \Gamma_{PR})$ on parameters of each model. However, the choice of priors is not trivial for a percolation model. Therefore, we choose to use the approximation from the Schwarz criterion which has the benefit of not requiring priors. Also note that because the dimensions of $\bm \Gamma_{ER}$ and $\bm \Gamma_{PR}$ are the same, the logarithm of the Bayes factor for our testing problem can be approximated by the log-likelihood difference, i.e. $ \log BF \approx 
      \log f(\bm g^\star, \hat{\bm \Gamma}_{ER}) - \log g(\bm g^\star, \hat{\bm \Gamma}_{PR})$. 

\subsection{Marginal Probability of Observations}
The computation of the probabilities of the observed networks under the two percolation models,  $\log f(\bm g^\star, \hat{\bm \Gamma}_{ER})$ and $\log g(\bm g^\star, \hat{\bm \Gamma}_{PR})$, is the key ingredient of the test. We now present a forward algorithm with particle filtering to approximate such quantities in our RG-HMM. Similar to the ML estimation, testing is also done conditional on the latent state $W(t_1) = 1, G(t_1) = g^\star(t_1)$ at the first observation time $t_1$. To simplify the notation, let $\bm X_m$ denote the pair $(W(t_m), G(t_m))$. Define $Z_m := P\Big( \bm G^\star_{2:m} =  \bm g^\star_{2:m}| \bm X_1 = \big(1, g^\star(t_1) \big)\Big)$ for each $m = 2, \cdots M$, and $Z_1 := 1$. Also let $q_m(x) = P( \bm G^\star_{m} = \bm g^\star_{m}| \bm X_m =x)$, $\eta_m(x) = P(\bm X_m =x|  \bm G^\star_{2:m-1} = \bm g^\star_{2:m-1}, \bm X_1 = (1, g^\star(t_1))$ for each $m = 2, \cdots M$, with the convention $\eta_2(x) = P(\bm X_2 =x| \bm X_1 = (1, g^\star(t_1))$. To alleviate the notational burden, we omit the condition at $t_1$ in the probability expression for the remainder of this section when no confusion is possible. With the recurrence relation (the derivation is provided in section 8 of the Supplement)
\begin{equation}
\begin{split}
    Z_m & =  Z_{m-1} \sum_x q_m(x)\eta_m(x) =  Z_{m-1} E(q_m(\bm X)), \bm X\sim \eta_m(\cdot)
\end{split}
\label{4.5}
\end{equation}
the probabilities $Z_M$ can be approximated recursively by replacing each $\eta_m(\cdot)$ in \eqref{4.5} with $\eta_m^B(\cdot)$, which is the particle approximation of the filtering distribution $\eta_m(\cdot)$. A forward algorithm with particle filtering is provided 
%in algorithm \ref{alg2} 
in section 9 of the Supplement to compute $Z_M^B$, i.e. the particle approximation of the marginal probability of observations. The time complexity is $\mathcal{O}(BM)$. Thus the full complexity to compare the two percolation regimes is $\mathcal{O}\big((B+H+\Psi)M \big)$.

\section{Simulation Studies}
In this section, we conduct simulations to evaluate the empirical performance of the proposed estimators described in Section 3 and the test described in Section 4. 

\subsection{Simulation Results for Estimation}
We first investigate the finite sample performance of the estimation procedure for networks with $N=20$ nodes simulated at $M =50$ observation time points, referred to as $\{t_m\}_{m=1}^{50}$, from the ER/PR percolation process with parameter setting $p = 0.7, q=0.3, \gamma=2$ and error rates $\alpha=0.03, \beta = 0.01$. We set the observation rate, denoted by $\kappa$, to be $0.6$, i.e. we set $t_m = m/0.6$. The observation rate is set to be smaller than the actual rate of change to mimic the scenario in application. For each ER and PR process, we generate 100 network time series with length 50 and obtain ML estimates of the model parameters from each data set. In the EM algorithm, we set $B = 50,000$ and initialize all parameters as $0.5$. The running time of one iteration on a high performance Linux computing cluster using 4 cores is approximately 10 minutes for the ER process and 20 minutes for the PR process under this setting. The mean and standard deviation of parameter estimates based on 100 replications for both the ER process and PR process are reported in Table \ref{t5.1}. 
\begin{table}[!htbp]
\small
\resizebox{\columnwidth}{!}{
\begin{tabular}{llllll}
\hline
& Birth Rate $p$ & Death Rate $q$ &Transition Rate $\gamma$ & Type-I Error $\alpha$ & Type-II Error $\beta$ \\\hline
Truth & 0.70  & 0.30 & 2 & 0.03 & 0.01 \\
ER-HMM   &0.680 (0.041)& 0.299 (0.063) & 1.746 (0.126)&  0.112 (0.021)& 0.037 (0.013)     \\
PR-HMM   &0.684 (0.035)& 0.289 (0.057) & 1.770 (0.113) & 0.118 (0.019) &  0.035 (0.013)  \\
\hline
\end{tabular}
}
\caption{Mean and standard deviation of parameter estimates based on $100$ simulations from ER/PR process with $B=50000$, $N=20$, $M=50$.}
\label{t5.1}
\end{table}
In Table \ref{t5.1}, we are seeing bias in the estimates, especially for type-I and type-II error rates, which is not unexpected given that we are approximating the distribution on a very large dimensional state space with only $B = 50,000$ particles. In principle, the larger $B$ is, the more accurate estimates will be. However, it comes with the trade-off between computational time and accuracy. 

\begin{figure}[!htb]
   \centering
    \includegraphics[width=0.65\textwidth]{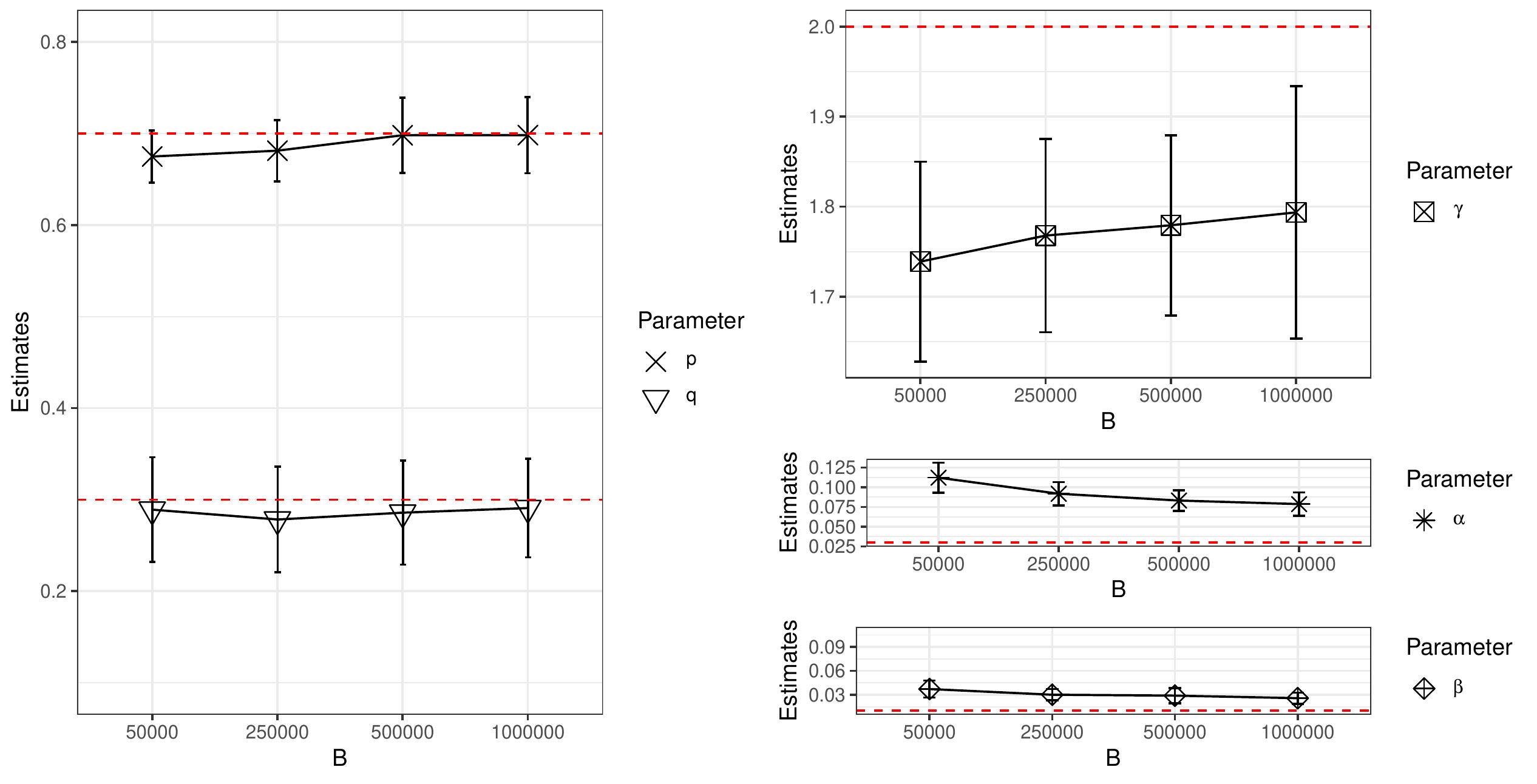}
   \caption{Mean and standard deviation of parameter estimates based on $20$ simulations from the ER process with varying $B$, $N=20$, $M=50$. %Visualization of Table \ref{t5.2}.
   }
   \label{plot_B}
 \end{figure}

\begin{figure}[!htb]
   \centering
    \includegraphics[width=0.65\textwidth]{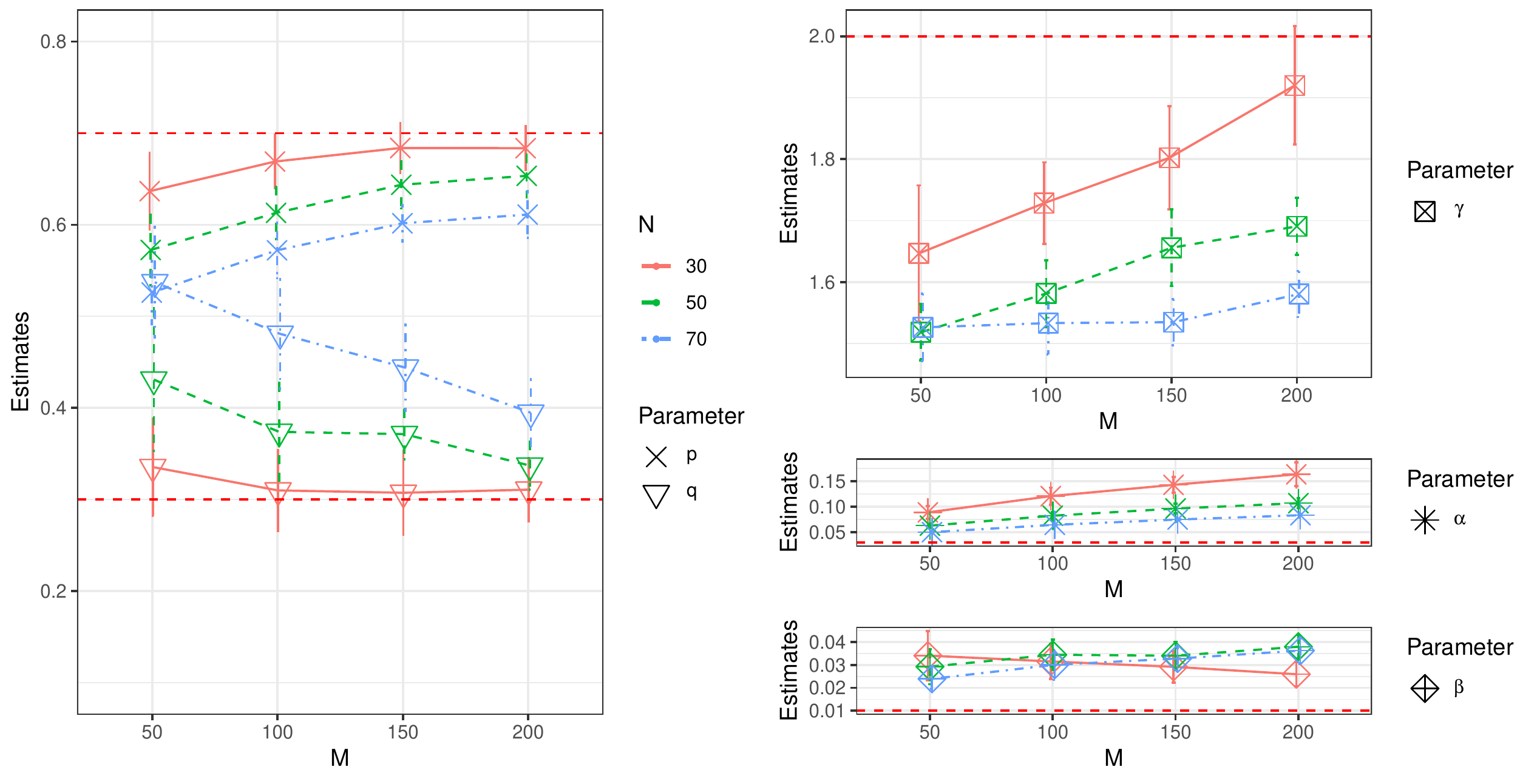}
   \caption{Mean and standard deviation of parameter estimates based on $20$ simulations from ER process with $B=50000$, $(N, M) \in \{30, 50, 70\} \times \{50, 100, 150, 200\}$. 
   %Visualization of Table \ref{t5.3}.
   }
   \label{plot_MN}
\end{figure}

We then perform a small simulation study for the ER process, as a very limited exploration of the performance of MLEs with respect to $B$ (the number of particles sampled at each observation moment), $M$ (the number of observation time points), and $N$ (network size), under the same model parameter setting as in the previous simulation study. 
%Table \ref{t5.2} 
Figure \ref{plot_B} shows the mean and standard deviation of parameter estimates from $20$ simulated network sequences with $N=20$ and $M=50$ using an increasing number of particles. 
%Table \ref{t5.3} 
Figure \ref{plot_MN}
shows the mean and standard deviation of parameter estimates based on $20$ simulations from an ER process with $B=50,000$ and $(N, M) \in \{30, 50, 70\} \times \{50, 100, 150, 200\}$. Numerical results are provided in section 10 of the Supplement. 

The main empirical findings are the following. 
From 
%Table \ref{t5.2}, 
Figure \ref{plot_B},
we can see that increasing $B$ could reduce bias in the estimates for type-I and type-II error rates, but such improvement is diminished as $B$ increases. In Figure \ref{plot_MN}, the performance of type-I error rate estimation improves as the size of the network increases, but it does not improve as the length of the network time series (i.e. $M$) increases, using a fixed number of particles. It is not surprising that as the number of observed networks is increasing in $M$, more particles are required to approximate the parameters, in order to reveal the true asymptotic behavior as $M$ increases. The type-II error rate estimate is not as sensitive to either $M$ or $N$, but rather to $B$. The performance of parameter estimates of $p$, $q$ and $\gamma$ are improved as $M$ increases. Additionally, for estimating these parameters from larger networks, the algorithm requires larger $M$, i.e. longer network time series with more observation time points, compared to that needed from smaller networks, in order to achieve similar performance. In summary, we can get more accurate estimates for error rates with larger $N$ and $B$, with the caveat that more particles are needed for larger $M$. For the estimates of $p$, $q$ and $\gamma$, improvement is seen as $M$ and $B$ increase, with the caveat that larger $M$ is needed for larger $N$.

\subsection{Simulation Results for Testing}
To assess the effectiveness of our testing framework using Bayes factor for discriminating between two percolation models, $ER$ and $PR$, we choose the rate of detection as the performance metric. For each simulated network sequence, we first compute its Bayes factor. We next identify it to be an ER sequence if the Bayes factor is greater than 1, otherwise we identify it as a PR sequence. Hence, the rate of detection for ER(PR) process is the percentage of ER(PR) sequences that are correctly identified as such among all simulated ER(PR) sequences. 

\begin{figure}[!htbp]
   \centering
  \includegraphics[width=0.6\textwidth]{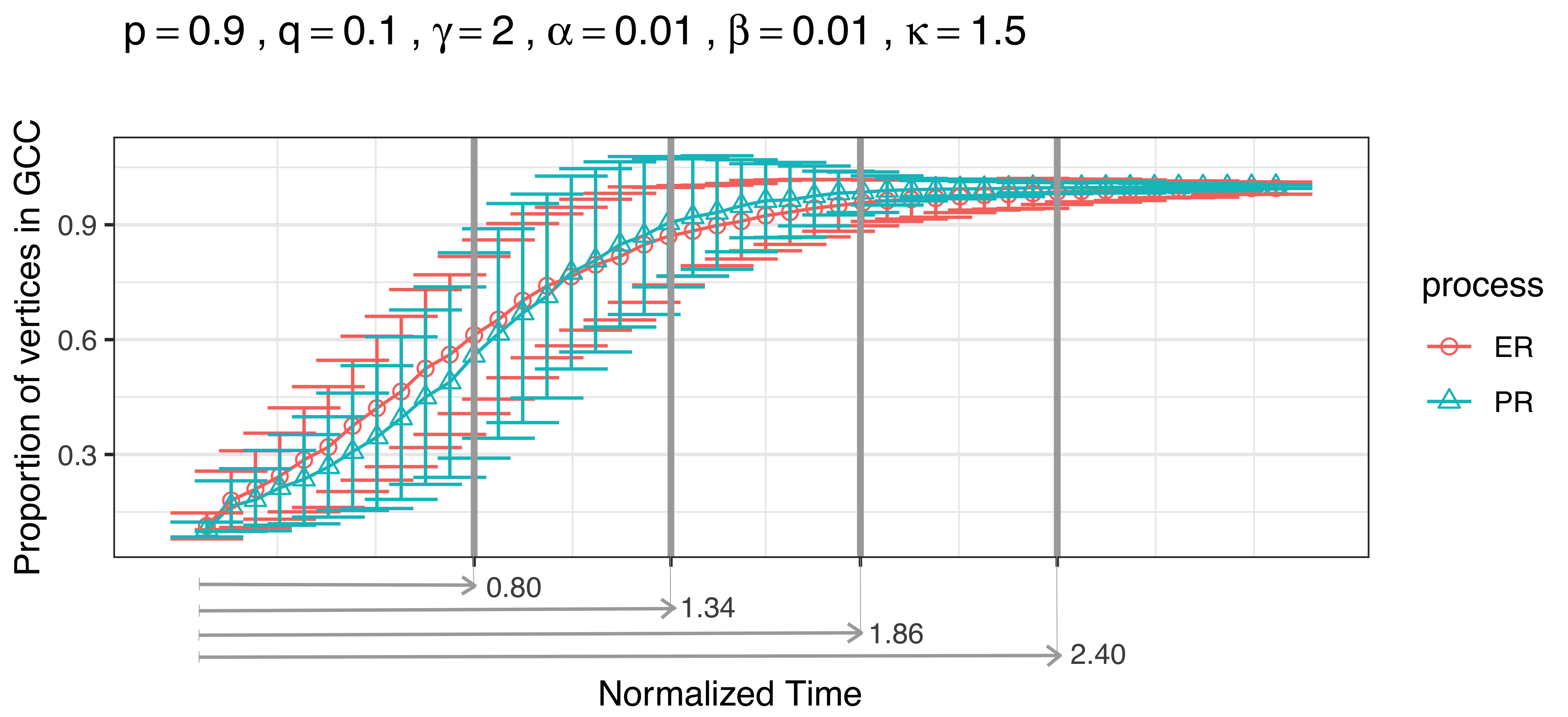}
   \caption{4 levels of $t^\prime_M$: $0.8, 1.34, 1.86, 2.4$. $t^\prime_M$ is the normalized observation duration time, which represents how much of the percolation curve we take for testing.}
   \label{t_norm}
 \end{figure}

A simulation study is designed to evaluate the effect of $B \in \{ 50,000; 250,000; 500,000\}$, $N \in \{10, 20, 30\}$, and $t^\prime_M= \{0.8, 1.34, 1.86, 2.4\}$ on the rate of detection for ER and PR using Bayes factor, where $t^\prime_M$ is the normalized observation duration time, which represents how much of the percolation curve we take for testing. Specifically, we define normalized observation time points as $t^\prime_m := \frac{\gamma t_m}{N}$. Since observations are made with regular intervals at rate $\kappa$, i.e. $t_m = m/\kappa$ for $m = 1, \cdots, M$, we have $t_m^\prime = \frac{m }{N (\kappa/\gamma)}$. Figure \ref{t_norm} illustrates the four levels of progression for the percolation curve we are considering for testing via arrowed lines. Each level of $t_M^\prime$ corresponds to a curve segment taken from the beginning of the curve to the gray line placed at the corresponding scaled time. 
   
\begin{figure}[!htb]
   \centering
   \includegraphics[width=0.48\textwidth]{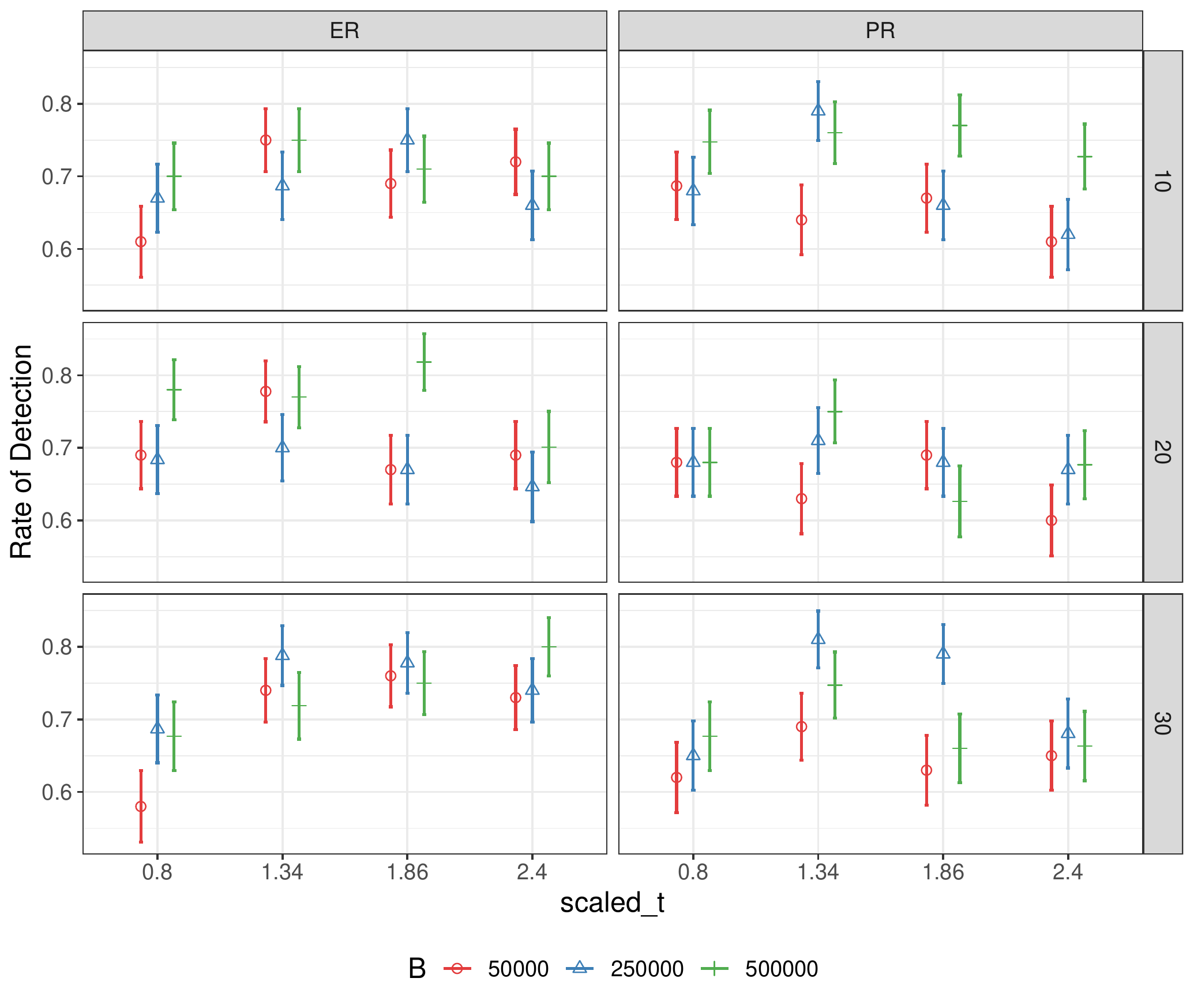}
    \includegraphics[width=0.48\textwidth]{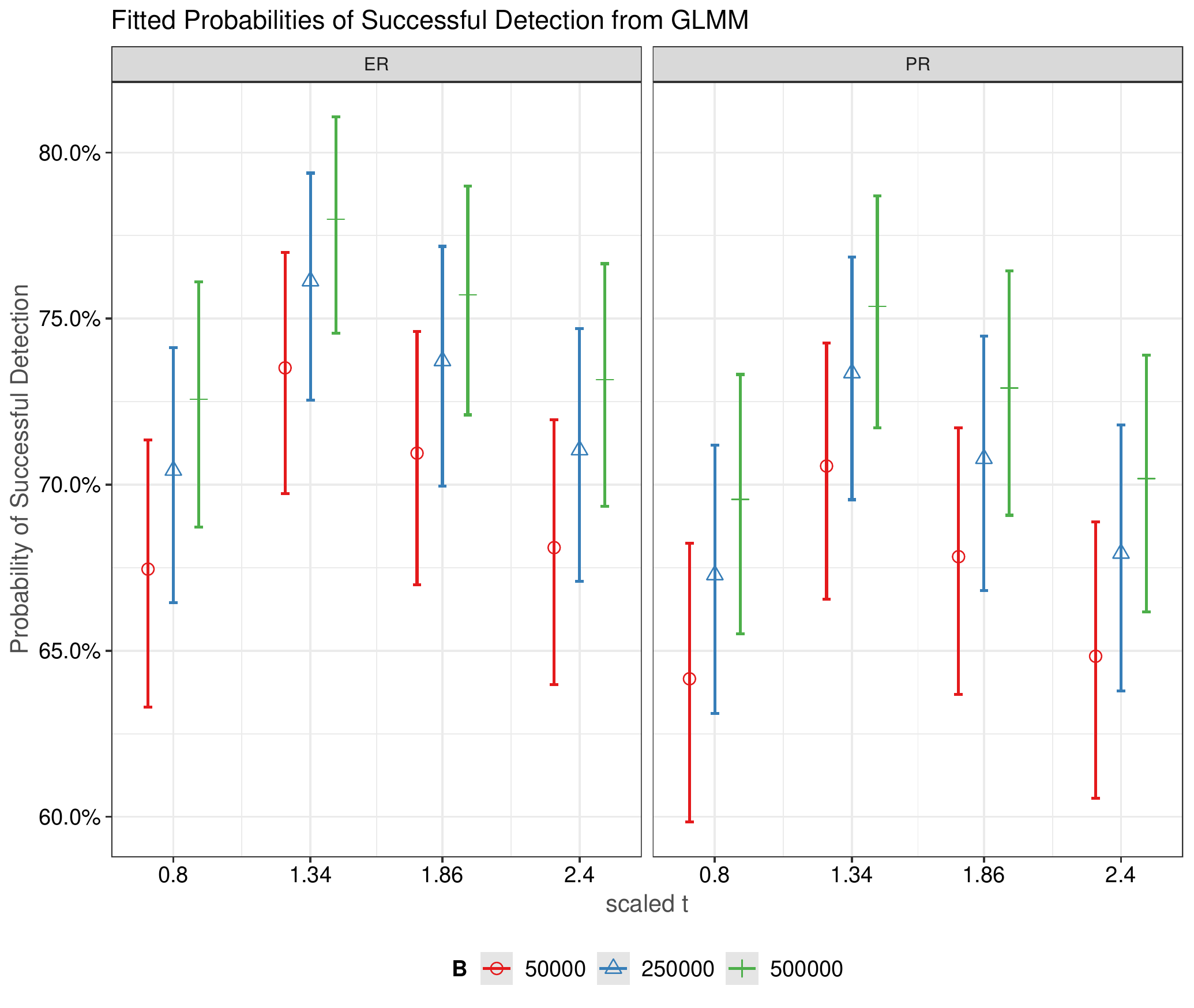}
   \caption{Results for the first simulation study. Left: rates of detection for ER and PR based on 100 replicates. Right: fitted probability of successful detection in GLMM.}
   \label{study1}
\end{figure}

For each combination of settings, we generate 100 ER sequences and 100 PR sequences with $p = 0.9, q=0.1, \gamma=2, \alpha=0.01$, and $\beta = 0.01$. We set the rate of observation, $\kappa$, to be 1.5 and conduct hypothesis testing on each of the simulated network sequences. Rates of detection for ER and PR, based on 100 replicates, along with estimated standard errors, are shown in Figure \ref{study1} (left). We then fit a generalized linear model with mixed effects on the testing results from the simulation study in order to analyze the relative contribution of variation in the rate of detection for ER/PR due to the change of values in $B$, $N$ and $t_M^\prime$. We set the seed used for generating the networks to be a random effect, since there might be some dependencies between the network sequences simulated using the same seeds. Additionally, those simulated from different seeds may be of a different level of difficulty to distinguish. The fitted probability of successful detection, as well as the estimates of the coefficients for the GLMM is shown in Figure \ref{study1} (right) and Figure \ref{study1-est}, respectively. The ANOVA table associated with the fitted model is provided in Table \ref{study1-anova}. The results show that the rate of detection is significantly affected by $B$ (number of particles), $t^\prime_M$ (how much percolation curve is observed for testing), but not $N$. And it's slightly easier for ER sequences to be identified as such than for PR sequences. 

We also conduct a second simulation study to evaluate the effect of observation rate $\kappa$ on the rate of detection for ER and PR using Bayes factor, which shows that rate of detection significantly increases by increasing $\kappa$ from $0.5$ to $1$, but stops to significantly increase as $\kappa$ further increases. Details of this simulation study is provided in 
%Figures \ref{study2}, \ref{study2-est} and Table \ref{study2-anova}. 
section 11 of the Supplement. 

In conclusion, our proposed testing framework can effectively distinguish between different percolation models using an observed sequence of noisy networks under proper settings. The size of the network does not affect the ability to discern between the two percolation regimes and such ability increases as $B$ (the number of particles) and $\kappa$ (the observation rate) increases. This coincides with our intuition that a better particle approximation for the latent distribution, as well as a more informed signal, would help in distinguishing between percolation models. 

%\begin{minipage}{\textwidth}
\noindent
  \begin{minipage}[b]{0.49\textwidth}
    \includegraphics[width=0.8\linewidth]{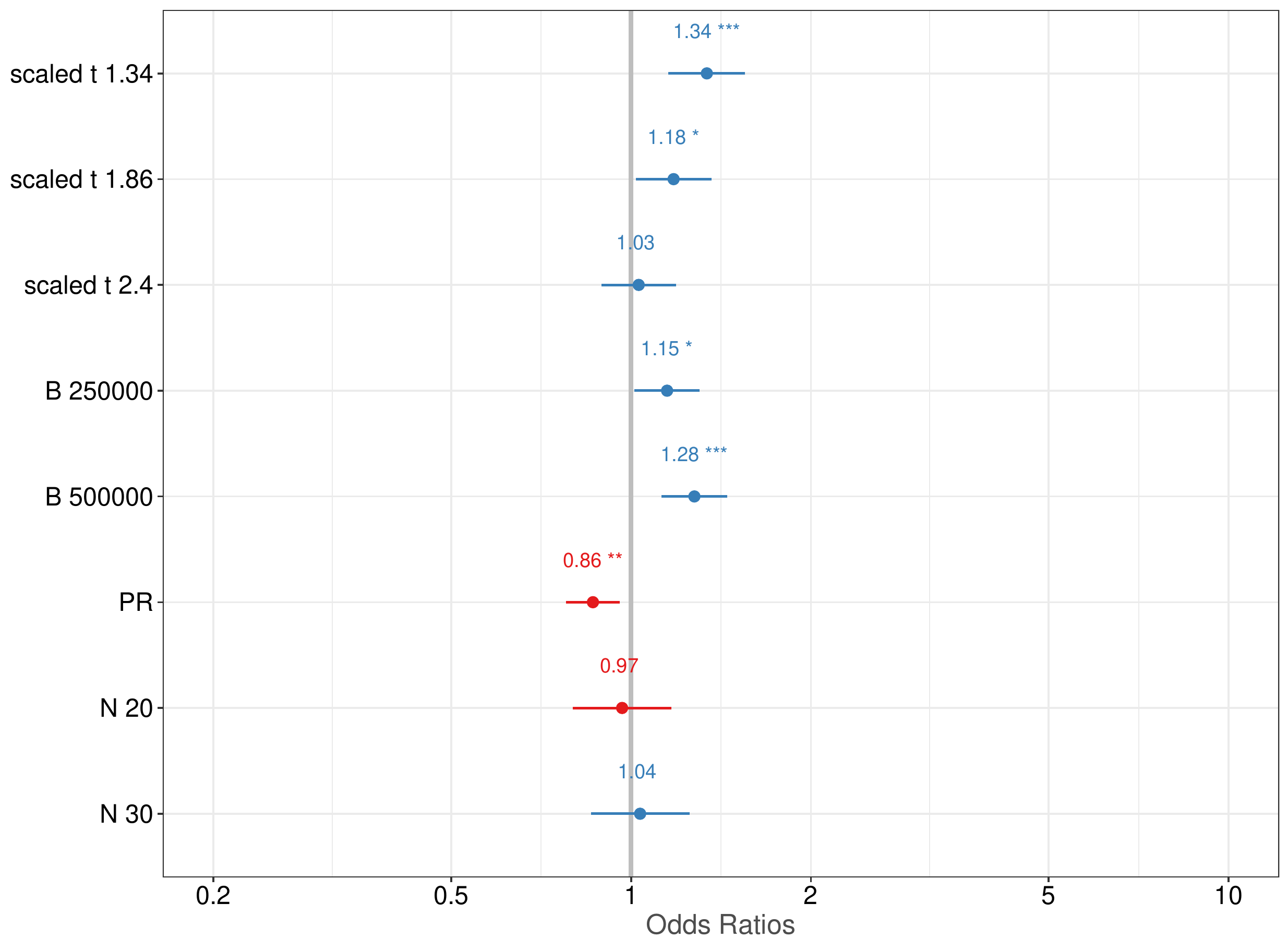}
    \captionof{figure}{Estimates of the coefficients from GLMM in the first simulation study.}
       \label{study1-est}
  \end{minipage}
  %\hfill
  \begin{minipage}[b]{0.49\textwidth}
        \small
        \begin{tabular}{lrrr}
          \hline
         & Chisq & Df & Pr($>$Chisq) \\ 
          \hline
        N & 0.52 & 2 & 0.7728 \\ 
          scaled\_t & 18.88 & 3 & 0.0003 \\ 
          process & 7.68 & 1 & 0.0056 \\ 
          B & 14.29 & 2 & 0.0008 \\ 
           \hline
        \end{tabular}
      \captionof{table}{ANOVA table for the first simulation study}
       \label{study1-anova}
    \end{minipage}

\section{Application to In Vivo Data during Human Seizures}
% \section{Real Data Analysis}
In this section, we show an example application of our methodology to electrocorticography data recorded from a human subject during three seizures.

\subsection{Functional Network Construction}

The data consist of invasive brain voltage recordings for 3 seizures from a patient with epilepsy. The entire array of electrodes consisted of 106 electrodes on the left hemisphere and 70 electrodes on the right hemisphere, with 1024 Hz sampling rate. For each seizure, we construct a time series of functional connectivity networks based on the voltage recordings in the following manner. We first band-pass filter the data between 4 and 50 Hz, and compute a bipolar reference by subtracting the activity of neighboring electrodes. 
%re-reference each channel to its adjacent channel on the same electrode shaft (bipolar re-referencing). 
Then we divide the re-referenced data into $1$ s windows and $0.5$ s overlap beginning $120$ s before seizure onset and ending $30$ s after seizure termination: the seizure onset and termination times were clinically determined. Finally, we construct a functional network for each $1$ s window by identifying significant cross-correlations between each pair of electrodes and controlling for multiple comparisons using the FDR ($q=0.5$). Details of this functional network construction procedure can be found in 
% prior work by Kramer et al. 
\cite{kramer2009network} and \cite{kramer2010coalescence}. 

Among the 3 seizures, two were clinically determined to be focal seizures with secondary generalization, and one was clinically determined to be a focal seizure without secondary generalization (i.e., the seizure activity remained local). We focus our analysis on the left hemisphere because each seizure begins in this hemisphere and we expect percolation to occur near the location of seizure onset as this region recruits other brain areas into the seizure dynamics. To that end, we construct functional networks using the $75$ channels on the left hemisphere of good signal quality.
We plot the size of the GCC over time, and edge density over time, for the resulting network time series from the first seizure in Figure \ref{s1}. Plots for the second and third seizures are provided in section 13 of the Supplement.

\subsection{Automatic Segment Finder}
Before we apply our estimation and testing framework to the constructed networks, we need first choose segments of the network time series consistent with percolation. 
%in connection with different stages of seizure, e.g. seizure onset and preceding termination.
Previous work (\cite{kramer2010coalescence}, \cite{martinet2020robust}) suggests sudden increases in the size of connected functional network communities occur at multiple times during a seizure (e.g., near seizure onset and termination), making nontrivial the choice of an appropriate time segment from the overall non-stationary signal. We developed a simple, automatic procedure for this purpose, consisting of roughly the following steps: 1) Select intervals of time that only contain network evolution in the regions of interest (ROIs) corresponding to periods of seizure evolution on clinical review of the data, as determined by a board-certified epileptologist (CJC) from the time intervals in which large dynamic communities appear (for details of the dynamic community detection procedure on network time series see \cite{martinet2020robust}). 2) Identify segments in the ROIs where both the size of GCC and density ramp up over time  and use them for testing. The details of this procedure are provided in section 12 of the Supplement. 
%Algorithm \ref{al3}, 
% The segments found for the first seizure using such procedure are 
Applying this procedure to the first seizure, we identify three segments, as 
shown in Figure \ref{s1} (bold curves). Results for the other two seizures are shown in section 13 of the Supplement. 
%All segments found for the first seizure after step 1-3 in this procedure are shown in Figure \ref{s1} with black-colored curves, among which the last curve is chosen for testing since it is the first segment right after seizure onset (step 4).

% Note that instead of using either GCC or density to segment the network time series, we can also use both of them to find the proper segment for testing. Specifically, we can first run the Automatic Segment Finder on the network time series with one metric. We can then run it again with the other metric, thus obtaining two segmented network time series. We can then take, as input, the overlapping of the two segments for testing, which is what we do in this analysis. The segments found for these seizures using such procedure are shown in Figure \ref{s1}-\ref{s3}.

\subsection{Estimation and Statistical Testing}
%We now use the constructed network time series to discriminate between two percolation models, ER and PR, in epileptic seizures.
% We now use the segments chosen from different ROIs of the constructed network time series for each seizure to test which one is a better fit between two competing percolation regimes, ER and PR. 
Having selected time segments of dynamic functional network evolution during the seizures, we now apply the statistical testing procedure to infer the percolation regime - ER or PR - for each segment. We take each chosen segment from the automatic segment finder
%Among all of the segments of the network time series found by the algorithm, we choose the one right after the seizure onset time 
as input for the statistical testing procedure under our RG-HMM framework, with $B = 500,000$. For each seizure, we report the mean and standard deviation of parameter estimates under the two percolation models, the likelihood evaluated at those estimates, and the estimated Bayes factors in log scale. The results for one of the chosen segments in seizure 1 are given in Table \ref{tab-s1}; see section 13 of the Supplement for results from other seizures.

%For each seizure, the mean and standard deviationof parameter estimates under the two percolation models, the likelihood evaluated at thoseestimates, and the estimated Bayes factor are given in 
%Table \ref{tab-s1} for seizure 1, Table \ref{tab-s2} for seizure 2, Table \ref{tab-s3} and Table \ref{tab-s3-2} for seizure 3 (two segments are chosen for the third seizure with one containing onset time the other laying right after onset time). 
%Table \ref{tab-s1} - \ref{tab-s3} (two segments are chosen for the third seizure with one containing onset time the other laying right after onset time). 

\begin{figure}[!htb] %13, 8
\centering 
 \includegraphics[width=\textwidth]{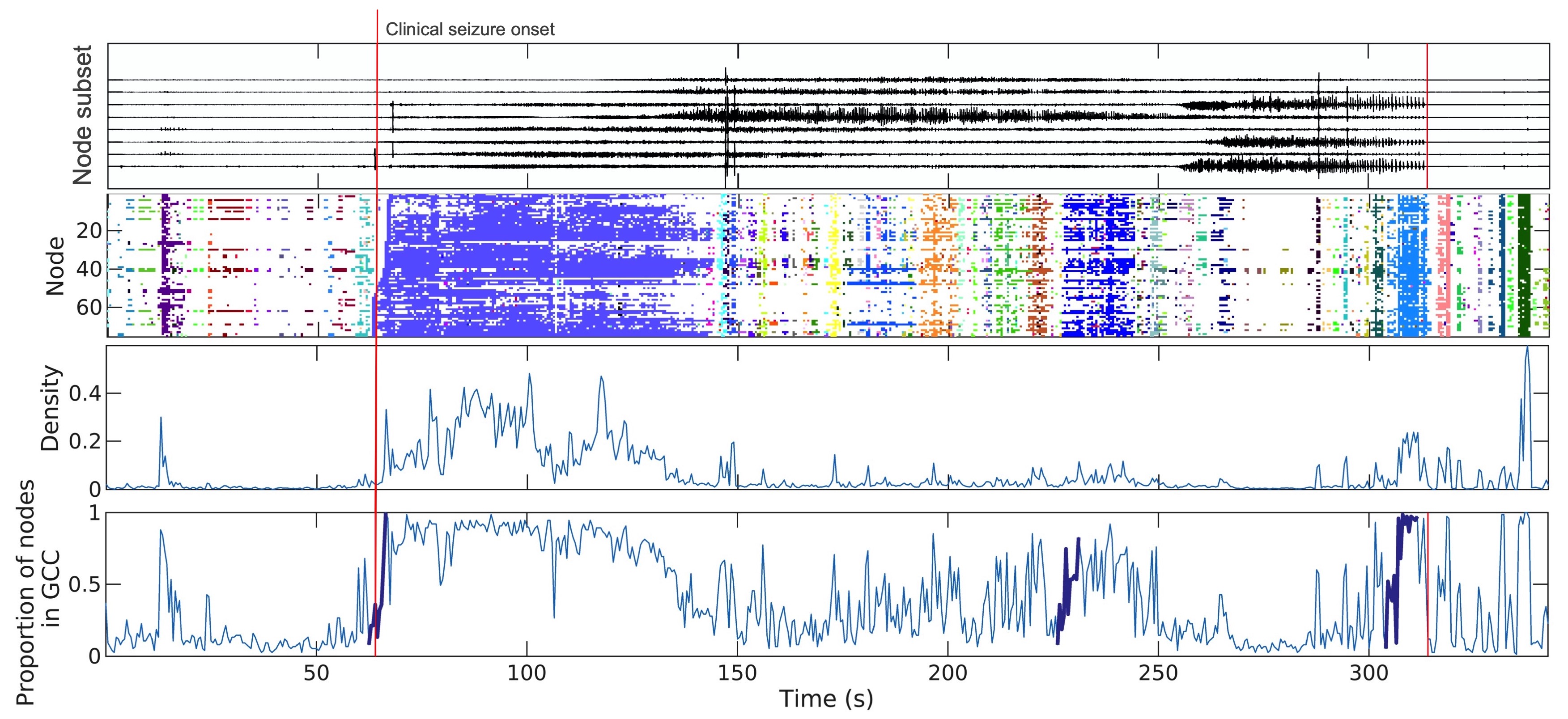}
\caption{Example selection of analysis segment for one seizure. Vertical lines represent clinically determined seizure onset and termination times. Top: Voltage time series recorded at eight electrodes on left brain hemisphere. Second row: Community membership (indicated by shades of gray) for each node over time; the three large dynamic communities in darker shades manifest ROIs. Third row: Network density over time. Bottom: Proportion of nodes in the GCC over time; the bold curves in ROIs are chosen by automatic segment finder for testing.}
\label{s1}
\end{figure}
The Bayes factors in log scale are positive for all segments from seizure 1 ($10.3$, $31.7$ and $62.0$, respectively); all segments from seizure 2 ($51.0$, $105.8$ and $7.1$, respectively); and 3 of 4 segments from seizure 3 ($10.9$, $61.5$, $15.1$ and $-5.0$, respectively). We conclude that Erdos-Renyi (ER) percolation is better supported by the data for the entire course of seizure 1 and seizure 2, as well as the early ictal stages of seizure 3, while product-rule (PR) percolation is better supported for the late ictal stage of seizure 3.

These preliminary results suggest that both types of percolation can occur in the dynamic functional networks of human seizures. For this patient,  though  each  seizure  begins  in  the  same  focal  location with the same recruitment patterns of type ER, the recruitment patterns differ preceding seizure termination, which are of type ER for seizure 1,2 and type PR for seizure 3. 
Clinically, seizure 3 differs from the other two seizures; seizure 3 remains spatially focal, while seizures 1 and 2 continue to propagate and only terminate after recruiting the entire brain.
% generalize to cover large areas of the brain.
These preliminary results suggest a different network dynamic for the two clinical seizure types. Whether local PR percolation (here, in the left hemisphere) correlates with focal seizure termination and prevents generalization (here, to the right hemisphere) requires further investigation. We note that identification of percolation type may also suggest different treatment strategies. For example, to prevent PR  percolation, the growing subnetworks that emerge could be separately targeted and prevented from joining.  Alternatively, to prevent ER percolation, effective treatment may require preventing the expansion of the emerging single connected component.

\begin{table}[!htb]
\centering
\resizebox{\columnwidth}{!}{
\begin{tabular}{lllllll}
\hline
   & $\hat p$         & $\hat q$      & $\hat \gamma$ & $\hat \alpha$   & $\hat \beta$ & $log lik$ \\ \hline
ER &   0.404 (0.112)&  0.248 (0.075)&  10.075 (1.423)&  0.068 (0.0005)& 0.435 (0.035)& -5636.822 (5.942)\\ \hline
PR & 0.468 (0.156)&  0.163 (0.072)&  9.819 (1.512)&  0.068 (0.0004)&  0.464 (0.032)& -5647.120 (6.850)\\ \hline
log(BF) & 10.297 (9.468))  &               &               &                 &               \\ \hline
\end{tabular}
}
\caption{Mean and standard deviation of parameter estimates for ER/PR process, log-likelihood estimates and Bayes factor (log scale) estimate based off of 10 trials for seizure 1 (first bold segment) on the left brain hemisphere. }
\label{tab-s1}
\end{table}

\section{Discussion}
We develop a class of random graph hidden Markov models (RG-HMMs) for characterizing percolation regimes in noisy, dynamically evolving networks in the presence of edge birth and edge death, as well as noise, under which exponential waiting times are used to model the time between the addition or removal of an edge. We present an EM algorithm with a particle filtering and sample path simulation scheme for estimating model parameters from a sequence of noisy networks observed only at a longitudinal subsampling of time points (i.e., not at every time step of edge change). We also provide a framework for statistical testing of competing hypotheses of percolation regimes, which involves calculation of the likelihoods evaluated at optimal estimates under each hypothesis, given the observed networks.

Furthermore, we establish the asymptotic property of the MLE for our RG-HMMs in its stationary period using asymptotic results for general HMMs with stationary Markov chains (\cite{bickel1998asymptotic}). Unfortunately, no proof is yet available for our model in its nonstationary period when networks evolve through the percolation phase transition. The most relevant work on asymptotic theory for HMMs with nonstationary Markov chains is in \cite{douc2001asymptotics}. This is an interesting subject for future research. 
%Nevertheless, from our numerical experience, bias in the MLE does not seem to compromise the ability to discern between competing models. It is important to note that although longer observational sequences are needed for an accurate estimate of model parameters, the ability to discern does not necessarily increase by observing more of the percolation curve. Instead, it is increased by observing network changes more frequently. 
% actually, when parameters are restricted to lie in some compact subset of the usual space $\Theta = \{(p, q, \gamma, \alpha,\beta)\,|\, p, q \in(0,1), \gamma \in(0, \infty), \alpha,\beta \in(0, 0.5)\}$, such CLT still holds without stationarity assumption. 

In the proposed model, we consider two canonical random graph models i.e., Erdos-Renyi and product-rule, representing two prototypical extremes of percolation. Notably, however, there is a wide spectrum of other random graph models, displaying a variety of percolation behaviors (e.g., \cite{riordan2011explosive}, Fig. 1). Our work may be extended to these other percolation regimes, given that our proposed RG-HMM is a flexible framework which may subsume a variety of random graph models. 

Our experience in the reported simulations and in working with empirical data sets is that the algorithm converges well. However, it is time-consuming, especially for larger networks requiring more particles for approximation. For example, when working with a 30 node network with 30 observation time points and 50,000 particles, one testing procedure, including ML estimations under two hypotheses and the calculation of Bayes factor, takes approximately 1 hour to run on a high performance Linux computing cluster using 4 cores with appropriate use of parallelization techniques for speedups. If we multiply the number of particles by 10, i.e. $B = 500,000$, the algorithm will take approximately $10$ hours. In principle, the larger $B$ is, the more accurate estimates will be. However, it comes with the trade-off between computational time and accuracy. While it is feasible to run our algorithm on even larger networks under manageable time with a relatively small particle size, we must interpret the results cautiously. In such cases, we recommend running multiple testing trials and taking an average of the approximated Bayes factor to reduce the high variance caused by a relatively small number of particles.

Given that recent work in network science, as it pertains to epilepsy, has shown an explosive density increase in functional connectivity networks in epilepsy patients during seizure onset, our work has the potential to be quite impactful for clinical neuroscience. To demonstrate the application to epileptic seizures, we apply our framework to assess, at a minimum, whether ER or PR percolation is better supported by the data for these particular seizures. We note that epilepsy is a complex and still incompletely understood disease. While it would be naive to expect that support for one type of percolation regime over another gleaned from our functional connectivity networks would offer significant insights into the etiology of this disease, we nevertheless are optimistic that results obtained through the application proposed here will be sufficiently suggestive as to inspire additional constructive thinking about the period of seizure onset, which may lead to improvements in treatment and disease management.

Several opportunities for extending the scope of this application remain. For example, we may apply our testing framework to seizure data from a group of patients and perform a meta-analysis to aggregate and contrast the findings from single seizures to identify patterns and interesting relationships that may come to light by comparing the test results against patient phenotype information. This phenotype information may include medical classification of severity of the disease, putative locations of seizure emergence and other clinically relevant measurements.
% This phenotype information may include medical classification of severity of the disease, putative locations of seizure emergence for those patients felt to suffer focal seizures, and any other biological measurements available and allowed under human subject restrictions. 

Lastly, although we have chosen to focus our attention on epileptic seizures, our class of models and our framework for parameter estimation and testing have the potential to be used outside of the realm of seizure data to better understand the emergence of organized structure in many other dynamic systems. With percolation theory omnipresent across a vast range of applications such as social networks, traffic networks, infectious disease networks, amino acid networks, and even understanding the spread of forest fires, computer viruses, etc. (\cite{saberi2015recent, bianconi2016percolation, stauffer2018introduction, islam2019universality}), our hope is that our methods may be used to better understand network dynamics across a wide range of fields.

\section*{Supplementary Materials}
Extended details can be found in the online Supplement.

\bibliographystyle{agsm}
\bibliography{main}

\end{document}

\def\spacingset#1{\renewcommand{\baselinestretch}%
{#1}\small\normalsize} \spacingset{1}

\if1\blind
{
  \title{\bf \Large Supplement to ``Inferring the Type of Phase Transitions Undergone in Epileptic Seizures Using Random Graph Hidden Markov Models for Percolation in Noisy Dynamic Networks"}
  \date{}
  \maketitle
} \fi

\if0\blind
{
  \title{\bf \Large Supplement to ``Inferring the Type of Phase Transitions Undergone in Epileptic Seizures Using Random Graph Hidden Markov Models for Percolation in Noisy Dynamic Networks"}
  \author{Xiaojing Zhu\thanks{These authors contributed equally.}, 
  Heather Shappell\footnotemark[1], 
  Mark A. Kramer, \\
  Catherine J. Chu, Eric D. Kolaczyk
  \hspace{.2cm}
  }
  \date{}
  \maketitle
} \fi

\section{Model Identifiability for RG-HMM \label{identifiability}}

In this section, we address the identifiability of RG-HMM. The model is identifiable if distinct values of $\bm \Gamma$ correspond to distinct probability distributions, i.e. if $\bm \Gamma_1 \neq \bm \Gamma_2$, then also $P_{\bm \Gamma_1} \neq P_{\bm \Gamma_2}$. We show that $f(g^\star)$ is identifiable if type-I error $\alpha < 0.5$ and type-II error $\beta < 0.5$.  

We will show that two sets of parameters $\bm \Gamma_1 = (p_1 ,q_1, \gamma_1, \alpha_1, \beta_1)$, and $\bm \Gamma_2 = (p_2 ,q_2, \gamma_2, \alpha_2, \beta_2)$ where $p_2 = q_1, q_2 = p_1, \alpha_2 = 1-\beta_1, \beta_2 = 1- \alpha_1$ will give the same probability distribution of the observed networks, written as $f\big (g^\star(t_1), \cdots, g^\star(t_M)\big)$. We will then assert that this is the only parameterization that will cause non-identifiability, thus we can make the model identifiable by restricting the parameter space: $\alpha <0.5 , \beta<0.5$. 

We now show that $f_{\bm \Gamma_1}\big(g^\star(t_1), \cdots, g^\star(t_M)\big) =f_{\bm \Gamma_2}\big(g^\star(t_1), \cdots, g^\star(t_M)\big)$, $\forall g^\star(t_1), \cdots, g^\star(t_M)$. Let $\bm x(t_m) = (g(t_m), w(t_m))$ denote the latent variables in the hidden layer, then conditional on $\bm x(t_1) = (g^\star(t_1), 1)$,
\begin{equation}
\begin{split}
        & \quad f_{\bm \Gamma_1}\big(g^\star(t_1), \cdots, g^\star(t_M)\big) \\
        &= \sum_{\bm x(t_1), \cdots, \bm x(t_M)}f_{\bm \Gamma_1}\big(g^\star(t_1), \cdots, g^\star(t_M), \bm x(t_1), \cdots, \bm x(t_M)\big) \\
        & = \sum_{\bm x(t_1), \cdots, \bm x(t_M)}
        f\big (\bm x(t_1)\big) 
         f_{\alpha_1, \beta_1}\big(g^\star(t_1)|g(t_1)\big)
        \prod_{m=2}^{M} f_{\alpha_1,\beta_1}\big(g^\star(t_m)\big \rvert g(t_{m})\big) f_{p_1, q_1, \gamma_1}\big(\bm x(t_m)\big \rvert \bm x(t_{m-1}) \big)\\ 
        & = \sum_{\bm x(t_2), \cdots, \bm x(t_M)}
        \prod_{m=2}^{M} f_{\alpha_1,\beta_1}\big(g^\star(t_m)\big \rvert g(t_{m})\big) f_{p_1, q_1, \gamma_1}\big(\bm x(t_m)\big \rvert \bm x(t_{m-1}) \big).
\end{split}
\label{D1}
\end{equation}
%the sum is over all possible true network variables at each observation time points. 
For any potential true network $g(t_m)$, there exists an `opposite' network $g^\prime(t_m)$ obtained by flipping all edges of $g(t_m)$ to non-edges and all non-edges to edges. Hence, if given the observed network $g^\star(t_m)$, $g(t_m)$ contains $a$ true edges, $b$ false non-edges, $c$ false edges and $d$ true non-edges, then  given $g^\star(t_m)$, $g^\prime(t_m)$ will contain $c$ true edges, $d$ false non-edges, $a$ false edges and $b$ true non-edges. Note that 
\begin{equation}
\begin{split}
 f_{\alpha_1, \beta_1}\big(g^\star(t_m)|g(t_m)\big) &= \alpha_1^c(1-\alpha_1)^d\beta_1^b(1-\beta_1)^a \\
  f_{\alpha_2, \beta_2}\big(g^\star(t_m)|g^\prime(t_m)\big) &= \alpha_2^a(1-\alpha_2)^b\beta_2^d(1-\beta_2)^c,
\end{split}
 \end{equation}
then we have $f_{\alpha_1, \beta_1}\big(g^\star(t_m)|g(t_m)\big) = f_{\alpha_2, \beta_2}\big(g^\star(t_m)|g^\prime(t_m)\big)$ if $\alpha_2 = 1-\beta_1, \beta_2 = 1- \alpha_1$.

Define $\bm x^\prime(t_m) = \big(g^\prime(t_m), w^\prime(t_m)\big)$, where $w^\prime(t_m) = 1- w(t_m)$. Next we show that $f_{p_1, q_1, \gamma_1}\big(\bm x(t_m)\big \rvert \bm x(t_{m-1}) \big) = f_{p_2, q_2, \gamma_2}\big(\bm x^\prime(t_m)\big \rvert \bm x^\prime(t_{m-1}) \big)$ if $p_1 = q_2$, $p_2 = q_1$, $\gamma_2 = \gamma_1$ for ER process. Then we have
%Recall that in Appendix \ref{Sample Path pdf},
\begin{equation}
\begin{split}
    f_{p_1, q_1, \gamma_1}\big(\bm x(t_m)\big \rvert \bm x(t_{m-1}) \big) = \sum_{\bm u} 
    \Big [ & e^{-\gamma_1(t_m - t_{m-1})} 
     \frac{(\gamma_1(t_m - t_{m-1})^R}{R!} \cdot
    \\ 
    & \prod_{r = 1}^{R}
f\big( g(\tau_r)\big \rvert w(\tau_{r}), g(\tau_{r-1})
\big) f\big( w(\tau_r)\big \rvert g(\tau_{r-1}), w(\tau_{r-1})\big) \Big],
\end{split}
\end{equation}
where the sum is over all possible sample paths connecting $\bm x(t_{m-1})$ and $\bm x(t_{m})$. Details for the pmf of a sample path are discussed in section \ref{Sample Path pdf}.
Let $\bm u^\prime$ be the `opposite' sample path of $\bm u$, which can be obtained by flipping all $(g(\tau_r), w(\tau_r))$ in $\bm u$ into $(g^\prime(\tau_r), 1-w(\tau_r))$. Hence if $\bm u$ is a sample path connecting $\bm x(t_{m-1})$ and $\bm x(t_{m})$, $\bm u^\prime$ will then be a sample path connecting $\bm x^\prime(t_{m-1})$ and $\bm x^\prime(t_{m})$. If we fix the rate parameter $\gamma_1 = \gamma_2$, then it is sufficient to show that $f\big( g(\tau_r)\big \rvert w(\tau_{r}), g(\tau_{r-1})
\big) f\big( w(\tau_r)\big \rvert g(\tau_{r-1}), w(\tau_{r-1}) = f\big( g^\prime(\tau_r)\big \rvert w\prime(\tau_{r}), g\prime(\tau_{r-1})
\big) f\big( w\prime(\tau_r)\big \rvert g\prime(\tau_{r-1}), w\prime(\tau_{r-1})$ for each $\tau_r$. Assume that $g(\tau_{r-1})$ contains $a$ edges and $b$ non-edges, then $g^\prime(\tau_{r-1})$ will contain $b$ edges and $a$ non-edges. Note that under ER process,
\begin{equation}
   f\big( g(\tau_r)\big \rvert w(\tau_{r}), g(\tau_{r-1})
\big) f\big( w(\tau_r)\big \rvert g(\tau_{r-1}), w(\tau_{r-1})  = 
\begin{cases} 
\frac{p_1}{b}, \text{ if } w(\tau_{r}) = 1, w(\tau_{r-1}) = 0 \\
\frac{1-q_1}{b}, \text{ if } w(\tau_{r}) = 1, w(\tau_{r-1}) = 1 \\ 
\frac{1-p_1}{a}, \text{ if } w(\tau_{r}) = 0, w(\tau_{r-1}) = 0 \\ 
\frac{q_1}{a}, \text{ if } w(\tau_{r}) = 0, w(\tau_{r-1}) = 1 \\ 
\end{cases}
\label{D4}
\end{equation}

\begin{equation}
   f\big( g^\prime(\tau_r)\big \rvert w\prime(\tau_{r}), g\prime(\tau_{r-1})
\big) f\big( w\prime(\tau_r)\big \rvert g\prime(\tau_{r-1}), w\prime(\tau_{r-1})  = 
\begin{cases} 
\frac{p_2}{a}, \text{ if } w\prime(\tau_{r}) = 1, w\prime(\tau_{r-1}) = 0 \\
\frac{1-q_2}{a}, \text{ if } w\prime(\tau_{r}) = 1, w\prime(\tau_{r-1}) = 1 \\ 
\frac{1-p_2}{b}, \text{ if } w\prime(\tau_{r}) = 0, w\prime(\tau_{r-1}) = 0 \\ 
\frac{q_2}{b}, \text{ if } w\prime(\tau_{r}) = 0, w\prime(\tau_{r-1}) = 1, \\ 
\end{cases}
\label{D5}
\end{equation}
then \eqref{D4} and \eqref{D5} are equivalent if $p_2 = p_1$ and $p_1 = p_2$. For PR process, it's not necessarily true since $f\big( g(\tau_r)\big \rvert w(\tau_{r}), g(\tau_{r-1})\big) $ depends on the sizes of connected components of $ g(\tau_r)$ and the relation between the connected components of $ g(\tau_r)$ and that of  $g^\prime(\tau_r)$ is intractable. 

Since each summand in \eqref{D1} under paramerterization $\bm \Gamma_1 = (p_1 ,q_1, \gamma_1, \alpha_1, \beta_1)$ will have one and only one match to a summand under the paramerterization $\bm \Gamma_2 = (p_2= q_1 ,q_2 = p_1, \gamma_1 = \gamma_2, \alpha_1 = 1-\beta_2, \beta_1 = 1-\alpha_2)$ by flipping the latent variables $\{\bm x(t_m)\}$ into $\{\bm x^\prime(t_m)\}$, then $f_{\bm \Gamma_1}\big(g^\star(t_1), \cdots, g^\star(t_M)\big) = f_{\bm \Gamma_2}\big(g^\star(t_1), \cdots, g^\star(t_M)\big )$. Therefore, the probability distributions of observed networks are the same under the two parameterization explained above, causing non-identifiability. By restricting the parameter space $\alpha <0.5 , \beta<0.5$, we can make the model identifiable.

\iffalse
\section{Properties of Embedded Discrete-time Markov Chain with Evenly-spaced Observation Times \label{discrete-time_mc}}
\begin{enumerate}[label=Property \arabic*.,itemindent=*]
  \item Assume that the observation times are evenly spaced in RG-HMM, then the discrete-time Markov chain embedded in the continuous-time percolation process in the hidden layer at discrete observation times is stationary.  
\end{enumerate}
\proof
Recall that in RG-HMM, we assume the latent network evolves according to the continuous-time percolation model, i.e. a network-valued bi-variate continuous-time Markov chain $\{W(t), G(t), t\geq 0\}$ taking on values in a finite set $\mathcal{X} = \{0, 1\} \times \{1, 2, 3, \cdots, 2^{N\choose 2}\}$, where $N$ is the number nodes in networks. To simplify the notation, let $X_m = X(t_m) = (W(t_m), G(t_m))$ be the r.v. pair at $t_m$, i.e. the hidden state at $m$-th observational time points. 

When the $M$ observation moments $t_1<t_2< \cdots < t_M$ are evenly spaced, i.e. $t_m - t_{m-1} = t_{m+1} - t_{m} = \Delta t, \forall \, m = 2, \cdots M-1$, the transition probability for the Markov chain $\{ X_m, m\geq 1 \}$ is 
$$P(X_m = j|X_{m-1} = i) = P(X(t_m) = j|X(t _{m-1}) = i) = P_{ij}(t_m - t_{m-1}) = P_{ij}(\Delta t)$$
for all $m$, which is independent of $m$. Note that $P_{ij}(t)$ is the probability that the Markov chain $\{X(t), t \geq 0\}$ presently in state $i$ will be in state $j$ after an additional time $t$. 

\fi

\section{Details on PMF of Feasible Sample Path \label{Sample Path pdf}}

The probability function of feasible sample path $\bm S_1, \bm V_1$ conditional on $G(t_1), W(t_1)$ is 
\begin{equation}
\begin{split}
& \qquad p_{sp} \Big( \bm s_1, \bm v_1 \big \rvert g(t_1), w(t_1)) \\
&= P\Big(
\bm S_1 =\big(g(\tau_1), g(\tau2),...,g(\tau_{R_1})\big), \bm V_1 =\big(w(\tau_1), w(\tau2),...,w(\tau_{R_1})\big)\big \rvert G(t_1) = g(t_1), \\
& \qquad \quad  W(t_1) = w(t_1) \Big) \cdot P(\tau_{R_1} \leq t_2 < \tau_{R_1 + 1} \big \rvert G(t_1) = g(t_1),W(t_1) = w(t_1),\bm S_1 = \bm s_1, \bm V_1 = \bm v_1)
 \\
& =\prod_{r = 1}^{R_1} P\Big(
G(\tau_r)=g(\tau_r),W(\tau_r) = w(\tau_r)\big \rvert G(\tau_{r-1}) = g(\tau_{r-1}),W(\tau_{r-1}) = w(\tau_{r-1})
\Big)  \\
& \qquad \cdot P(\tau_{R_1} \leq t_2 < \tau_{R_1 + 1} \big \rvert G(t_1) = g(t_1),W(t_1) = w(t_1),\bm S_1 = \bm s_1, \bm V_1 = \bm v_1)
 \\
& = \prod_{r = 1}^{R_1} f\Big(
g(\tau_r),w(\tau_r)\big \rvert g(\tau_{r-1}), w(\tau_{r-1})
\Big)  \\
& \qquad \cdot P(\tau_{R_1} \leq t_2 < \tau_{R_1 + 1} \big \rvert G(t_1) = g(t_1),W(t_1) = w(t_1),\bm S_1 = \bm s_1, \bm V_1 = \bm v_1)
 \\
 & =  \prod_{r = 1}^{R_1}
g\Big( g(\tau_r)\big \rvert w(\tau_{r}), g(\tau_{r-1})
\Big) h\Big( w(\tau_r)\big \rvert g(\tau_{r-1}), w(\tau_{r-1})
\Big) \\
& \qquad \cdot P(\tau_{R_1} \leq t_2 < \tau_{R_1 + 1} \big \rvert G(t_1) = g(t_1),W(t_1) = w(t_1),\bm S_1 = \bm s_1, \bm V_1 = \bm v_1),
 \\
\end{split}
\end{equation} 
where 
\begin{equation}
\begin{split}
& \qquad 
P(\tau_{R_1} \leq t_2 < \tau_{R_1 + 1} \big \rvert G(t_1) = g(t_1),W(t_1) = w(t_1),\bm S_1 = \bm s_1, \bm V_1 = \bm v_1)\\
&= P({R_1}\text{ transitions between }(t_1, t_2])  \\
& = \exp (-\gamma (t_2 -t_1)) 
\frac{(\gamma (t_2 -t_1))^{R_1}}{R_1!}
\qquad \text{since } R_1 \sim poisson (\gamma(t_2-t_1)).
\end{split}
\end{equation} 
Therefore, 
\begin{equation}
\begin{split}
& \qquad p_{sp} \Big( \bm s_1, \bm v_1 \big \rvert g(t_1), w(t_1)) \\
&= \exp (-\gamma (t_2 -t_1)) 
\frac{(\gamma (t_2 -t_1))^{R_1}}{R_1!}
 \cdot 
\prod_{r = 1}^{R_1}
g\Big( g(\tau_r)\big \rvert w(\tau_{r}), g(\tau_{r-1})
\Big) h\Big( w(\tau_r)\big \rvert g(\tau_{r-1}), w(\tau_{r-1})
\Big),
\end{split}
\end{equation}
where $\bm s_1 = (g(\tau_1), g(\tau_2),...,g(\tau_R)), \bm v_1 = (w(\tau_1), w(\tau_2),...,w(\tau_R)), \tau_0 = t_1$,  $g(\cdot\big \rvert\cdot)$ is the conditional pmf of r.v. $G(\tau_2)$ given $\big(W(\tau_2), G(\tau_1)\big)$, and $h(\cdot\big \rvert\cdot)$ is the conditional pmf of r.v. $W(\tau_2)$ given $\big(G(\tau_1), W(\tau_1)\big)$. $\tau_1,\tau_2$ are two arbitrary consecutive transition time points, $\tau_1<\tau_2$. More details on $g(\cdot\big \rvert\cdot)$ and $h(\cdot\big \rvert\cdot)$on are provided in section \ref{transition prob.} of this Supplement. 

Generally, for $m=2, 3,...,M$, the pmf of feasible sample path $\bm S_{m-1}, \bm V_{m-1}$ that brings $g(t_{m-1})$ to $g(t_{m})$ conditional on $G(t_{m-1}), W(t_{m-1})$ is
\begin{equation}
\begin{split}
& \qquad p_{sp} \Big( \bm s_{m-1}, \bm v_{m-1} \big \rvert g(t_{m-1}), w(t_{m-1})) \\
&= \exp (-\gamma (t_{m} -t_{m-1})) 
\frac{(\gamma (t_{m} -t_{m-1}))^{R_{m-1}}}{R_{m-1}!} \\
& \qquad
 \cdot 
\prod_{r = 1}^{R_{m-1}}
g\Big( g(\tau_r^{(m-1)})\big \rvert w(\tau_{r}^{(m-1)}), g(\tau_{r-1}^{(m-1)})
\Big) h\Big( w(\tau_r^{(m-1)})\big \rvert g(\tau_{r-1}^{(m-1)}), w(\tau_{r-1}^{(m-1)})
\Big),  
\end{split}
\label{sample_path}
\end{equation}
where $\bm s_{m-1} = (g(\tau_1^{(m-1)}), g(\tau_2^{(m-1)}),...,g(\tau_{R_{m-1}}^{(m-1)})), \bm v_{m-1} =(w(\tau_1^{(m-1)}), w(\tau_2^{(m-1)}),...,w(\tau_{R_{m-1}}^{(m-1)}))$ is the sample path from $g(t_{m-1})$ to $g(t_m)$ between $t_{m-1}$, and $t_{m}$; $(\tau_1^{(m-1)}, ..., \tau_{R_{m-1}}^{(m-1)})$ are the $R_{m-1}$ transition time points between $t_{m-1}$ and $t_{m}$; $\tau_0^{(m-1)} = t_{m-1}$.

\section{Transition Probability for Percolation Models \label{transition prob.}}
Recall that $\mathcal{X} = \{0, 1\} \times \{1, 2, 3, \cdots, 2^{N\choose 2}\}$, where $N$ is the number of nodes in networks. The one-step transition probability for the network-valued Markov chain, $\{\big(G(t),W(t)\big)\in \mathcal{X}, t_1\leq t \leq t_M\}$, from $(g_1,w_1)$ to $(g_2,w_2)$ in a single step is 
\begin{equation}
\begin{split}
&P \Big(
\big(G(\tau_2),W(\tau_2)\big) = \big(g_2,w_2\big)\big \rvert\big(G(\tau_1), W(\tau_1)\big)  =  \big(g_1,w_1\big) 
\Big)\\
&= P\Big(
G(\tau_2)=g_2,W(\tau_2) = w_2\big \rvert G(\tau_1) = g_1,W(\tau_1) = w_1
\Big) \\
&= P\Big(
G(\tau_2)=g_2\big \rvert W(\tau_2)=w_2,G(\tau_1)=g_1,W(\tau_1)=w_1
\Big) 
\cdot \\
& \qquad \qquad \qquad \qquad\qquad \qquad \qquad P\Big(
W(\tau_2)=w_2\big \rvert G(\tau_1)=g_1,W(\tau_1)=w_1
\Big) \\
&= P\Big(
G(\tau_2)=g_2\big \rvert W(\tau_2)=w_2,G(\tau_1)=g_1
\Big) 
\cdot
P\Big(
W(\tau_2)=w_2\big \rvert G(\tau_1)=g_1,W(\tau_1)=w_1
\Big) \\ 
&\qquad (\text{since }G(\tau_2) \independent W(\tau_{1}) \big \rvert G(\tau_{1}), W(\tau_{2})),
\end{split}
\end{equation}
where $\tau_1,\tau_2$ are two arbitrary consecutive transition time points, $\tau_1<\tau_2$. The first probability is specified by the percolation model, specifically by the way of choosing edge to add or delete, and the second probability is given in Table 1.
%\ref{tab1}.

We can rewrite the one-step transition probability as 
\begin{equation}
\label{1.14}
f(g_2,w_2\big \rvert g_1,w_1) = g (g_2\big \rvert w_2,g_1) \cdot h(w_2\big \rvert g_1,w_1),
\end{equation}
where $f(\cdot\big \rvert\cdot)$ is the conditional pmf of rvs $\big(G(\tau_2),W(\tau_2)\big)$ given $\big(G(\tau_1),W(\tau_1)\big)$, $g(\cdot\big \rvert\cdot)$ is the conditional pmf of rv. $G(\tau_2)$ given $\big(W(\tau_2), G(\tau_1)\big)$, and $h(\cdot\big \rvert\cdot)$ is the conditional pmf of rv. $W(\tau_2)$ given $\big(G(\tau_1), W(\tau_1)\big)$. $\tau_1,\tau_2$ are two arbitrary consecutive transition time points, $\tau_1<\tau_2$.

Specifically, $g(g_2\big \rvert w_2,g_1)$ is parameter-free, only depend on edge change rule in the percolation model.
\begin{equation}
\begin{split}
h(w_2\big \rvert g_1,w_1) & = h_{p,q}(w_2\big \rvert g_1,w_1) \\
& = w_2 \bm 1_{\{g_1 = \text{Empty} \}} + 
(1-w_2) \bm 1_{\{g_1 = \text{Complete} \}} + \\
&
\big(p^{w_2}(1-p)^{1-w_2} \bm 1_{\{w_1 = 0 \}} +
      (1-q)^{w_2}q^{1-w_2}  \bm 1_{\{w_1 = 1 \}}
 \big)
\bm 1_{\{g_1 \neq \text{Empty or Complete}  \}}.
\end{split}
\end{equation}

\section{Proof of Missing Information Principal\label{MissingInformationPrincipal}}
Recall that $\bm \theta = (p, q, \gamma)$, we prove below that 
\begin{equation}
     S(\bm \theta; \bm w_{2:M},\bm g_{2:M})  =  E \big[ S(\bm \theta; \bm w_{2:M},  \bm g_{2:M}, \bm S, \bm V)\big \rvert\bm W_{2:M} = \bm w_{2:M}, \bm G_{2:M} =\bm g_{2:M} \big] 
\end{equation}
\proof
\begin{equation}
    \begin{split}
        Right & = \sum_{\bm s,\bm v} \frac{\partial}{\partial \bm \theta} \log f_{\bm \theta} (\bm w, \bm g, \bm s, \bm v) \cdot f(\bm s, \bm v|\bm w, \bm g) \\
        & = \sum_{\bm s,\bm v} \frac{\partial}{\partial \bm \theta} \left( \log f_{\bm \theta} (\bm s, \bm v \,|\,\bm w, \bm g) + \log f_{\bm \theta} (\bm w, \bm g)  \right) \cdot f(\bm s, \bm v|\bm w, \bm g) \\
        & = \sum_{\bm s,\bm v} \frac{\partial}{\partial \bm \theta} \log f_{\bm \theta} (\bm s, \bm v \,|\,\bm w, \bm g) \cdot f(\bm s, \bm v|\bm w, \bm g) +  \sum_{\bm s,\bm v} \frac{\partial}{\partial \bm \theta} \log f_{\bm \theta} (\bm w, \bm g) \cdot f(\bm s, \bm v|\bm w, \bm g) \\
        & = 0 +  \frac{\partial}{\partial \bm \theta} \log f_{\bm \theta} (\bm w, \bm g) \sum_{\bm s,\bm v}  f(\bm s, \bm v|\bm w, \bm g) \\
        & =  0 +  \frac{\partial}{\partial \bm \theta} \log f_{\bm \theta} (\bm w, \bm g) \cdot 1 = Left
    \end{split}
\end{equation}
Notice that $ \sum_{\bm s,\bm v} \frac{\partial}{\partial \bm \theta} \log f_{\bm \theta} (\bm s, \bm v \,|\,\bm w, \bm g) \cdot f(\bm s, \bm v|\bm w, \bm g) = \sum_{\bm s, \bm v} \frac{\frac{\partial}{\partial \bm \theta} f_{\bm \theta}(\bm s, \bm v | \bm w, \bm g) }{f_{\bm \theta}(\bm s, \bm v | \bm w, \bm g)} \cdot f(\bm s, \bm v | \bm w, \bm g) = \sum_{\bm s, \bm v} \frac{\partial}{\partial \bm \theta} f_{\bm \theta}(\bm s, \bm v | \bm w, \bm g) = \frac{\partial}{\partial \bm \theta} \sum_{\bm s, \bm v} f_{\bm \theta}(\bm s, \bm v | \bm w, \bm g) = \frac{\partial}{\partial \bm \theta} 1 = 0$

%\section{Details on Simulating Sample Paths by Metropolis-Hasting Algorithm \label{MH algorithm}}

\section{Particle Filtering Procedure\label{subsection3.3}}
The particle filtering procedure is as follows:
\begin{enumerate}
    \item Start from the first observed network $g^\star(t_1)$. Set $G(t_1) = g^\star(t_1)$ and $W(t_1) = 1$.
    \item 
    To create $B$ number of particles at observation time $t_2$, denoted by $\{w^b(t_2), g^b(t_2)\}_{b=1}^B$, repeat the following for $b = 1,\cdots,B$.
    \begin{enumerate}
    \item Start from $W(t_1) = 1, G(t_1) = g^\star(t_1)$ and simulate according to a continuous time ER/PR process up until time $t_2-t_1$. Then set $w^b(t_2), g^b(t_2)$ equal to the current binary variable value and network in the simulated chain.
    \end{enumerate}
\item For $m = 2, \cdots, M-1$, repeat the following steps.

  \begin{enumerate}
      \item For each one of the simulated pairs in $\{(w^b(t_m), g^b(t_m))\}_{b=1}^B$, calculate the conditional probability $p_b  := P(G^\star(t_m) = g^\star(t_m)
      \big \rvert G(t_m) = g^b(t_m)  )$ with current parameter estimates. 
      \item To create $B $ number of particles at time $t_{m+1}$, repeat the following for $b = 1,\cdots, B$.
      \begin{enumerate}
          \item Sample an index $\eta$ from $1,2,...,B$ with probability proportional to $p_1, p_2,...,p_B$.
          \item Start from $W(t_m) = w^\eta(t_m), G(t_m) = g^\eta(t_m)$ and simulate according to a continuous time ER/PR process up until time $t_{m+1}-t_m$. Then set $w^b(t_{m+1}), g^b(t_{m+1})$ equal to the current binary variable value and network in the simulated chain.
      \end{enumerate}
  \end{enumerate}

%\item To create $\Psi$ ancestral lines, or particle path, denoted by $\{ \bm w^\psi_{2:M}, \bm g^\psi_{2:M}\}_{\psi = 1}^\Psi$, as samples from the conditional distribution of $\bm W_{2:M}, \bm G_{2:M}$ given the observed networks $\bm g_{1:M}^\star$, we first calculate the conditional probability at the last observation time $p_b  := P(G^\star(t_M) = g^\star(t_M)
%      \big \rvert G(t_M) = g^b(t_M)  )$ with current parameter estimates, for $b = 1,\cdots, B$, then repeat the following for $\psi = 1,...,\Psi$. 
%  \begin{enumerate}
%    \item Sample an index $\eta$ from $1,2,...,B$ with probability proportional to $p_1, p_2,...,p_B$.
%     \item Trace back the ancestral line for the $\eta$-th particle at time $t_M$ (i.e. $(w^\eta(t_M), g^\eta(t_M)))$, and set the ancestral line as one sample from the conditional distribution of $\bm W_{2:M}, \bm G_{2:M}$, given $\bm g^\star_{2:M}$.
%  \end{enumerate}

\item 
To obtain $\Psi$ samples from the conditional distribution of $\bm W_{2:M}, \bm G_{2:M}$ given the observed networks $\bm g_{1:M}^\star$, we sample $\Psi$ ancestral lines, or particle paths, denoted by $\{ \bm w^\psi_{2:M}, \bm g^\psi_{2:M}\}_{\psi = 1}^\Psi$ with the following procedure. We first calculate the conditional probability at the last observation time $p_b  := P(G^\star(t_M) = g^\star(t_M)
      \big \rvert G(t_M) = g^b(t_M)  )$ with current parameter estimates, for $b = 1,\cdots, B$, then repeat the following for $\psi = 1,...,\Psi$. 
  \begin{enumerate}
    \item Sample an index $\eta_M$ from $1,2,...,B$ with probability proportional to $p_1, p_2,...,p_B$.
     \item Trace back the ancestral line for the $\eta_M$-th particle, i.e. $(w^{\eta_M}(t_M), g^{\eta_M}(t_M))$, at time $t_M$. In other words, identify the ancestral particle index for the $\eta_M$-th particle at time $t_M$, and call it $\eta_{M-1}$, then identify the ancestral particle index for $\eta_{M-1}$-th particle at time $t_{M-1}$, and call it $\eta_{M-2}$, and continue to do this backward for $\eta_{m}$-th particle at time $t_{m}$, $m=M-2, \cdots, 2$. Note that at each observation moment, each particle has exactly one parent. Then set the ancestral line, i.e. $\{(w^{\eta_{m}}(t_m), g^{\eta_{m}}(t_m)) \}_{m=2}^M$ as one sample from the conditional distribution of $\bm W_{2:M}, \bm G_{2:M}$, given $\bm g^\star_{2:M}$. 
  \end{enumerate}

\end{enumerate}
The particle filtering allows us to sample $\Psi$ network and binary variable sequences that are more likely to generate the observed networks in the error process. It generates $B$ possible latent variables at each observation time, thus effectively reducing the space from which we will sample from for the particle paths. It also allows us to sample true network and binary sequences without actually having the transition probabilities in closed form.
%in the next step. 

\section{Simulating the Sample Path}
The inner expectations 
%in \eqref{gamma}, \eqref{p}, and \eqref{q} 
in formulas for $\hat \gamma$, $\hat p$ and $\hat q$ (equation (9)-(11) in the main papar) 
are taken with respect to the conditional distribution of sample path $\bm S, \bm V$ that connects the given true sequences $\bm w_{2:M}, \bm g_{2:M}$ at the observation times in the hidden layer. In order to estimate those expectations, we propose to sample from the conditional distributions by simulating paths using an MCMC method. We draw upon the work of %Snijders et al. 
 \cite{snijders2010maximum} where draws are generated by the Metropolis-Hastings algorithm, using a proposal distribution consisting of small changes in a sample path that brings one network to the next.
 
 Specifically, for each $m=2, \cdots, M$, we need to generate a sample path $\bm s_{m-1}, \bm v_{m-1}$ that brings $(w(t_{m-1}), g(t_{m-1}))$ to $(w(t_{m}), g(t_{m}))$. The probability function of the sample path, conditional on $(w(t_{m-1}), g(t_{m-1}))$ is given in \eqref{sample_path}. To simplify the notation, denote the sample path and its conditional probability function by $\bm u$ and $p(\bm u)$, respectively. The acceptance ratio in the Metropolis-Hasting algorithm for a current state $\bm u$ and a proposed state $\bm {\tilde u}$ is 
 \begin{equation}
     \min\{1, \frac{p(\bm {\tilde u}) q(\bm u|\bm {\tilde u}) }{ p(\bm u) q(\bm {\tilde u}|\bm u )}\}
 \end{equation}
 where $q(\bm {\tilde u}|\bm u )$ is the proposal distribution under which a new candidate sample path $\bm {\tilde u}$ is proposed by making small changes in the current sample path $\bm u$. The small changes consist of three operations: paired deletion, paired insertion and permutation. Paired deletion involves removing a pair of same edge changes in sample path. For example, if sample path $\bm u$ consists of adding an edge at one time and deleting the same edge after, then a new sample path $\bm {\tilde u}$ can be obtained by removing such pair of same edge changes from $\bm u$ since the two edge changes offset each other producing no change on the network. Similarly, paired insertion involves adding a pair of same edge changes in a current sample path, and permutation involves permuting all edge changes in a current sample path. All of these operations are done in such a way that the proposed new sample path $\bm {\tilde u}$ is still a feasible path with only one edge change at a time connecting the two networks at the consecutive observation times in the hidden layer. More discussion of this MCMC procedure can be found in %Snijders et al. 
 \cite{snijders2010maximum}.

\section{Proof of Theorem 3.1
%\ref{theorem} 
\label{appendex_proof}}
\begin{customthm}{3.1}
Let $\pi_{\bm \Gamma}(\bm x)$ be the stationary distribution of a Markov chain in RG-HMM $\{\bm X(t_m), G^\star(t_m)\}$ and $\mathcal{L}_0$ be the limiting covariance matrix of 
%$m^{-1/2} \frac{\partial}{\partial \bm{\Gamma}} \log p_{\bm \Gamma}( G^\star(t_1), \cdots,  G^\star(t_m))$. 
$m^{-1/2} \frac{\partial}{\partial \bm{\Gamma}} \log p_{\bm \Gamma}( \bm G^\star_{1:m})$. 
Assume (C1) that $t_m - t_{m-1} = t_{m+1} - t_{m} = \Delta t, \forall \, m \geq 2$, and stationarity is reached at $t_1$, (C2) that $\forall \, \bm x \in \mathcal{X}$, $\bm \Gamma \rightarrow \pi_{\bm \Gamma}(\bm x)$ has two continuous derivatives in some small neighborhood of $\bm \Gamma_0$, and (C3) that $\mathcal{L}_0$ is nonsingular. Then $m^{1/2}(\bm{\hat \Gamma} - \bm{ \Gamma_0}) \xrightarrow{p} N(0, \mathcal{L}_0^{-1})$ as $m \rightarrow \infty$.
\label{theorem}
\end{customthm}
\proof
%Bickel et al.
\cite{bickel1998asymptotic} provide a central limit theorem for the MLE of a general HMM with stationary Markov chain under 6 regularity conditions (A1)-(A6). Our approach is to show that (A1)-A(6) hold for our RG-HMM under assumptions (C1)-(C3). 

Let $P_{ab}(t)$ denote the probability that the Markov chain $\{\bm X(t), t \geq 0\}$ in our RG-HMM presently in state $a$ will be in state $b$ after an additional time $t$, $P_{ab}$ denote the one-step transition probability that the process will next enter state $b$ when it leaves state $a$ (details in Appendix \ref{transition prob.}), $P^n_{ab}$ denote the $n$-step transition probability.
%and $q_{ab}$ denote the transition rate of continuous-time Markov chain, i.e. $q_{ab} = \gamma P_{ab}$. 

Under (C1), the $M$ observation moments $t_1<t_2< \cdots < t_M$ are evenly spaced at the stationary period of Markov chain, then the transition probability for the discrete-time Markov chain $\{ \bm X(t_m), m\geq 1 \}$ in our RG-HMM is $P(\bm X(t_m) = b|\bm X(t _{m-1}) = a) = P_{ab}(t_m - t_{m-1}) = P_{ab}(\Delta t)$, $\forall \, m$, which is independent of $m$. Hence, (C1) ensures that $\{\bm X(t_m)\}$ is a stationary Markov chain with homogeneous transition probability matrix $\{\alpha_{\bm \Gamma_0}(a, b)\} = \{P_{ab}(\Delta t)\}$. We now restate (A1)-(A6) and show that they all hold for our RG-HMM $\{\bm X(t_m), G^\star(t_m)\}$. 

(A1) The transition probability matrix $\{\alpha_{\bm \Gamma_0}(a, b)\}$ is ergodic, that is, irreducible and aperiodic. 

(A2) For all $a$, $b \in \mathcal{X}$ and $g^\star \in \{1, 2, ..., 2^{N \choose 2}\}$, the maps $\bm \Gamma \rightarrow \alpha_{\bm \Gamma}(a, b)$, $\bm \Gamma \rightarrow \pi_{\bm \Gamma}(a)$ and $\bm \Gamma \rightarrow g_{\bm \Gamma}(g^\star| a)$ have two continuous derivatives in some neighborhood of $\bm \Gamma_0$.

(A3) Write $\bm \Gamma = ( \Gamma_1, \cdots, \Gamma_d) \equiv (p, q, \gamma, \alpha, \beta)$. There exists a $\delta > 0$ such that (i) for all $1 \leq i \leq d$ and all $a$, 
\begin{equation}
    E_{\bm \Gamma_0}\Big[\sup_{|\bm \Gamma - \bm \Gamma_0| < \delta}|\frac{\partial}{\partial \Gamma_i} \log g_{\bm \Gamma}(G^\star|a)|^2\Big] < \infty;
\end{equation}
(ii) for all $a \leq i, j \leq d$ and all $a$, 
\begin{equation}
    E_{\bm \Gamma_0}\Big[\sup_{|\bm \Gamma - \bm \Gamma_0| < \delta}|\frac{\partial^2}{\partial \Gamma_i \partial \Gamma_j} \log g_{\bm \Gamma}(G^\star|a)|\Big] < \infty;
\end{equation}
(iii) for $j=1, 2$, all $1 \leq i_l \leq d$, $l = 1, \cdots, j$ and all $a$, 
\begin{equation}
   \int \sup_{|\bm \Gamma - \bm \Gamma_0| < \delta}|\frac{\partial^j}{\partial \Gamma_{i_1} \partial \Gamma_{i_j}} g_{\bm \Gamma}(g^\star|a)| \nu (dg^\star) < \infty.
\end{equation}

(A4) There exists a $\delta >0$ such that with 
\begin{equation}
    \rho(g^\star) = \sup_{|\bm \Gamma - \bm \Gamma_0| < \delta} \max_{a,b \in \mathcal{X}} \frac{g_{\bm \Gamma }(g^\star | a)}{g_{\bm \Gamma }(g^\star | b)},
\end{equation}
$P_{\bm \Gamma_0}(\rho(G^\star(t_m)) = \infty | \bm X(t_m) = a) < 1$ for all $a$.

(A5) $\bm \Gamma_0$ is an interior point of the set $\Theta$, which is the set to which $\bm \Gamma$ belong. 

(A6) The MLE is strongly consistent, which requires (i) (A1); (ii) for all $a$, $b$ and $g^\star$, the map $\bm \Gamma \rightarrow \alpha_{\bm \Gamma}(a, b)$ and the map $\bm \Gamma \rightarrow g_{\bm \Gamma}(g^\star|a)$ are continuous on $\Theta$; (iii) for each $a$, $E_{\bm \Gamma_0}|\log g_{\bm \Gamma_0}(G^\star|a)| < \infty$; (iv) for each $a$ and $\bm \Gamma$ there is a $\delta >0$ such that $E_{\bm \Gamma_0}\Big[ \sup_{|\bm \Gamma^\prime - \bm \Gamma|<\delta}(\log g_{\bm \Gamma^\prime}(G^\star|a))^+\Big] < \infty$; (v) for each $\bm \Gamma$ such that the laws $P_{\bm \Gamma}$ and $P_{\bm \Gamma_0}$ agree, $\bm \Gamma = \bm \Gamma_0$. 

In our RG-HMM, all states in the Markov chain $\{\bm X(t_m)\}$ communicate with each other as any two networks can be connected by a network path that leads the one to the other, hence irreducible. Note that $|\mathcal{X}| < \infty$, then $\{\bm X(t_m)\}$ is positive recurrent as every irreducible Markov chain with a finite state space is positive recurrent. Also note that the transition probability 
\begin{equation}
    \alpha_{\bm \Gamma_0}(a, b) = P_{ab}(\Delta t) =  \sum_{n=0}^\infty P_{ab}^n \cdot  \exp (-\gamma \Delta t) 
\frac{(\gamma \Delta t)^{n}}{n!} \in (0, +\infty),
\label{G5}
\end{equation}
for all $a, b \in \mathcal{X}$, $p, q \in (0, 1)$ and $\gamma \in (0, +\infty)$. Let $\alpha^{(k)}_{\bm \Gamma_0}(a, b)$ denote the $k$-step transition probability of $\{\bm X(t_m)\}$, then the matrix $\{\alpha^{(k)}_{\bm \Gamma_0}(a, b)\}$ can be calculated by multiplying the matrix $\{\alpha_{\bm \Gamma_0}(a, b)\}$ by itself $k$ times, yielding positive diagonal elements, which means that each state can visit back itself in any number of transition steps, hence it is aperiodic. Therefore, (A1) holds, which also assures the existence of unique stationary distribution for $\{\bm X(t_m)\}$. 

Note that $P_{ab}$ is given in Appendix \ref{transition prob.} (example with 3-node network is provided in the end of the proof), and the matrix $\{P_{ab}^n\}$ can be calculated by multiplying the matrix $\{P_{ab}\}$ by itself $n$ times, thus we have 
\begin{equation}
    P_{ab}^n = \sum_{i=1}^{k(n)} c_{0i} p^{c_{1i}} q^{c_{2i}} (1-p)^{c_{3i}} (1-q)^{c_{4i}},
\label{G6}
\end{equation}
where $\{c_{ji}\}_{j=0}^4$ and $k(n)$ are constant depending on the values of $a$, $b$ and $n$. $c_{0i}$ is the term corresponding to the product of probabilities of random graph process, thus $c_{0i}\leq 1$. $k(n)$ can be interpreted as the number of feasible paths from $a$ to $b$ with length $n$, and each path contains $c_{1i} + c_{4i}$ edge additions and $c_{2i} + c_{3i}$ edge deletions, thus $0\leq c_{1i}, c_{2i}, c_{3i}, c_{4i} \leq n$.
Also 
%from \eqref{3.2}, 
we have 
\begin{equation}
    g_{\bm \Gamma}(g^\star|a) = \alpha^c(1-\alpha)^d \beta^b (1-\beta)^a,
\label{G7}
\end{equation}
where $a$, $b$, $c$, $d$ are constant depending on the true network $a$ and observed network $g^\star$.

From the analytical form of $\alpha_{\bm \Gamma_0}(a, b)$ and $ g_{\bm \Gamma}(g^\star|a)$ given in \eqref{G5}, \eqref{G6} and \eqref{G7}, along with assumption (C2), identifiability addressed in Appendix \ref{identifiability} and finite state space $\mathcal{X}$, we can verify that when $p, q \in (0, 1)$, $\gamma > 0$ and $\alpha, \beta \in (0, 0.5)$, (A2)-(A6) all hold. (A3) holds since the derivatives of $g_{\bm \Gamma}(g^\star|a)$ w.r.t $\alpha$ and $\beta$ are bounded in some neighborhood of $\Gamma_0$, and the expectation and integral involve finite summation of finite terms, thus finite as well. (A4) holds since $g_{\bm \Gamma}(g^\star|a) \neq 0$ with $\alpha, \beta \in (0, 0.5)$, and $\rho (g^\star)<\infty$ for all $g\star$ and $a$. (A5) holds since $\Theta$ is an open set, $\Theta = \{(p, q, \gamma, \alpha,\beta)\,|\, p, q \in(0,1), \gamma \in(0, \infty), \alpha,\beta \in(0, 0.5)\}$. (A6)-(i)(iii)(iv)(v) all hold since $g_{\bm \Gamma}(g^\star|a)$ are bounded in some neighborhood of $\Gamma_0$, and the model is identifiable under set $\Theta$. 

As it is hard to verify analytically the continuity regarding $\bm \Gamma \rightarrow \pi_{\bm \Gamma}(a)$, we make it the assumption (C2). We can verify that (A2) and (A6)-(ii) also hold under (C2). To see this is so, we show, as an example, that $\alpha_{\bm \Gamma_0}(a, b)$ is continuous in $p\in (0, 1)$ and the first and second derivatives of $\alpha_{\bm \Gamma_0}(a, b)$ w.r.t $p$ are continuous in some neighborhood of $p_0$. The continuity w.r.t other parameters $q, \gamma, \alpha, \beta$ can be verified using a similar argument. Define 
\begin{equation}
    f_n(p): =  P_{ab}^n \cdot  \exp (-\gamma \Delta t) 
\frac{(\gamma \Delta t)^{n}}{n!} =  \exp (-\gamma \Delta t) \frac{(\gamma \Delta t)^{n}}{n!} 
      \cdot
    \sum_{i=1}^{k(n)} c_{0i} p^{c_{1i}} q^{c_{2i}} (1-p)^{c_{3i}} (1-q)^{c_{4i}},
\end{equation}
\begin{equation}
    f(p) :=  \alpha_{\bm \Gamma_0}(a, b) = \sum_{n=0}^\infty f_n(p).
\end{equation}
Note that for any $\gamma$, $p$ and $n$, $|f_n(p)| \leq  M_n := \exp (-\gamma \Delta t) \frac{(\gamma \Delta t)^{n}}{n!}$, and $\sum_{n=1}^\infty M_n = 1< \infty $, then by Weierstrass M-test $\sum_{n=1}^\infty f_n(p)$ converges uniformly to $f(p)$ in $p \in (0, 1)$. Note that $f_n(p)$ is continuous, then $f(p)$ is continuous in $(0, 1)$. Under uniform convergence, we can interchange the order of summation and differentiation, which gives 
\begin{equation}
\begin{split}
    f^\prime(p) =  \sum_{n=0}^\infty f^\prime_n(p) 
    & =  \sum_{n=0}^\infty 
    \exp (-\gamma \Delta t) \frac{(\gamma \Delta t)^{n}}{n!} 
      \cdot
    \sum_{i=1}^{k(n)} c_{0i} p^{c_{1i}} q^{c_{2i}} (1-p)^{c_{3i}} (1-q)^{c_{4i}}( \frac{c_{1i}}{p} - \frac{c_{3i}}{1-p}). \\
\end{split}
\end{equation}
For $p \in (p_0-\epsilon, p_0 + \epsilon)$, we have 
\begin{equation}
\begin{split}
 \sum_{n=0}^\infty |f^\prime_n(p)| 
    & \leq  \sum_{n=0}^\infty 
    \exp (-\gamma \Delta t) \frac{(\gamma \Delta t)^{n}}{n!} 
      \cdot
    \sum_{i=1}^{k(n)} c_{0i} p^{c_{1i}} q^{c_{2i}} (1-p)^{c_{3i}} (1-q)^{c_{4i}}(\left| \frac{c_{1i}}{p}\right| + \left |\frac{c_{3i}}{1-p}\right|) \\
    & \leq  \sum_{n=0}^\infty 
    \exp (-\gamma \Delta t) \frac{(\gamma \Delta t)^{n}}{n!} 
      \cdot
    \sum_{i=1}^{k(n)} c_{0i} p^{c_{1i}} q^{c_{2i}} (1-p)^{c_{3i}} (1-q)^{c_{4i}}(\left| \frac{n}{p}\right|+\left|\frac{n}{1-p}\right|) \\
    & \leq \sum_{n=0}^\infty 
    \exp (-\gamma \Delta t) \frac{(\gamma \Delta t)^{n}}{n!} 
      \cdot (| \frac{n}{p}| + |\frac{n}{1-p}|) P_{ab}^n \\
      & \leq (\frac{1}{|p|} + \frac{1}{|1-p|}) \sum_{n=0}^\infty 
    \exp (-\gamma \Delta t) \frac{(\gamma \Delta t)^{n}}{n!} 
      \cdot n \cdot 1 < \infty,\\
\end{split}
\end{equation}
so again by Weierstrass M-test $\sum_{n=1}^\infty f_n^\prime (p)$ converges uniformly to $f^\prime(p)$ in some neighborhood of $p_0$. Then we have for $p \in (p_0-\epsilon, p_0 + \epsilon)$
\begin{equation}
\begin{split}
    f^{\prime \prime}(p) = \sum_{n=0}^\infty f^{\prime \prime }_n(p) &=  \sum_{n=0}^\infty 
    \exp (-\gamma \Delta t) \frac{(\gamma \Delta t)^{n}}{n!} 
      \cdot
    \sum_{i=1}^{k(n)} c_{0i} p^{c_{1i}} q^{c_{2i}} (1-p)^{c_{3i}} (1-q)^{c_{4i}}\left[( \frac{c_{1i}}{p} - \frac{c_{3i}}{1-p})^2 \right. \\
    & \qquad \qquad \qquad \qquad \qquad \qquad  \qquad \qquad \qquad \qquad  - \left .\frac{c_{1i}}{p^2} - \frac{c_{3i}}{(1-p)^2}\right],
\end{split}
\end{equation}
\begin{equation}
\begin{split}
  \sum_{n=0}^\infty |f^{\prime \prime }_n(p)| &\leq  \sum_{n=0}^\infty 
    \exp (-\gamma \Delta t) \frac{(\gamma \Delta t)^{n}}{n!} 
      \cdot
    \sum_{i=1}^{k(n)} c_{0i} p^{c_{1i}} q^{c_{2i}} (1-p)^{c_{3i}} (1-q)^{c_{4i}} \left | ( \right.\frac{c_{1i}}{p} - \frac{c_{3i}}{1-p})^2 \\
    & \qquad \qquad \qquad \qquad \qquad \qquad  \qquad \qquad \qquad \qquad  - \frac{c_{1i}}{p^2} - \frac{c_{3i}}{(1-p)^2}\left.\right | \\
    & \leq  ( \frac{1}{p^2} + \frac{1}{(1-p)^2} + \frac{2}{p(1-p)}) \sum_{n=0}^\infty 
    \exp (-\gamma \Delta t) \frac{(\gamma \Delta t)^{n}}{n!}\cdot n^2 + \\
    & \qquad \qquad 
     ( \frac{1}{p^2} + \frac{1}{(1-p)^2}) \sum_{n=0}^\infty 
    \exp (-\gamma \Delta t) \frac{(\gamma \Delta t)^{n}}{n!} \cdot n < \infty,
\end{split}
\end{equation}
hence $\sum_{n=1}^\infty f_n^{\prime\prime} (p)$ converges uniformly to $f^{\prime\prime}(p)$ in some neighborhood of $p_0$. Since $f_n^{\prime}(p)$ and $f_n^{\prime\prime}(p)$ are continuous in some neighborhood of $p_0$, and they are all uniformly convergent, then $f^{\prime}(p)$ and $f^{\prime\prime}(p)$ are continuous in some neighborhood of $p_0$.

\begin{example} 
\label{3-node_example}
The one-step transition matrix $\{P_{ab}\}$ for networks with $3$ nodes under ER model is shown in \eqref{pab_mat}, where the state space is $\mathcal{X} = \{0, 1\} \times \{1, 2, \cdots, 8\}) \setminus \{(0, 8), (1, 1)\}$ as complete network cannot be obtained by deleting an edge and empty network cannot be obtained by adding an edge. And the mapping between labels and networks is given in Figure \ref{mapping}.

\begin{equation}
\{P_{ab}\} = \begin{smallmatrix}
& (0, 1) & (0,2) & (0, 3) & (0, 4) & (0, 5) & (0, 6) & (0, 7) & (1, 2) & (1, 3) & (1, 4) & (1, 5) & (1, 6) & (1, 7)  & (1, 8)\\
(0, 1) & 0 & 0 & 0 & 0 & 0 & 0 & 0 &\frac{1}{3}  & \frac{1}{3} & \frac{1}{3} & 0 & 0 & 0 & 0  \\\hline
(0, 2) & 1-p & 0 & 0 & 0 & 0 & 0 & 0 &0  & 0 & 0 & \frac{1}{2}p & 0 & \frac{1}{2}p & 0  \\
(0, 3) & 1-p & 0 & 0 & 0 & 0 & 0 & 0 &0  & 0 & 0 & \frac{1}{2}p & \frac{1}{2}p & 0 & 0  \\
(0, 4) & 1-p & 0 & 0 & 0 & 0 & 0 & 0 &0  & 0 & 0 & 0 & \frac{1}{2}p & \frac{1}{2}p & 0  \\
\hline
(0, 5) & 0 & \frac{1}{2}(1-p)  & \frac{1}{2}(1-p) & 0 & 0 & 0 & 0 &0  & 0 & 0 & 0 & 0 & 0 & p  \\
(0, 6) & 0 & 0 & \frac{1}{2}(1-p)  & \frac{1}{2}(1-p) & 0 & 0 & 0 &0  & 0 & 0 & 0 & 0 & 0 & p  \\
(0, 7) & 0 & \frac{1}{2}(1-p)  & 0 & \frac{1}{2}(1-p) & 0 & 0 &0  & 0 & 0 & 0 & 0 & 0 & 0 & p  \\\hline
(1, 2) & q & 0 & 0 & 0 & 0 & 0 & 0 &0  & 0 & 0 & \frac{1}{2}(1-q) & 0 & \frac{1}{2}(1-q) & 0  \\
(1, 3) & q & 0 & 0 & 0 & 0 & 0 & 0 &0  & 0 & 0 & \frac{1}{2}(1-q) & \frac{1}{2}(1-q) & 0 & 0  \\
(1, 4) & q & 0 & 0 & 0 & 0 & 0 & 0 &0  & 0 & 0 & 0 & \frac{1}{2}(1-q) & \frac{1}{2}(1-q) & 0  \\\hline
(1, 5) & 0 & \frac{1}{2}q  & \frac{1}{2}q & 0 & 0 & 0 & 0 &0  & 0 & 0 & 0 & 0 & 0 & 1-q  \\
(1, 6) & 0 & 0 & \frac{1}{2}q  & \frac{1}{2}q & 0 & 0 & 0 &0  & 0 & 0 & 0 & 0 & 0 & 1-q  \\
(1, 7) & 0 & \frac{1}{2}q  & 0 & \frac{1}{2}q & 0 & 0 &0  & 0 & 0 & 0 & 0 & 0 & 0 & 1-q  \\\hline
(1, 8) & 0 & 0 & 0 & 0 &\frac{1}{3}  & \frac{1}{3} & \frac{1}{3} & 0 & 0 & 0 & 0 & 0 & 0 & 0   \\
\end{smallmatrix}
\label{pab_mat}
\end{equation}

\begin{figure}[H]
    \centering
\hspace{-3in}
\vspace{-0.1in}
\begin{tikzpicture}[auto, node distance=1.1cm,  state/.style={circle, draw, minimum size=0.3cm}]
  \node[state](1) {\tiny A};
  \node[state](2) [below left of=1] {\tiny B};
  \node[state](3) [below right of=1] {\tiny C};
\hspace{1.0in} 
  \node[state](1) {\tiny A};
  \node[state](2) [below left of=1] {\tiny B};
  \node[state](3) [below right of=1] {\tiny C};
  \path[every node/.style={font=\sffamily\small}]
    (1) edge node {} (2);
\hspace{1.0in} 
  \node[state](1) {\tiny A};
  \node[state](2) [below left of=1] {\tiny B};
  \node[state](3) [below right of=1] {\tiny C};
  \path[every node/.style={font=\sffamily\small}]
    (1) edge node {} (3);
\hspace{1.0in} 
  \node[state](1) {\tiny A};
  \node[state](2) [below left of=1] {\tiny B};
  \node[state](3) [below right of=1] {\tiny C};
  \path[every node/.style={font=\sffamily\small}]
    (2) edge node {} (3);
\end{tikzpicture}\\
1 \hspace{0.8in} 2  \hspace{0.8in} 3\hspace{0.8in} 4 \\~\\
\hspace{-3in}
\vspace{-0.1in}
\begin{tikzpicture}[auto, node distance=1cm, state/.style={circle, draw, minimum size=0.3cm}]
  \node[state](1) {\tiny A};
  \node[state](2) [below left of=1] {\tiny B};
  \node[state](3) [below right of=1] {\tiny C};
    \path[every node/.style={font=\sffamily\small}]
    (1) edge node {} (2)
    (1) edge node {} (3);
\hspace{1.0in} 
  \node[state](1) {\tiny A};
  \node[state](2) [below left of=1] {\tiny B};
  \node[state](3) [below right of=1] {\tiny C};
  \path[every node/.style={font=\sffamily\small}]
    (1) edge node {} (3)
    (2) edge node {} (3);
\hspace{1.0in} 
  \node[state](1) {\tiny A};
  \node[state](2) [below left of=1] {\tiny B};
  \node[state](3) [below right of=1] {\tiny C};
  \path[every node/.style={font=\sffamily\small}]
    (1) edge node {} (2)
    (2) edge node {} (3);
\hspace{1.0in} 
  \node[state](1) {\tiny A};
  \node[state](2) [below left of=1] {\tiny B};
  \node[state](3) [below right of=1] {\tiny C};
  \path[every node/.style={font=\sffamily\small}]
    (2) edge node {} (3)
    (1) edge node {} (2)
    (1) edge node {} (3);
\end{tikzpicture}\\
5 \hspace{0.8in} 6  \hspace{0.8in} 7\hspace{0.8in} 8
\caption{All possible network values for $G(t)$ with $N = 3$ nodes indexed from $1$ to $8$ in state space.}
\label{mapping}
\end{figure}
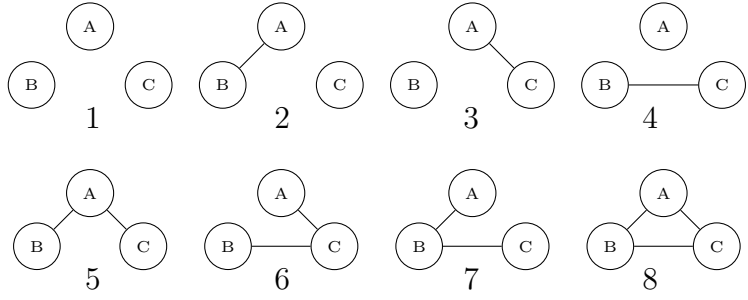

\end{example}

% \section{Large-sample Approximations for Bayes Factors}
% %Kass and Raftery 
% \cite{kass1995bayes} proposed two large-sample approximations for Bayes factors, shown as follows:
% \begin{enumerate}
%     \item Laplace's Method 
%     \begin{equation}
%             B_{fg} \approx \frac{f(\bm g^\star, \hat{\bm \Gamma}_{ER})}
%     {g(\bm g^\star, \hat{\bm\Gamma}_{ER})} \cdot 
%     \frac{(2\pi)^{d_f/2} p_f(\hat {\bm \Gamma}_{ER}) \Delta_{\bm \Gamma_{ER}}^{-1/2}}
%     {(2\pi)^{d_g/2} p_g(\hat  {\bm \Gamma}_{PR}) \Delta_{\bm \Gamma_{ER}}^{-1/2}},
%     \end{equation}
%     where $d_f$ and $d_g$ are the dimensions of $\bm \Gamma_{ER}$ and $\bm \Gamma_{PR}$, $\Delta_{\bm \Gamma_{ER}}$ and $\Delta_{\bm \Gamma_{PR}}$ are the information determinants for estimating $\bm \Gamma_{ER}$ and $\bm \Gamma_{PR}$ from $\bm g^\star$, and $\hat{\bm\Gamma}_{ER}$ and $\hat{\bm\Gamma}_{PR}$ are the MLEs under $H_f$ and $H_g$, respectively. 
%     \item The Schwarz Criterion
%     \begin{equation}
%     \log B_{fg} \approx 
%       \log f(\bm g^\star, \hat{\bm \Gamma}_{ER}) - \log g(\bm g^\star, \hat{\bm \Gamma}_{PR}) - \frac{1}{2}(d_f -d_g)\log(n),
%     \end{equation}
%       where $d_f$ and $d_g$ are the dimensions of $\bm \Gamma_{ER}$ and $\bm \Gamma_{PR}$, $\hat{\bm\Gamma}_{ER}$ and $\hat{\bm\theta}_{PR}$ are the MLEs under $H_f$ and $H_g$, respectively, and $n$ is the sample size. 
% \end{enumerate}

\section{Derivation of Recurrence Relation in Section 4.2  
%\eqref{4.5}
}
Define $Z_m := P\Big( \bm G^\star_{2:m} =  \bm g^\star_{2:m}| \bm X_1 = \big(1, g^\star(t_1) \big)\Big)$ for each $m = 2, \cdots M$, and $Z_1 := 1$. Also let $q_m(x) = P( \bm G^\star_{m} = \bm g^\star_{m}| \bm X_m =x)$, $\eta_m(x) = P(\bm X_m =x|  \bm G^\star_{2:m-1} = \bm g^\star_{2:m-1}, \bm X_1 = (1, g^\star(t_1))$ for each $m = 2, \cdots M$, with conventions $\eta_2(x) = P(\bm X_2 =x| \bm X_1 = (1, g^\star(t_1))$. To alleviate the notational burden, we omit the condition at $t_1$ in the probability expression when no confusion is possible.
\begin{equation}
\begin{split}
    Z_m & = P( \bm G^\star_{2:m} =  \bm g^\star_{2:m}) = 
    P(\bm G^\star_{2:m-1} = \bm g^\star_{2:m-1})
    P( \bm G^\star_{m} = \bm g^\star_{m}| \bm G^\star_{2:m-1} = \bm g^\star_{2:m-1}) \\
    & = Z_{m-1} \sum_x  P( \bm G^\star_{m} = \bm g^\star_{m}, \bm X_m =x|\bm G^\star_{2:m-1} =  \bm g^\star_{2:m-1}) \\
    & = Z_{m-1} \sum_x  P( \bm G^\star_{m} = \bm g^\star_{m}| \bm X_m =x, \bm G^\star_{2:m-1} = \bm g^\star_{2:m-1})
    P(\bm X_m =x|  \bm G^\star_{2:m-1} = \bm g^\star_{2:m-1}) \\
    & = Z_{m-1} \sum_x  P( \bm G^\star_{m} = \bm g^\star_{m}| \bm X_m =x)
    P(\bm X_m =x|  \bm G^\star_{2:m-1} = \bm g^\star_{2:m-1}) \\
    & = Z_{m-1} \sum_x q_m(x)\eta_m(x)\\ & =  Z_{m-1} E(q_m(\bm X)), \bm X\sim \eta_m(\cdot)
\end{split}
\end{equation}

\section{A Forward Algorithm with Particle Filtering}
We now present a forward algorithm with particle filtering 
%in algorithm \ref{alg2} 
to compute $Z_M^B$, i.e. the particle approximation of the marginal probability of observations.
\begin{algorithm}[!htb]
\SetAlgoLined
\DontPrintSemicolon
 \KwIn{$\bm g^\star = \big (g^\star(t_1),g^\star(t_2),..., g^\star(t_M) \big)$, $B$}
 \KwOut{$Z^B_M$}
 Set $Z_1^B = 1$, sample $\xi_2^b \overset{iid}{\sim} \eta_2(\cdot) = P\big(\bm X_2 = \cdot\, |\, \bm X_1 = (1, g(t_1))\big)$, $b = 1, \cdots, B$, and set $\eta_2^B(\cdot) = \frac{1}{B} \sum_{b=1}^B \bm 1\{\xi_2^b = \cdot\}$ \\
 For $m = 2, \cdots M$: set
   \begin{equation*}
  Z_{m}^B \gets Z_{m-1}^B\sum_x q_m(x)\eta_m^B(x)  =  Z_{m-1}^B  \frac{1}{B}\sum_{b=1}^B q_m(\xi_m^b)
   \end{equation*}
  Sample 
  \begin{equation}
      \xi_{m+1}^b\overset{iid}{\sim} \eta_{m+1}^B(\cdot) = \sum_{b=1}^B \frac{q_m(\xi_m^b) P(\bm X_{m+1}=\cdot \,|\,\bm X_{m} = \xi_m^b)}{ \sum_{i=1}^B q_m(\xi_m^b)},\, b = 1, \cdots, B \tag{$\star$}
\label{star}
  \end{equation}
  and set $\eta_{m+1}^B(\cdot) =\sum_{b=1}^B \bm 1\{\xi_{m+1}^b = \cdot\} $
\caption{Forward Algorithm with Particle Filtering}
\label{alg2}
\end{algorithm}
The time complexity is $\mathcal{O}(BM)$. Most of the computation is in the simulation of \eqref{star}, which can be done as 
%in step 3 of section \ref{subsection3.3}. 
in step 3, section 5 of this Supplement.

\section{Numerical Simulation Results for Estimation}

\begin{table}[H]
\small
\centering
\begin{tabular}{lrrrrr}
  \hline
 B & $\hat p$ & $\hat q$ & $\hat \gamma$& $\hat \alpha$ & $\hat \beta$ \\ 
  \hline
 50,000 & 0.675 (0.029) & 0.289 (0.057) & 1.739 (0.111) & 0.112 (0.019) & 0.037 (0.011) \\ 
 250,000 & 0.681 (0.033) & 0.278 (0.058) & 1.768 (0.107) & 0.092 (0.015) & 0.030 (0.007) \\ 
500,000 & 0.698 (0.041) & 0.286 (0.057) & 1.779 (0.100)& 0.083 (0.013) & 0.029 (0.010) \\ 
 1,000,000 & 0.698 (0.042) & 0.291 (0.054) & 1.794 (0.140) & 0.079 (0.015) & 0.026 (0.007) \\
   \hline
\end{tabular}
\caption{Numerical results for Figure 5 in the paper. Mean and standard deviation of parameter estimates based on $20$ simulations from the ER process with varying $B$, $N=20$, $M=50$.  }
\label{t5.2}
\end{table}

\begin{table}[H]
\small
\centering
\begin{tabular}{llrrrrr}
  \hline
 $N$ & $M$ & $\hat p$ & $\hat q$ & $\hat \gamma$ & $\hat \alpha$ & $\hat \beta$ \\ 
  \hline
 30 & 50 & 0.637 (0.042) & 0.335 (0.053) & 1.647 (0.111) & 0.089 (0.013) & 0.034 (0.011) \\ 
  & 100 & 0.669 (0.029) & 0.310 (0.044)& 1.728 (0.066)& 0.121 (0.017)& 0.032 (0.008)\\ 
     & 150 & 0.684 (0.027) & 0.307 (0.046)& 1.802 (0.084)& 0.143 (0.015) & 0.029 (0.007)\\ 
     & 200 & 0.684 (0.024)& 0.311 (0.034)& 1.920 (0.096)& 0.163 (0.023)& 0.026 (0.005)\\ 
    \hline
  50 & 50 & 0.572 (0.038) & 0.431 (0.078) & 1.518 (0.046) & 0.063 (0.005) & 0.029 (0.008) \\ 
  & 100 & 0.613 (0.029) & 0.374 (0.053) & 1.581 (0.054) & 0.082 (0.007) & 0.034 (0.007) \\ 
  & 150 & 0.644 (0.026) & 0.371 (0.027) & 1.656 (0.062) & 0.096 (0.009) & 0.034 (0.006) \\ 
  & 200 & 0.653 (0.023) & 0.337 (0.026) & 1.691 (0.046) & 0.107 (0.010) & 0.038 (0.005) \\ 
   \hline
 70 & 50 & 0.526 (0.035) & 0.537 (0.061) & 1.526 (0.055) & 0.050 (0.003) & 0.024 (0.004) \\ 
  & 100 & 0.572 (0.030) & 0.481 (0.060) & 1.533 (0.051) & 0.064 (0.004) & 0.030 (0.007) \\ 
  & 150 & 0.602 (0.020) & 0.444 (0.047) & 1.535 (0.038) & 0.075 (0.004) & 0.033 (0.005) \\ 
  & 200 & 0.611 (0.025) & 0.395 (0.037) & 1.580 (0.037) & 0.083 (0.005) & 0.036 (0.006) \\ 
   \hline
\end{tabular}
\caption{Numerical results for Figure 6 in the paper. Mean and standard deviation of parameter estimates based on $20$ simulations from ER process with $B=50000$, $(N, M) \in \{30, 50, 70\} \times \{50, 100, 150, 200\}$.}
\label{t5.3}
\end{table}

\section{Supplementary Simulation Results for Testing}
\begin{figure}[H]
   \centering
   \includegraphics[width=0.45\textwidth]{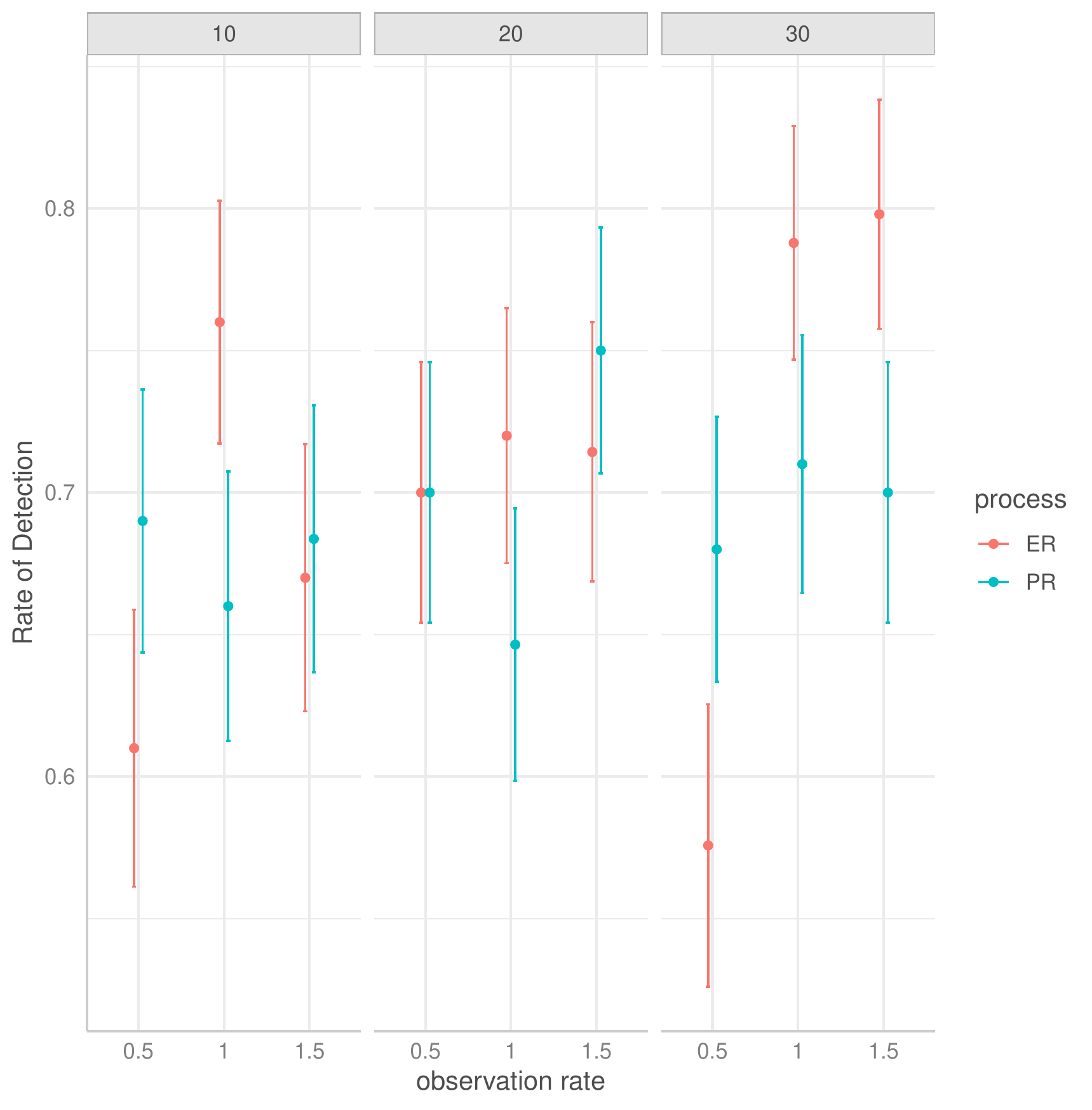}
    \includegraphics[width=0.45\textwidth]{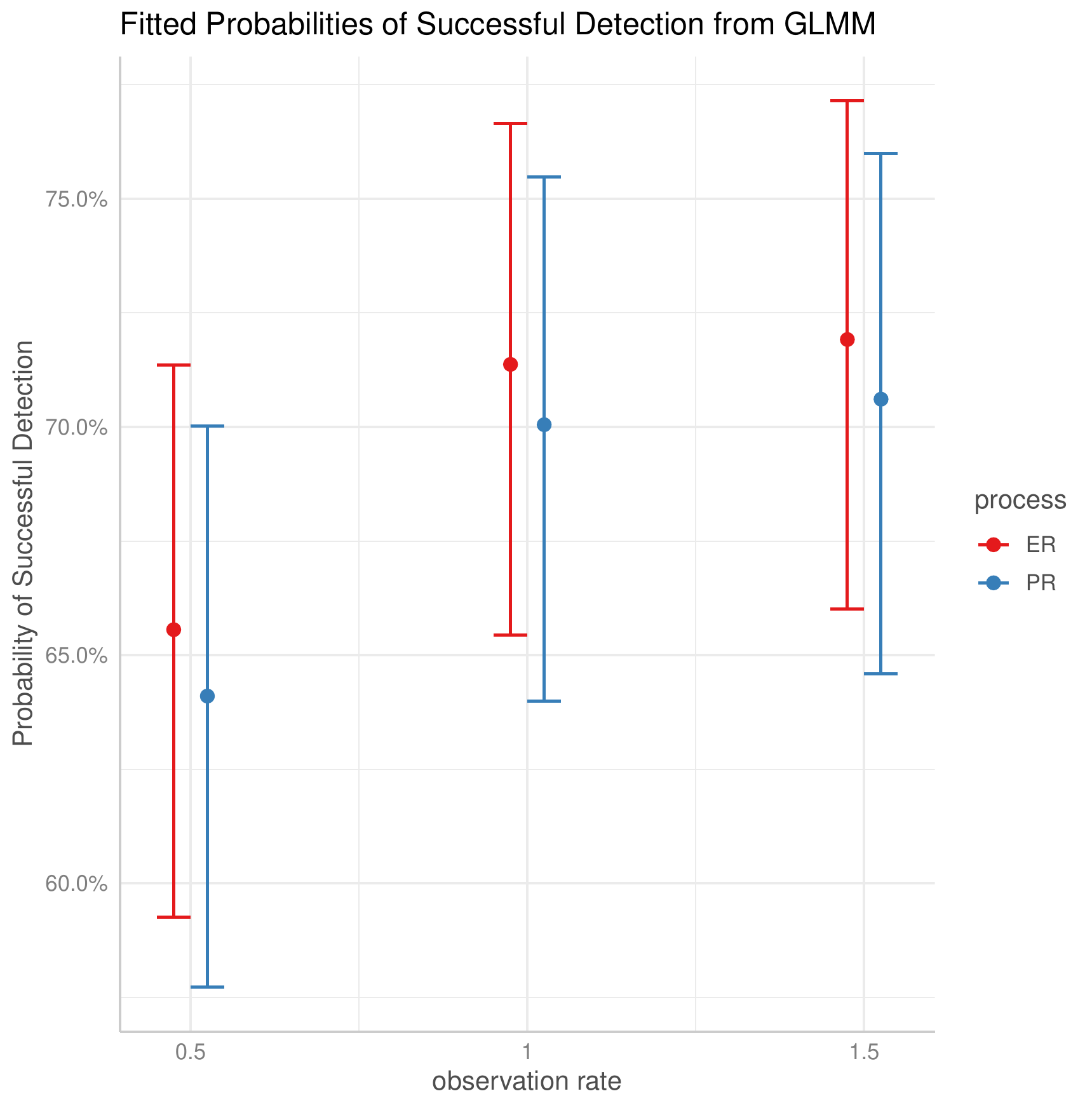}
   \caption{Results for the second simulation study. Left: rates of detection for ER and PR based on 100 replicates. Right: fitted probability of successful detection in GLMM.}
   \label{study2}
   \end{figure}

A second simulation study is designed to evaluate the effect of observation rate $\kappa$ on the rate of detection for ER and PR using Bayes factor. We set $B=500,000$, while keeping the rate of change fixed at 2 ($\gamma=2$). We let $\kappa$ take on values in $\{0.5, 1, 1.5\}$ and $N$ take on values in $\{10, 20, 30\}$. For each combination of settings, we generate 100 ER sequences and 100 PR sequences. We then conduct hypothesis testing on each sequence. We also fit a GLMM to analyze the testing results. The simulation and GLMM results are shown in Figures \ref{study2}, \ref{study2-est} and Table \ref{study2-anova}. We can see that the observation rate significantly affect the rate of detection. The rate of detection significantly increases by increasing $\kappa$ from $0.5$ to $1$, but stops to significantly increase as $\kappa$ further increases.

%\begin{minipage}{\textwidth}
\noindent
  \begin{minipage}[b]{0.49\textwidth}
    \includegraphics[width=0.6\linewidth]{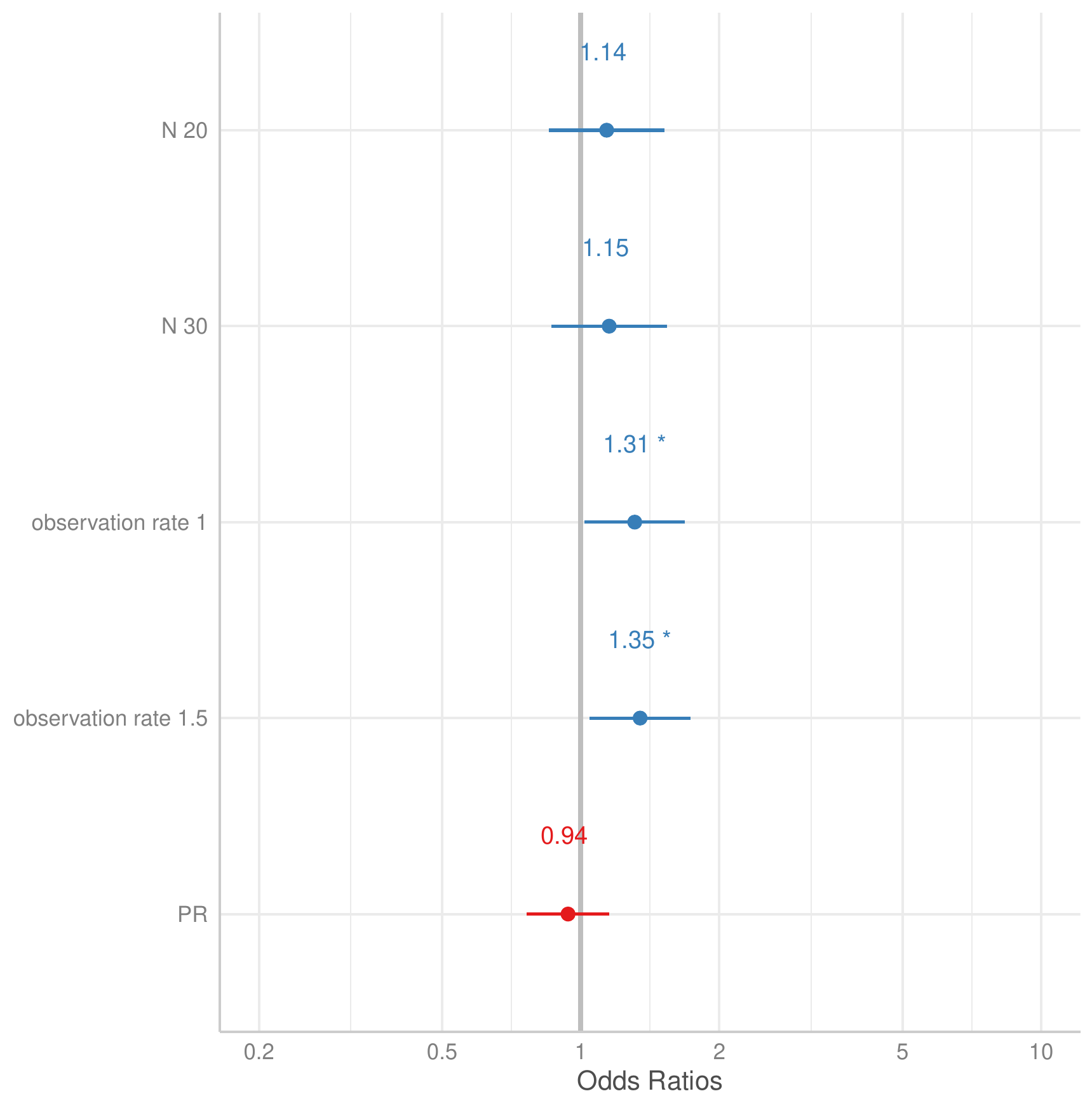}
    \captionof{figure}{Estimates of the coefficients from GLMM in the second simulation study.}
       \label{study2-est}
  \end{minipage}
  %\hfill
  \begin{minipage}[b]{0.49\textwidth}
        \small
        \begin{tabular}{lrrr}
          \hline
         & Chisq & Df & Pr($>$Chisq) \\ 
          \hline
        N & 1.14 & 2 & 0.5668 \\ 
          process & 0.36 & 1 & 0.5458 \\ 
          observation\_rate & 6.61 & 2 & 0.0368 \\ 
          \hline
        \end{tabular}
      \captionof{table}{ANOVA table for the second simulation study}
       \label{study2-anova}
    \end{minipage}
%\vspace{10pt}
%\end{minipage}

\section{Automatic Segment Finder}

% Before we apply our estimation and testing framework to the constructed networks, we need first choose a segment of the network time series where percolation is most likely to occur. The choice of the time series segment is nontrivial and requires finding the stationary period (i.e. the period where the network evolution is reminiscent of percolation process) out of the non-stationary signal. An automatic procedure, shown in Algorithm \ref{al3}, named as automatic segment finder, is proposed for objectively choosing the proper segment for testing. This algorithm involves the following steps:

Before we apply our estimation and testing framework to the constructed networks, we need first choose a segment of the network time series consistent with percolation in connection with different stages of seizure, e.g.  seizure onset and preceding termination. The choice of an appropriate time series segment from the overall non-stationary signal is nontrivial. We developed a simple, automatic procedure for the purpose, shown in Algorithm \ref{al3}. This procedure consists of the following steps:
\begin{enumerate}
    \item Subset the time series by windowing so that it only contains network evolution in the region of interest (ROI).
    \item 
    Split the time series into different low-high-low segments based on GCC/density of the network time series. Each low-high-low segment starts when the size of the GCC/density is below low level and ends when it reaches low level for the first time after reaching high level. The two levels are set to be the first and third quartiles of the size of the GCC/density over time.
    \item For each low-high-low segment, trim off the tail with decreased GCC/density, which is from when it reaches high level for the last time to the end. 
    % \item Choose the segment right after or containing seizure onset for testing.
    \item Use the identified low-high segments for testing.
\end{enumerate}
The purpose of step 2-3 is to automatically identify all segments where both the size of GCC and density ramp up in the ROI. There are several ways proposed to choose the ROIs in a network time series. One way is to choose the segment from 50 s before seizure onset to 50 s after seizure onset. The second way is via the subjective determination of an expert. The third way is through dynamic community detection on network time series (\cite{martinet2020robust}), and to choose the region where the large dynamic community appears. We used the third in our analysis since it is the most objective approach and yields reliable ROIs assessed by epileptologist.  

\begin{algorithm}[!htb]
\SetAlgoLined
\DontPrintSemicolon
\scriptsize
\KwIn{Network time series $\bm g_{1:M}^\star = [g^*(t_1), \cdots, g^*(t_M)]$, ROI $=[t^\prime, t^{\prime\prime}]$, metric on networks: GCC or Density }
\KwOut{A network time series segment $\bm g_{a:b}^\star = [g^*(t_a), g^*(t_{a+1}) \cdots, g^*(t_b)], 1\leq a < b \leq M$}
Initialize: $set = \{\} \text{ an empty set to store segmenting indices}$

$\bm c_{1:M} \in R^M \gets \text{GCC}(\bm g_{1:M}^\star)/\text{Density}(\bm g_{1:M}^\star)$ \Comment*[r]{Calculate size of GCC/density over time}

$l, r \gets \underset{l,r \in 1:M}{\arg\max} (t_r-t_l) \text{ s.t. } t_l,t_r \in $ ROI  \Comment*[r]{Take a subsection of time series}

$Q_1 \gets \text{first quartile of } \bm c_{l:r}$

$Q_3 \gets \text{third quartile of } \bm c_{l:r}$

$anchor \gets l$
\Comment*[r]{Start segmenting time series with sliding window} 

\While{$\bm c_{anchor} > Q_1$}{ $anchor \gets anchor+1$
 \Comment*[r]{Anchor the left point of a potential segment} %at the first time point below $Q_1$}
}
$set = \{set, anchor\}$

\While{not finished segmenting $\bm c_{l:r}$}{
     $i \gets 1$ \\
     \While{$\bm c_{anchor+i} < Q_3 \And anchor+i < r$ }
     {
            $i \gets i+1$
            \Comment*[r]{Grow the segment until it reaches $Q_3$}
     }
          \While{$\bm c_{anchor+i} > Q_1 \And anchor+i < r$}
     {
            $i \gets i+1$
            \Comment*[r]{Keep growing the segment until it reaches $Q_1$}
     }
     
     $anchor \gets anchor + i$
      \Comment*[r]{Anchor the right point of the grown segment} 
      
     $set = \{set, anchor\}$
 }
 
$a, b^\prime \gets \text{a pair of consecutive indices in } set \text{ s.t. } \tilde t \in [a, b^\prime]$ 

$b \gets b^\prime$

\While{$\bm c_b < Q_3$}{ 
    $b \gets b - 1$ \Comment*[r]{Trim off the tail of segment}
} 
\caption{\textsc{Automatic Segment Finder}}
\label{al3}
\end{algorithm}

Note that instead of using either GCC or density to segment the network time series, we can also use both of them to find the proper segment for testing. Specifically, we can first run the Automatic Segment Finder on the network time series with one metric. We can then run it again with the other metric, thus obtaining two segmented network time series. And then we can take, as input, the overlapping of the two segments for testing, which is what we do in this analysis.

\section{Results for Seizure 1, 2 and 3 on the Left Hemisphere}
%%%%%%%%%%%%%%%%
%% Seizure 1 - left
%%%%%%%%%%%%%%%%%
\begin{figure}[H] %13, 8
\centering 
  \includegraphics[width=\textwidth]{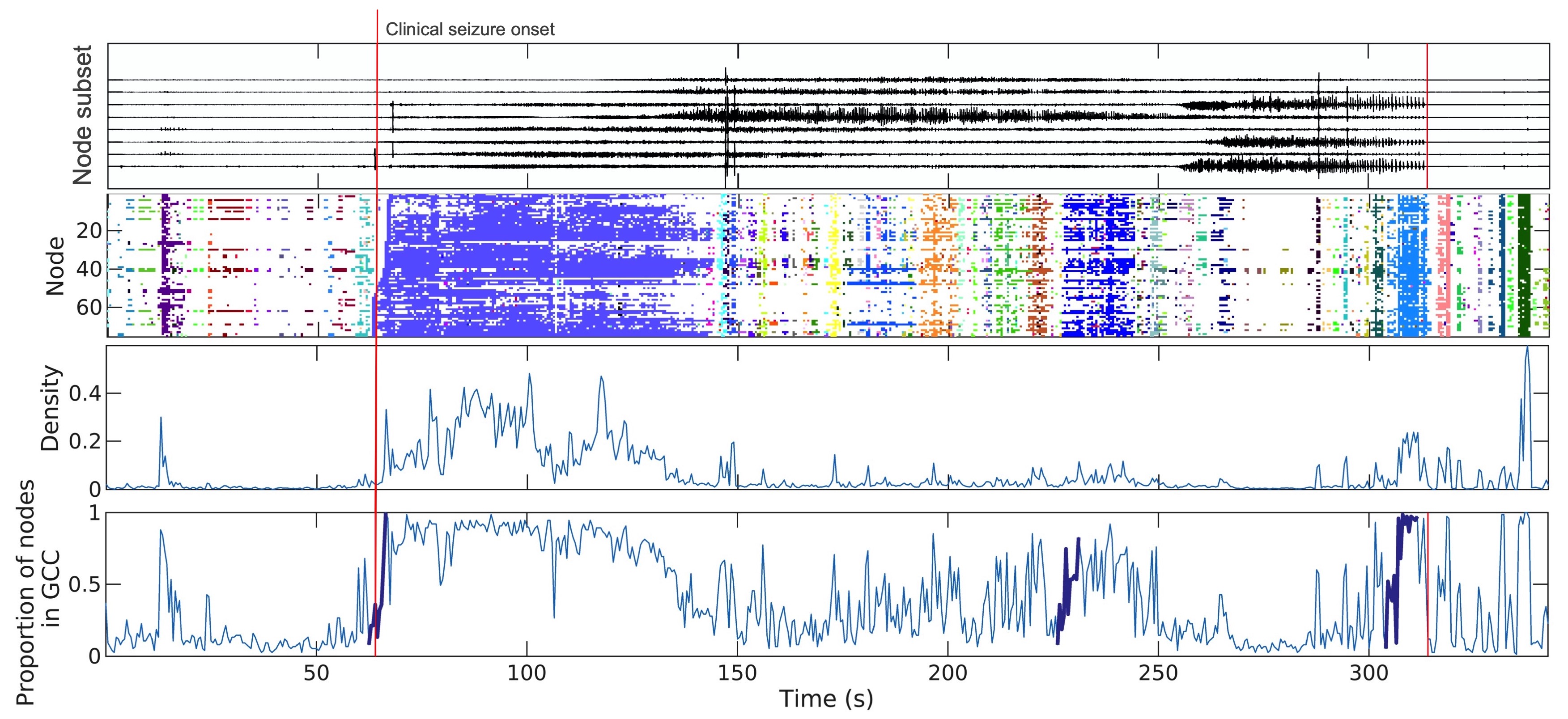}
\caption{Seizure 1: left hemisphere. Two vertical lines represent seizure onset and terminations times, respectively. Top: Voltage time series recorded at eight electrodes on left brain hemisphere. Second row: Community membership (indicated by colors) for each node over time; the large dynamic communities manifest ROIs. Third row: Network density over time. Bottom: Proportion of nodes in the GCC over time; the bold curves in ROIs are chosen by automatic segment finder for testing.}
\label{s1}
\end{figure}

\begin{table}[H]
\centering
\resizebox{\columnwidth}{!}{
\begin{tabular}{lllllll}
\hline
   & $\hat p$         & $\hat q$      & $\hat \gamma$ & $\hat \alpha$   & $\hat \beta$ & $log lik$ \\ \hline
ER &   0.404 (0.112)&  0.248 (0.075)&  10.075 (1.423)&  0.068 (0.0005)& 0.435 (0.035)& -5636.822 (5.942)\\ \hline
PR & 0.468 (0.156)&  0.163 (0.072)&  9.819 (1.512)&  0.068 (0.0004)&  0.464 (0.032)& -5647.120 (6.850)\\ \hline
log(BF) & 10.297 (9.468))  &               &               &                 &               \\ \hline
\end{tabular}
}
\caption{Mean and standard deviation of parameter estimates for ER/PR process, log-likelihood estimates and Bayes factor (log scale) estimate based off of 10 trials for seizure 1 on the left brain hemisphere (first bold segment).}
\label{tab-s1}
\end{table}

\begin{table}[H]
\centering
\resizebox{\columnwidth}{!}{
\begin{tabular}{lllllll}
\hline
   & $\hat p$         & $\hat q$      & $\hat \gamma$ & $\hat \alpha$   & $\hat \beta$ & $log lik$ \\ \hline
ER &  0.580 (0.111)&  0.279 (0.110)&  6.480 (0.823)&  0.033 (0.0004)&  0.273 (0.047)& -4222.710 (15.563)\\ \hline
PR &  0.625 (0.092)&  0.250 (0.134)&  6.010 (0.348)&  0.033 (0.0004)&  0.336 (0.040)& -4254.414 (11.118)\\ \hline
log(BF) & 31.704 (15.141) &               &               &                 &               \\ \hline
\end{tabular}
}
\caption{Mean and standard deviation of parameter estimates for ER/PR process, log-likelihood estimates and Bayes factor (log scale) estimate based off of 10 trials for seizure 1 on the left brain hemisphere (second bold segment).}
\label{tab-s1-2}
\end{table}

\begin{table}[H]
\centering
\resizebox{\columnwidth}{!}{
\begin{tabular}{lllllll}
\hline
   & $\hat p$         & $\hat q$      & $\hat \gamma$ & $\hat \alpha$   & $\hat \beta$ & $log lik$ \\ \hline
ER &  0.647 (0.051)&  0.138 (0.055)&  10.037 (0.631)& 0.128 (0.0007)&  0.214 (0.024)& -16058.532 (24.298)\\ \hline
PR & 0.642 (0.037)&  0.141 (0.078)&  7.567 (1.209)&  0.129 (0.0004)&  0.228 (0.026)& -16120.504 (26.966)\\ \hline
log(BF) &61.971 (33.687)  &               &               &                 &               \\ \hline
\end{tabular}
}
\caption{Mean and standard deviation of parameter estimates for ER/PR process, log-likelihood estimates and Bayes factor (log scale) estimate based off of 10 trials for seizure 1 on the left brain hemisphere (third bold segment).}
\label{tab-s1-3}
\end{table}

%%%%%%%%%%%%%%%%
%% Seizure 2 - left
%%%%%%%%%%%%%%%%%
\begin{figure}[H] %13, 8
\centering 
  \includegraphics[width=\textwidth]{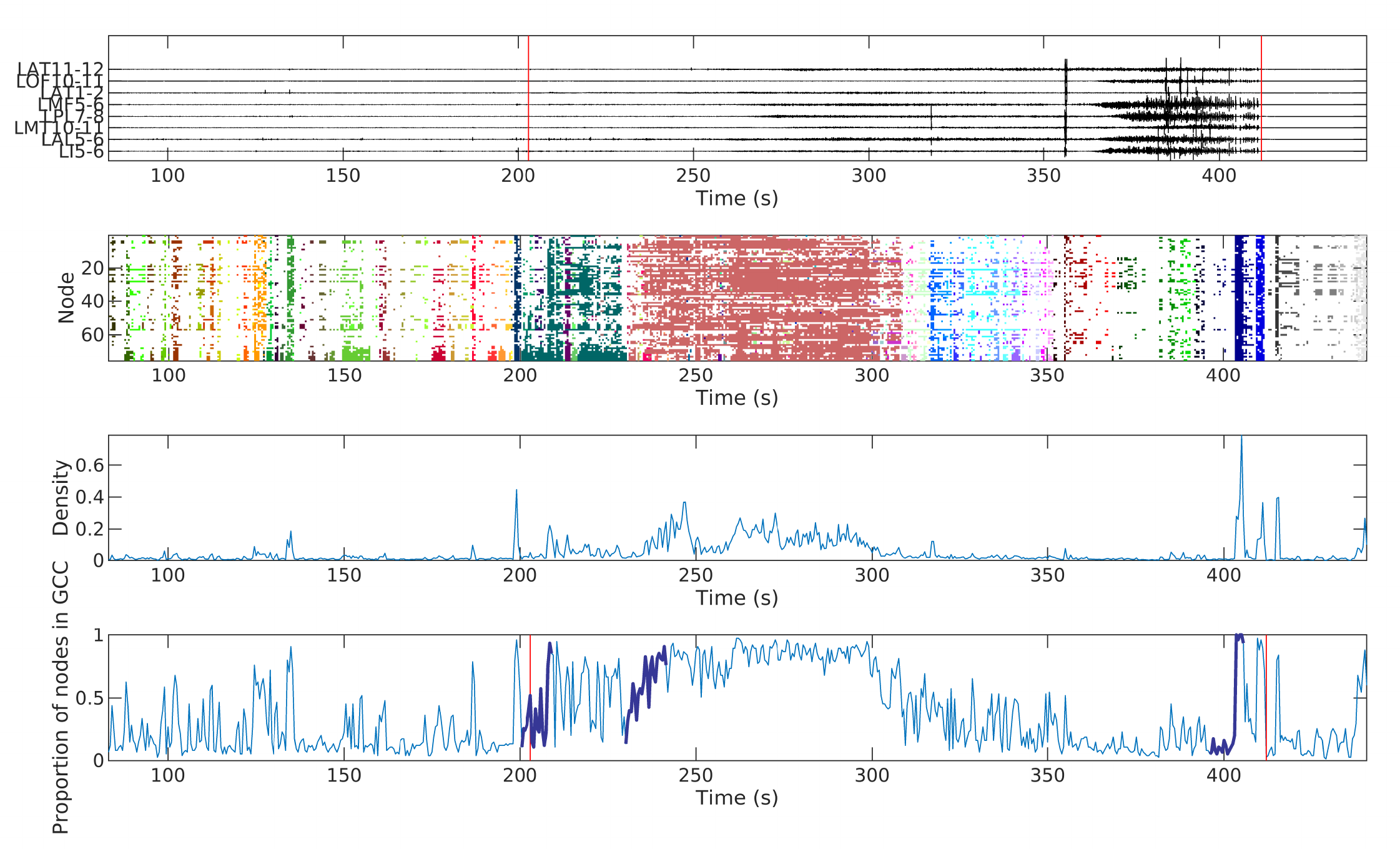}
\caption{Seizure 2: left hemisphere. Two vertical lines represent seizure onset and terminations times, respectively. Top: Voltage time series recorded at eight electrodes on left brain hemisphere. Second row: Community membership (indicated by colors) for each node over time; the large dynamic communities manifest ROIs. Third row: Network density over time. Bottom: Proportion of nodes in the GCC over time; the bold curves in ROIs are chosen by automatic segment finder for testing.}
\label{s2}
\end{figure}

\begin{table}[H]
\centering
\resizebox{\columnwidth}{!}{
\begin{tabular}{lllllll}
\hline
   & $\hat p$         & $\hat q$      & $\hat \gamma$ & $\hat \alpha$   & $\hat \beta$ & $log lik$ \\ \hline
ER &   0.573 (0.051)&  0.269 (0.075)&  6.041 (0.223)&  0.047 (0.0004)&  0.262 (0.019)& -9226.559 (25.313)\\ \hline
PR &  0.587 (0.057)&  0.288 (0.103)&  5.732 (0.242)&  0.047 (0.0001)&  0.279 (0.020)& -9277.567 (21.869)\\ \hline
log(BF) & 51.008 (31.198)  &               &               &                 &               \\ \hline
\end{tabular}
}
\caption{Mean and standard deviation of parameter estimates for ER/PR process, log-likelihood estimates and Bayes factor (log scale) estimate based off of 10 trials for seizure 2 on the left brain hemisphere (first bold segment).}
\label{tab-s2}
\end{table}

\begin{table}[H]
\centering
\resizebox{\columnwidth}{!}{
\begin{tabular}{lllllll}
\hline
   & $\hat p$         & $\hat q$      & $\hat \gamma$ & $\hat \alpha$   & $\hat \beta$ & $log lik$ \\ \hline
ER &   0.522 (0.049)& 0.258 (0.067)&  8.846 (0.596)&  0.081 (0.0007)&  0.354 (0.031)& -18375.956 (52.389)\\ \hline
PR &  0.545 (0.068)&  0.277 (0.088)&  7.159 (0.852)&  0.081 (0.0009)&  0.410 (0.037)& -18481.718 (40.650)\\ \hline
log(BF) & 105.762 (73.922)  &               &               &                 &               \\ \hline
\end{tabular}
}
\caption{Mean and standard deviation of parameter estimates for ER/PR process, log-likelihood estimates and Bayes factor (log scale) estimate based off of 10 trials for seizure 2 on the left brain hemisphere (second bold segment).}
\label{tab-s2-2}
\end{table}

\begin{table}[H]
\centering
\resizebox{\columnwidth}{!}{
\begin{tabular}{lllllll}
\hline
   & $\hat p$         & $\hat q$      & $\hat \gamma$ & $\hat \alpha$   & $\hat \beta$ & $log lik$ \\ \hline
ER &   0.533 (0.026)&  0.141(0.038)&  8.292(0.548)&  0.099 (0.0001)&  0.150 (0.021)& -16234.315 (13.354)\\ \hline
PR & 0.575 (0.079)&  0.174 (0.082)& 7.783 (0.568)&  0.099 (0.0001)&  0.177 (0.025)& -16241.436 (15.226)\\ \hline
log(BF) & 7.119 (19.931)  &               &               &                 &               \\ \hline
\end{tabular}
}
\caption{Mean and standard deviation of parameter estimates for ER/PR process, log-likelihood estimates and Bayes factor (log scale) estimate based off of 10 trials for seizure 2 on the left brain hemisphere (third bold segment).}
\label{tab-s2-3}
\end{table}

%%%%%%%%%%%%%%%%
%% Seizure 3 - left
%%%%%%%%%%%%%%%%%
\begin{figure}[H] %13, 8
\centering 
 \includegraphics[width=\textwidth]{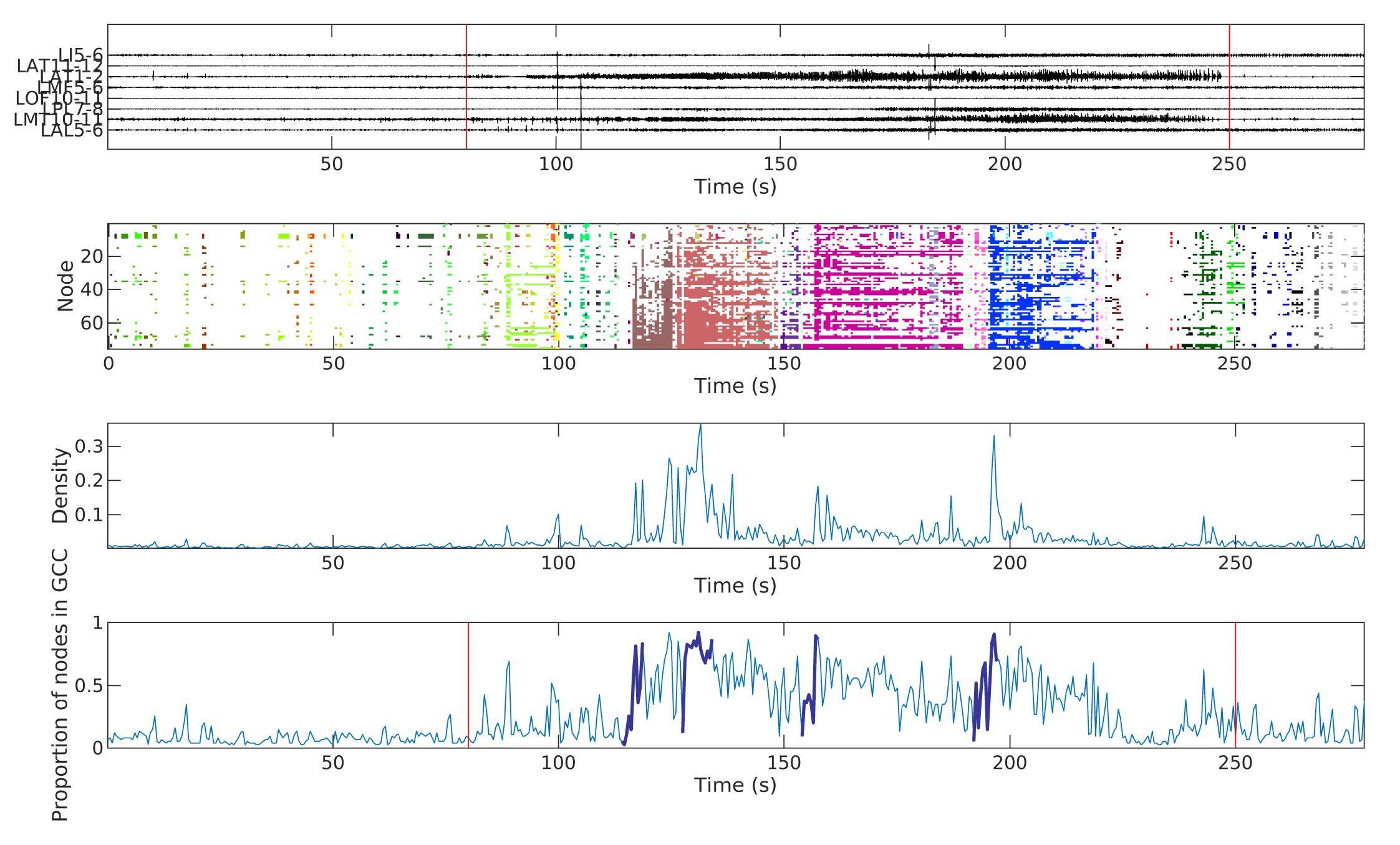}
\caption{Seizure 3: left hemisphere. Two vertical lines represent seizure onset and terminations times, respectively. Top: Voltage time series recorded at eight electrodes on left brain hemisphere. Second row: Community membership (indicated by colors) for each node over time; the large dynamic communities manifest ROIs. Third row: Network density over time. Bottom: Proportion of nodes in the GCC over time; the bold curves in ROIs are chosen by automatic segment finder for testing.}
\label{s3}
\end{figure}

\begin{table}[H]
\centering
\resizebox{\columnwidth}{!}{
\begin{tabular}{lllllll}
\hline
   & $\hat p$         & $\hat q$      & $\hat \gamma$ & $\hat \alpha$   & $\hat \beta$ & $log lik$ \\ \hline
ER &   0.651 (0.042) &  0.166 (0.084) & 6.367 (0.625) & 0.055 (0.0003) & 0.200 (0.027) & -5388.566 (9.834)\\ \hline
PR &  0.665 (0.040) &  0.128 (0.079) &  5.722 (0.341) &  0.055 (0.0002) &  0.225 (0.036) & -5399.512 (7.227)\\ \hline
log(BF) &  10.945 (12.786)  &               &               &                 &               \\ \hline
\end{tabular}
}
\caption{Mean and standard deviation of parameter estimates for ER/PR process, log-likelihood estimates and Bayes factor (log scale) estimate based off of 10 trials for seizure 3 on the left brain hemisphere (first bold segment).}
\label{tab-s3}
\end{table}

\begin{table}[H]
\centering
\resizebox{\columnwidth}{!}{
\begin{tabular}{lllllll}
\hline
   & $\hat p$         & $\hat q$      & $\hat \gamma$ & $\hat \alpha$   & $\hat \beta$ & $log lik$ \\ \hline
ER &   0.824 (0.064) & 0.091 (0.036) & 12.746 (1.522) & 0.198 (0.0006) & 0.268 (0.023) & -18120.573 (25.416)\\ \hline
PR &  0.682 (0.112) & 0.134 (0.077) & 7.442 (2.783) & 0.201 (0.0007) & 0.219 (0.071) & -18182.074 (22.634)\\ \hline
log(BF) &  61.503 (25.172)  &               &               &                 &               \\ \hline
\end{tabular}
}
\caption{Mean and standard deviation of parameter estimates for ER/PR process, log-likelihood estimates and Bayes factor (log scale) estimate based off of 10 trials for seizure 3 on the left brain hemisphere (second bold segment).}
\label{tab-s3-2}
\end{table}

\begin{table}[H]
\centering
\resizebox{\columnwidth}{!}{
\begin{tabular}{lllllll}
\hline
   & $\hat p$         & $\hat q$      & $\hat \gamma$ & $\hat \alpha$   & $\hat \beta$ & $log lik$ \\ \hline
ER &   0.462 (0.078)&  0.217(0.125&  7.629 (0.643)&  0.053 (0.0003)&  0.185 (0.027)& -4102.365 (17.419)\\ \hline
PR &  0.535 (0.126)&  0.241(0.151)& 6.961 (0.616)&  0.053 (0.0004) & 0.259 (0.070)& -4117.433 (5.860)\\ \hline
log(BF) &  15.068 (20.097)  &               &               &                 &               \\ \hline
\end{tabular}
}
\caption{Mean and standard deviation of parameter estimates for ER/PR process, log-likelihood estimates and Bayes factor (log scale) estimate based off of 10 trials for seizure 3 on the left brain hemisphere (third bold segment).}
\label{tab-s3-3}
\end{table}

\begin{table}[H]
\centering
\resizebox{\columnwidth}{!}{
\begin{tabular}{lllllll}
\hline
   & $\hat p$         & $\hat q$      & $\hat \gamma$ & $\hat \alpha$   & $\hat \beta$ & $log lik$ \\ \hline
ER &   0.629 (0.091)& 0.154 (0.070)& 7.700 (0.586)& 0.076 (0.0003)&  0.179 (0.059)& -6836.452 (6.823)\\ \hline
PR &  0.674 (0.035)& 0.119 (0.051)& 6.660 (0.569)& 0.076 (0.0003)&  0.248 (0.031)& -6831.450 (7.429)\\ \hline
log(BF) &  -5.002 (7.224)  &               &               &                 &               \\ \hline
\end{tabular}
}
\caption{Mean and standard deviation of parameter estimates for ER/PR process, log-likelihood estimates and Bayes factor (log scale) estimate based off of 10 trials for seizure 3 on the left brain hemisphere (fourth bold segment).}
\label{tab-s3-4}
\end{table}

\bibliographystyle{agsm}
\bibliography{main_supp}